  \documentclass{ucalgarythesis}
  
  \usepackage[utf8]{inputenc}
  \usepackage{amssymb,amsthm,amsmath}
  \usepackage[hidelinks]{hyperref}
  \usepackage[usenames,dvipsnames]{color}
  
  \usepackage{graphicx}
  \usepackage{graphics}
  \usepackage{caption}
  \usepackage{subcaption}
  \usepackage{svg}
  \usepackage{url}
  \usepackage{siunitx}
  \usepackage[numbers,sort&compress]{natbib}
  \usepackage{multirow}
  \usepackage{listings}
  \usepackage{floatrow}

  \theoremstyle{plain}

  \theoremstyle{definition}

  \usepackage{tikz}

\newcommand{\doubleH}{%
  \mathrel{\text{%
    \tikz[baseline=-0.5ex] \draw[line width=0.15ex] (0,-0.5ex) -- (1ex,-0.5ex) (0,0.5ex) -- (1ex,0.5ex) (0,1ex) -- (0,-1ex) (1ex,1ex) -- (1ex,-1ex);
  }}%
}

\begin{document}



  \title{
   Realization of an electromechanical nano-string device
   \\ \bigskip
   }
   
  \author{Armin Tabesh}
  \thesis{Thesis}
  \dept{Graduate Program in Physics and Astronomy}
  \degree{Master of Science}
  \gradyear{2023}
  \monthname{SEPTEMBER}

  \frontmatter           
  \makethesistitle       



  \begin{thesisabstract}  
    Electromechanics is the field of studying the interaction between microwave resonators and mechanical oscillators. It has been an interesting topic in the recent decade due to its numerous potential applications in science and technology, including ground-state cooling of macroscopic objects, quantum sensing, quantum memory, and quantum transduction. This thesis presents a comprehensive analysis of a project focused on simulating, designing, and modeling electromechanical devices with the ultimate objective of achieving their successful implementation.
    Through the thesis, after an overview of the theoretical model of electromechanics, I will introduce our design for the device and explain how we simulated it to optimize its characteristics. Next, I will discuss the nanofabrication process we have developed for the device, along with the fundamental aspects of the characterization method and setup. Subsequently, I will present the theoretical model I have developed based on electromechanics. This model has significant potential to open up new avenues for future research, building upon the foundation laid by the current project.

  \end{thesisabstract}



  \chapter{Preface}
    In January 2021, amidst the pandemic, I joined Dr. Barzanjeh's group as a master's student. At that time, the group was in its early stages, and I was the first student to join the team. This unique opportunity allowed me to establish the laboratory from the ground up, involving tasks such as installing equipment and purchasing devices. I took on various responsibilities, including building RF cables, wiring the dilution fridge, configuring electronics for microwave measurements, and fine-tuning optical tables, among other crucial duties.
    
My research project focused on the theoretical exploration of tripartite opto-electro-mechanical systems, involving the design and simulation of an electromechanical device. Additionally, there was a potential plan to fabricate and establish a measurement setup to characterize the device in the future.
To start my research, I conducted an extensive review of relevant literature, gaining a deep understanding of the theoretical model essential for the project. Thereafter, I started learning the fundamentals of COMSOL, familiarizing myself with the simulation of optical, optomechanical, and electromechanical systems.
Once the electromechanical devices were designed, I performed simulations of their structure, successfully extracting crucial electromechanical characteristics. Subsequently, I devoted myself to learning fabrication techniques, intending to independently fabricate the device. With the assistance of Dr. Trong Ngo (postdoc), we collaborated to develop nanofabrication recipes tailored for fabricating nanostring and mechanical resonators. 
    
During my Master's program, I have been involved in other related projects. I simulated a photonic crystal made out of Yttrium Iron Garnet (YIG) and optimized the geometry using COMSOL LiveLink with MATLAB to find a set of geometrical parameters that yield a mode frequency in a desired range with a high quality factor. The photonic crystal has been fabricated by a collaborator, and its characterization is still an ongoing project in our lab. The significance of this project relies on the magnetic properties of YIG. We are aiming to achieve a tripartite opto-magno-mechanical interaction out of this project.\\
Furthermore, I developed the theoretical model of a hybrid system comprising microwave, magnonic, and mechanical subsystems. The main idea behind this project was to investigate the impact of coupling a magnonic subsystem with an electromechanical system, leading to the generation of photon-magnon entanglement.

Apart from theoretical investigations, I took an active role in building and advancing microwave measurement setups, which have been extensively utilized for various projects within the group. For instance, in collaboration with Prof. Can-Ming Hu, we were tasked with characterizing a tunable microwave isolator at cryogenic temperatures. I was responsible for preparing the measurement setup and conducting the essential measurements. The result of this work as a paper is currently under review. 

This thesis mainly represents my research on the electromechanical device. I have tried to gather the information and knowledge I gained through my research (regarding theory, simulation, fabrication, and characterization methods) and present it in an effective manner, so it can help a non-familiar reader with a basic knowledge of quantum optics with learning the material and possibly continuing the research, in a shorter time and with much less effort than I made.\\
The thesis does not cover optical and optomechanical analysis/simulation that I have done along the main topic of my research, but since it is relevant to electromechanical systems, the last chapter of the thesis represents my theoretical analysis of tripartite electro-magno-mechanical systems.
    


  \chapter{Acknowledgements}  
  A master’s program in physics is not a straightforward journey, particularly if its commencement is accompanied by experiencing living abroad for the first time, in the middle of a pandemic. Now that I am close to the end of this journey, I look back to the path I paved and realize I could never be here without the support of all the individuals who helped and supported me.
  
    First of all, I thank Shabir, who believed in me and gave me the opportunity of learning. He provided me with ample resources I needed to nurture and grow and taught me what a real researcher is like.
    
    Before starting graduate studies, to qualify for a master’s student position in the quantum field, I had to build a solid background in physics and quantum mechanics/science, which was not possible without the support and help of Dr. Ali Sadeghi and Dr. Shant Shahbazian. They taught younger me how to be a good physics student and how to look at the world from a physicist’s perspective.
    
    Everyone who has experienced nanofabrication understands its complexity and knows that without good mentors, learning the methods and fabricating devices can be a tedious job full of blind trials and errors. I appreciate the presence of Gustavo de Oliviera Luiz, Scott Munro, Elham Zohari, and Trong Ngo in nanoFAB as they undeniably taught me a great deal about nanofabrication.
     
    Completing a master’s degree is not only studying and conducting research. A graduate student usually deals with administrative processes which can be disturbing, time-consuming, and occasionally stressful. I thank Jo-Anne Brown, Marni Farrant, Yanmei Fei, and Naomi Tanner for all their assistance and guidance as they significantly facilitated these procedures.
    
    I have been so lucky to have kind and gracious labmates, who made working in the lab an enjoyable experience. I thank each and every one of them for making a friendly and supportive environment and a close-knit family.
    
    Living and studying in Canada, while my parents, Mehri and Mohammad Ali, and my, brother, Aryan, live on the other side of the world has been an extreme emotional challenge. Although it has been more difficult for my family, they have always fully supported me, from the day I made my decision until now. I am grateful to have such an incredible supportive family.
    
    And finally, the only person who has accompanied me throughout the whole journey, in ease and difficulty, in hopefulness and hopelessness, and had my back whenever I was in the lowest, is my beloved wife, Maryam. I have been and always will be grateful for having such a wonderful partner in my life.

    Today, after a long way full of toughnesses, I find myself at the peak and appreciate all the supports I received to get here. I am staring at my next destination, a higher peak, which appears much more challenging. Deep inside, I am confident that I can make it, with having these incredible supporters by my side.


  \chapter[Dedication]{}
  \begin{dedication}
     \emph{To curiosity and ambition.}
  \end{dedication}
  

  \begin{singlespace}   

  \renewcommand\contentsname{Table of Contents}
  \cleardoublepage\phantomsection
  \addcontentsline{toc}{chapter}{\contentsname}
  \tableofcontents

  \renewcommand{\listfigurename}{List of Figures and Illustrations}
  \cleardoublepage\phantomsection
  \addcontentsline{toc}{chapter}{\listfigurename}
  \listoffigures

  \renewcommand{\listtablename}{List of Tables}
  \cleardoublepage\phantomsection
  \addcontentsline{toc}{chapter}{\listtablename}
  \listoftables
  
  \chapter{List of Symbols, Abbreviations and Nomenclature}      
  \begin{tabbing}
    Symbol or abbreviation~~~~~\= \ \ \ \ \   \parbox{8in}{Definition}\\

    \addsymbol \mbox{$\omega_c$}: {The mode (angular) frequency of cavity}
    \addsymbol \mbox{$\omega_p$}: {The (angular) frequency of externally applied pump}
    \addsymbol \mbox{$\Delta = \omega_c-\omega_p$}: {Pump detuning with respect to the cavity frequency}
    \addsymbol \mbox{$\kappa_\text{in}$}: {Intrinsic damping rate of cavity}
    \addsymbol \mbox{$\kappa_\text{ex}$}: {Extrinsic damping rate of cavity}
    \addsymbol \mbox{$\kappa = \kappa_\text{in}+\kappa_\text{ex}$}: {Total damping rate of cavity}
    \addsymbol \mbox{$\Omega$}: {The mode (angular) frequency of mechanical oscillator}
    \addsymbol \mbox{$\gamma$}: {Damping rate of the mechanical oscillator}
    \addsymbol \mbox{$m_\text{eff}$}: {Effective mass of the mechanical oscillator}
    \addsymbol \mbox{$x_\text{ZPF}$}: {Zero-point fluctuations of the mechanical oscillator}
    \addsymbol \mbox{$\hat{a}$}: {Photon annihilation operator}
    \addsymbol \mbox{$\hat{b}$}: {Phonon annihilation operator}
    \addsymbol \mbox{$g_0$}: {Single-photon opto/electromechanical coupling rate}
    \addsymbol \mbox{$\bar{n}_c$}: {Average number of photons in the cavity ($\bar{n}_c = \langle\hat{a}^\dagger\hat{a}\rangle$)}
    \addsymbol \mbox{$g = \sqrt{\bar{n}_c}g_0$}: {Enhanced optomechanical coupling rate}
    \addsymbol \mbox{$\mathcal{C}_0 = g_0^2/\kappa\gamma$}: {Single-photon cooperativity}
    \addsymbol \mbox{$\mathcal{C} = \bar{n}_c\mathcal{C}_0 = g^2/\kappa\gamma$}: {Enhanced cooperativity}
    \addsymbol \mbox{$\chi_a(\omega)$}: {The inverse of the free microwave/optical susceptibility}
    \addsymbol \mbox{$\chi_b(\omega)$}: {The inverse of the free mechanical susceptibility}
    \addsymbol \mbox{$\chi_{b, \text{opt}}(\omega)$}: {The correction to the inverse of the free microwave/optical susceptibility, introduced by the optomechanical interaction}
    \addsymbol \mbox{$\gamma_\text{opt}$}: {Optomechanical damping rate}
    \addsymbol \mbox{$\vec{E}(\vec{r})$}: {Electric field at the position $\vec{r}$}
    \addsymbol \mbox{$\vec{D}(\vec{r})$}: {Electric displacement field at the position $\vec{r}$}
    \addsymbol \mbox{$\varepsilon(\vec{r})$}: {Electric permittivity at the position $\vec{r}$}
    \addsymbol \mbox{$\vec{Q}(\vec{r})$}: {Spatial displacement of the structure at the position $\vec{r}$}
    \addsymbol \mbox{$L$}: {Total inductance of the microwave resonator}
    \addsymbol \mbox{$C_m$}: {Capacitance of the capacitor with the oscillating plate}
    \addsymbol \mbox{$C_s$}: {Stray capacitance in the microwave resonator}

  \end{tabbing}
  
  \end{singlespace}     
  
%

  \mainmatter           
    
  
  \chapter{Introduction}
\label{chp:Introduction}

An electromagnetic wave possesses momentum \cite{serway2004modern, hecht2012optics}, that enables the wave to transfer a portion of its momentum to objects that interacts with, particularly when it undergoes reflection or scattering. When an electromagnetic wave impinges upon the surface of an object, it exerts a force that manifests as pressure on the object's surface. This phenomenon is commonly referred to as radiation pressure which plays a crucial role in the interaction between a cavity that confines an electromagnetic field and a mechanical object or a mechanical oscillator \cite{cavityOptomechanics}.

\begin{figure}[!h]
    \centering
     \begin{subfigure}[b]{0.47\textwidth}
         \centering
         \includegraphics[width=\textwidth]{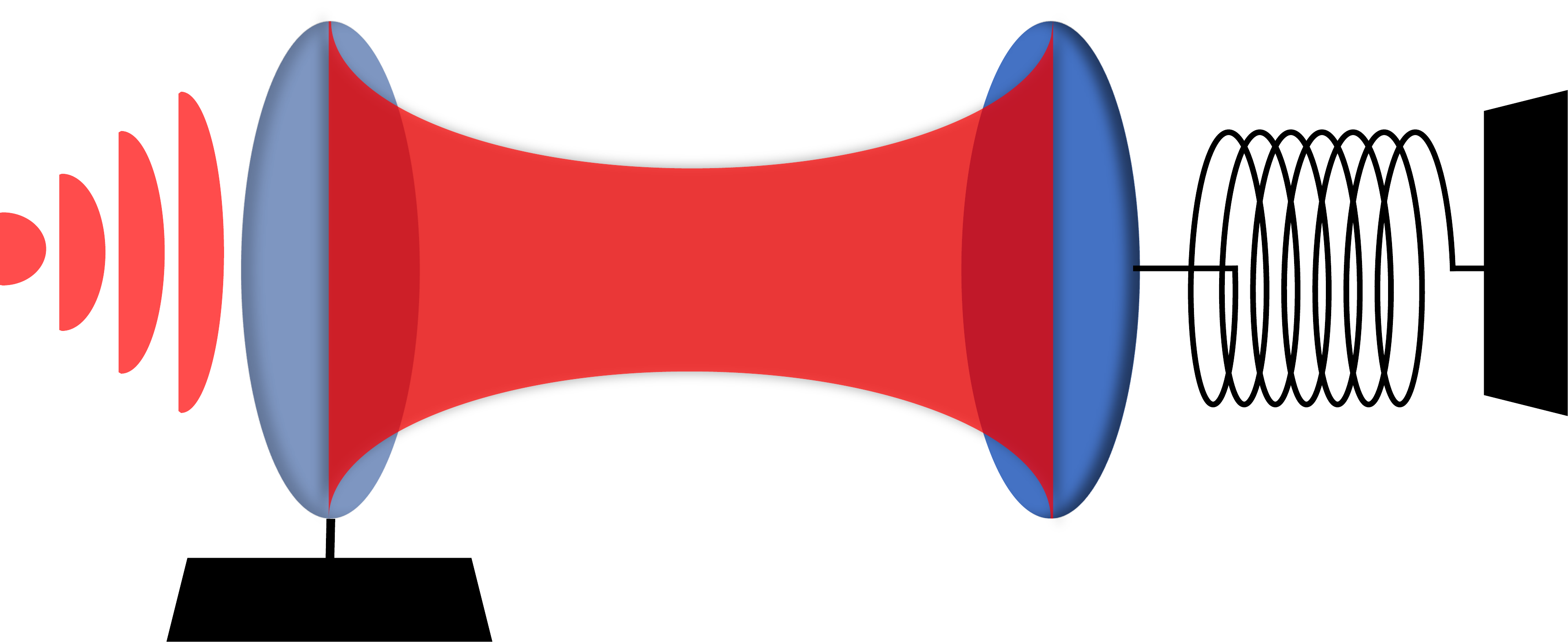}
         \caption{}
         \label{fig:optomechanics_schematic}
     \end{subfigure}
     \hfill
     \begin{subfigure}[b]{0.47\textwidth}
         \centering
         \includegraphics[width=\textwidth]{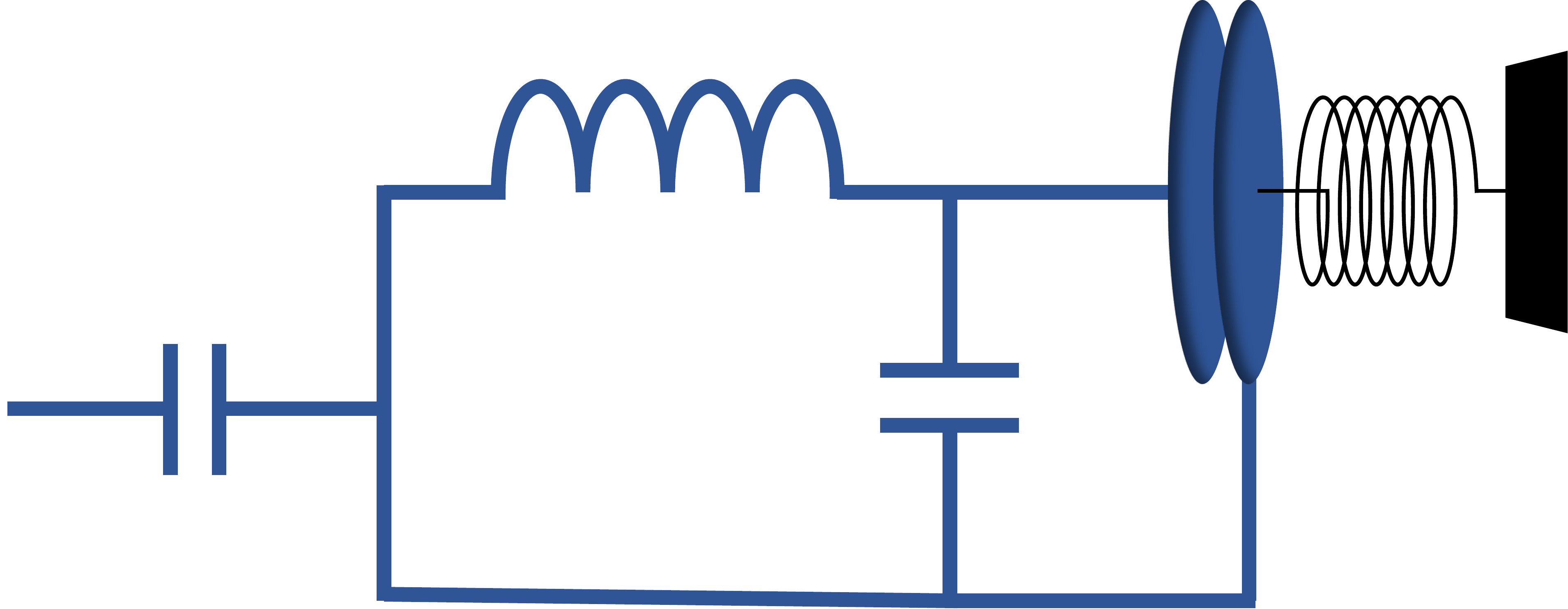}
         \caption{}
         \label{fig:electromechanics_schematic}
     \end{subfigure}
     \caption[Optomechanics and electromechanics schemes]{(a) An optical cavity with one partially reflective fixed and one perfectly reflective oscillating mirror makes a cavity optomechanical system. In such a system, the optical field trapped in the cavity can interact with the mechanical motion of the oscillating plate, due to radiation pressure. (b) An LC circuit (microwave resonator) connected to a capacitor with one oscillating plate makes a cavity electromechanical system. In such a system, the microwave field trapped in the resonator can interact with the mechanical motion of the oscillating plate, due to radiation pressure.}
     \label{fig:opto/electro-mechanics_scheme}
\end{figure}

To clarify further, let us consider a Fabry-Pérot cavity arrangement comprising a partially reflective mirror that remains fixed and a perfect mirror capable of oscillating around its resting position (see Figure \ref{fig:optomechanics_schematic}). This configuration encompasses an optical resonator, known as the cavity, and a mechanical oscillator represented by the oscillating mirror. The frequency of the optical mode within the cavity depends on its length, a characteristic that can be modified by the mirror's oscillatory motion. Now, let us imagine introducing an external optical field (pump laser) into the cavity through the partially reflective mirror. This injected field exerts a force on the mirror, resulting from the radiation pressure it experiences. Consequently, the mirror undergoes oscillations around its equilibrium position. These oscillations affect the electromagnetic fields present within the cavity by modifying the mode frequency of the cavity. 

This dynamic interplay between the mirror's oscillations and the electromagnetic fields is commonly known as optomechanical interaction \cite{cavityOptomechanics}. Such an interaction, primarily reliant on radiation pressure, can manifest independently of the cavity's presence. However, the cavity significantly alters this interaction by confining and capturing the electromagnetic field between its mirrors. By incorporating the cavity, the optomechanical interaction experiences significant enhancements, yielding intriguing possibilities such as amplified nonlinearity \cite{Burgwal2023, Shahidani:14} or improved frequency conversion capabilities \cite{arnold2020converting}.

The term `optics' typically refers to the study of electromagnetic fields within the frequency range of a few hundred Tera-Hertz (THz), for example around $200$ THz in the telecommunications band. A similar interaction to optomechanics exists for electromagnetic fields in the microwave frequency range, approximately $10$ GHz. The term `optomechanics' is sometimes used to describe this interaction in the microwave domain since the theoretical models for both frequency regimes are largely similar. However, to provide greater clarity and specificity, the term `electromechanics' is often employed when discussing microwave fields, given that microwave electromagnetic fields are typically generated and manipulated using electrical circuits \cite{pozar2011microwave, montgomery1987principles}.

To gain a better understanding of electromechanical systems, let us consider a microwave resonator, which can be explained by an LC circuit (as shown in Figure \ref{fig:electromechanics_schematic}). An LC circuit can function as a microwave resonator capable of storing microwave electromagnetic fields at a specific frequency. If we connect a capacitor with one of its plates capable of oscillating (thus serving as a mechanical oscillator) to the circuit, we create a resonant electromechanical system. In this setup, the electromagnetic field induced within the circuit exerts a force on the oscillating plate of the capacitor. Consequently, the plate undergoes oscillatory motion, thereby modulating the capacitance of the circuit and influencing its mode frequency. Once again, we observe a reciprocal interaction between the electromagnetic field and the mechanical oscillator. However, it is important to note that the frequency of the electromagnetic field and the implementation of the systems differ in the case of electromechanics compared to optomechanics.

\section{Electromechanics versus optomechanics}
As we will see in detail in the next chapter, according to the quantum mechanical model that describes a system of an electromagnetic cavity (either optical or microwave) and a mechanical oscillator, the term indicating the linear electromechanical or optomechanical interaction reads \cite{cavityOptomechanics},
\begin{equation}
    \hat{\mathcal{H}}_\text{int} = \hbar g_0 \sqrt{\bar{n}_c} \left(\hat{a} + \hat{a}^\dagger\right)\left(\hat{b} + \hat{b}^\dagger\right),
    \label{eq:interaction-Hamiltonian}
\end{equation}
where $\hbar$ is the reduced Planck's constant, $g_0$ is called the single-photon coupling rate, $\bar{n}_c$ is the average number of photons in the cavity that are injected in by an external pump, and $\hat{a}$ and $\hat{b}$ are the annihilation operators of the cavity and the mechanical resonators, respectively.
We will show, in the analysis, that by tuning the external pump to the frequency $\omega_p = \omega_c - \Omega$ with $\omega_c$ and $\Omega$ being the angular frequencies of the cavity (or resonator) and the mechanical oscillator, under specific conditions including resolved side-bands condition which we will define later, the Hamiltonian can be written in the form of a beam-splitter interaction given by
\begin{equation}
    \hat{\mathcal{H}}_\text{int} \simeq \hbar g_0\sqrt{\bar{n}_c} \left(\hat{a}\hat{b}^\dagger + \hat{a}^\dagger\hat{b}\right).
\end{equation}
An external pump with such tuning is called a red-detuned pump (as the frequency is below the frequency of the cavity), and the interaction describes photon-phonon exchange: annihilating a phonon while creating a photon with the same quantum properties, and vice versa, which can be used for photon transduction.\\
In contrast, under the same conditions, by tuning the external pump to $\omega_p = \omega_c+\Omega$, the interaction Hamiltonian can be approximated by,
\begin{equation}
    \hat{\mathcal{H}}_\text{int} \simeq \hbar g_0 \sqrt{\bar{n}_c}\left(\hat{a}\hat{b} + \hat{a}^\dagger\hat{b}^\dagger\right).
\end{equation}
The pump used in this context is referred to as a blue-detuned pump, meaning its frequency is higher than that of the cavity. This specific choice of pump frequency enables a parametric interaction that has the potential to generate entanglement between photons and phonons. The beam-splitter and parametric interactions can lead to signal transduction and entanglement generation and numerous other applications in the optical and microwave regimes, as we will discuss later.

Now, let us take another look at Equation (\ref{eq:interaction-Hamiltonian}). As mentioned before, $g_0$ is called the single-photon coupling rate. It indicates the rate of quanta exchange between the electromagnetic and mechanical subsystems when there is only one photon in the cavity. There is also a $\sqrt{\bar{n}_c}$ multiplier in the equation which amplifies the interaction term as the average number of the photons in the cavity, $\bar{n}_c$, increases. This multiplier indicates the role of the cavity in improving the interaction by keeping the intracavity photons between the confinement mirrors. We define the enhanced coupling rate as $g = \sqrt{\bar{n}_c}g_0$, showing the effective excitation exchange rate between the cavity and the mechanical oscillator. At this point, we introduce two more parameters that appear in the equations of motion that describe the evolution of the system: $\gamma$, the mechanical, and $\kappa$, the cavity damping rates. Damping rates indicate the rate of energy exchange between each subsystem and the environment. Clearly, for an opto/electromechanical interaction to be effective, it is crucial that the coupling rate surpasses the damping rates. Otherwise, the excitations of photons or phonons will dissipate into the surrounding heat bath and be lost before any exchange can occur with the other subsystem. The regime in which the single-photon coupling rate exceeds the damping rates is referred to as the `Ultra-strong coupling regime'. This regime exhibits several intriguing behaviors \cite{liao2013correlated, kronwald2013full, nunnenkamp2011single, rabl2011photon, komar2013single, akram2010single, kronwald2013optomechanically}, but its realization has not been achieved thus far. Nevertheless, as previously mentioned, we can enhance the coupling by employing pumping techniques and increasing the number of photons within the cavity. This allows us to attain a regime known as the `strong coupling regime', where the condition $g=g_0\sqrt{\bar{n}_c}>\kappa,\gamma$ is met. In this regime, the coupling between the subsystems becomes significantly amplified.

As detailed in Appendix (\ref{appendix:input-output-theorem}), it is possible to enhance the average number of photons within the cavity and, correspondingly, increase the coupling strength by employing a more powerful external pump. However, this approach of increasing the pump power and subsequently raising the photon number in the cavity comes with a drawback. It leads to heating effects within the system, thereby causing an increase in the damping rate of both the cavity and the mechanical resonator, for example, due to two-photon absorption \cite{sang2009applications} in the optical domain or interacting with Two-Level Systems (TLS) \cite{pappas2011two, gao2008semiempirical, macha2010losses} in the microwave domain. Consequently, the heating phenomenon results in elevated noise levels across all subsystems, ultimately leading to the degradation or destruction of quantum effects. 

We saw that cavity optomechanics and cavity electromechanics are similar in fundamental concepts and functionalities. However, there are also distinctions due to the extreme frequency difference between microwave and optical regimes. To understand one of the distinctions, we first need to investigate the noise properties of such systems. The relation between the average number of noise quanta, $\bar{n}$, in a bosonic field of angular frequency $\omega$ which is in equilibrium with a heat bath with temperature $T$ is defined as the Bose-Einstein distribution \cite{greiner2012thermodynamics},
\begin{equation}
    \bar{n} = \frac{1}{\exp\left[\hbar\omega/k_BT\right]-1},
\end{equation}
where $k_B$ is the Boltzmann constant. Table (\ref{tab:thermal-noise-quanta}) shows the values of $\bar{n}$ for different frequencies and temperatures.
\begin{table}[]
    \centering
    \begin{tabular}{c||c|c|c|c|}
          & $300$ kHz & $4$ MHz & $10$ GHz & $200$ THz\\\hline\hline
          $7$ mK & $4.9\times10^2$ & $3.6\times10^{1}$ & $2.2\times10^{-30}$ & $7.8\times10^{-593208}$\\\hline
          $4$ K & $2.8\times10^5$ & $2.1\times10^4$ & $7.9$ & $7.7\times10^{-1039}$ \\\hline
          $300$ K & $2.1\times10^7$ & $1.6\times10^6$ & $6.3\times10^2$ & $1.4\times10^{-14}$ \\\hline
    \end{tabular}
    \caption[Thermal noise quanta]{\textbf{Thermal noise quanta} - The number of the noise quanta decreases as the frequency of the field increases or the temperature decreases.}
    \label{tab:thermal-noise-quanta}
\end{table}
As the numbers show, optical frequency systems will not be affected by the thermal noise of the environment, even at room temperature. Consequently, it is possible to perform quantum optical experiments at room temperature, without using any cryogenic systems. However, in the case of microwave systems, reliable measurements cannot be performed at room temperature due to the substantial number of thermal noise quanta present (approximately 630 quanta noise at 10 GHz). To establish a suitable environment for quantum experiments in the microwave domain, cryogenic environments are essential. By utilizing cryogenics, not only do we have superconductivity, but also thermal noise can be effectively eliminated, reducing the average number of noise quanta to a few or even less than one. This ensures a more controlled and reliable experimental setup for microwave-based investigations. For mechanical systems, however, the noise properties are even worse than microwaves. There are optomechanical and electromechanical systems with mechanical frequencies from $\sim100$ kHz \cite{kleckner2011optomechanical} to $\sim5$ GHz \cite{riedinger2016non}, and for any frequency within this range, the noise level of room temperature is not sufficiently low. Therefore, we unavoidably need cryogenic systems to take advantage of noise suppression and observe quantum effects.\\

Another distinction between electromechanics and optomechanics arises from the properties of current quantum processors. The most advanced quantum processors that have been built and developed so far are based on superconducting qubits \cite{huang2020superconducting, gambetta2017building, clarke2008superconducting, kjaergaard2020superconducting, krantz2019quantum} which operate in the microwave frequency regime. Because of the similarity in platform and frequency, superconducting qubits are naturally more compatible with microwave systems than optical systems. Meaning that it is more feasible and practical to couple a superconducting qubit to electromechanical systems rather than optomechanical ones.

\section{Applications}
Opto/electromechanics has a wide range of applications in science and technology \cite{barzanjeh2022optomechanics}. For instance, it has been shown that one can generate quantum states of motion on microscopic and macroscopic mechanical oscillators employing opto/electromechanical interaction \cite{mancini1998optomechanical, ashkin1978trapping, teufel2011sideband}. Such experiments can be used for deeper investigations of the fundamentals of quantum mechanics.

Furthermore, opto/electromechanical systems can be harnessed for ultra-sensitive classical or quantum sensing \cite{li2021cavity, michimura2020quantum, barzanjeh2022optomechanics, motazedifard2021ultraprecision, arcizet2006high, torovs2020quantum, cohen2006torsional, feng2007very, wu2014dissipative, kaviani2022remote, kaviani2020optomechanical}. Mechanical oscillators couple to the environment typically well, and on the other hand, manipulating and measuring optical/electrical fields and systems are feasible using highly developed equipment. Opto/electromechanics makes extremely precise measurements possible by bringing the aforementioned advantages of mechanical and optical/microwave systems. As a specific example of the application of optomechanics in quantum sensing, one can mention the Laser Interferometer Gravitational-wave Observatory (LIGO) \cite{aasi2015advanced} project. Optomechanics played a significant role in the adjustments, stabilization, and measurements in LIGO \cite{cahillane2022review}.

\begin{figure}[h]
    \centering
    \includegraphics[width=0.5\textwidth]{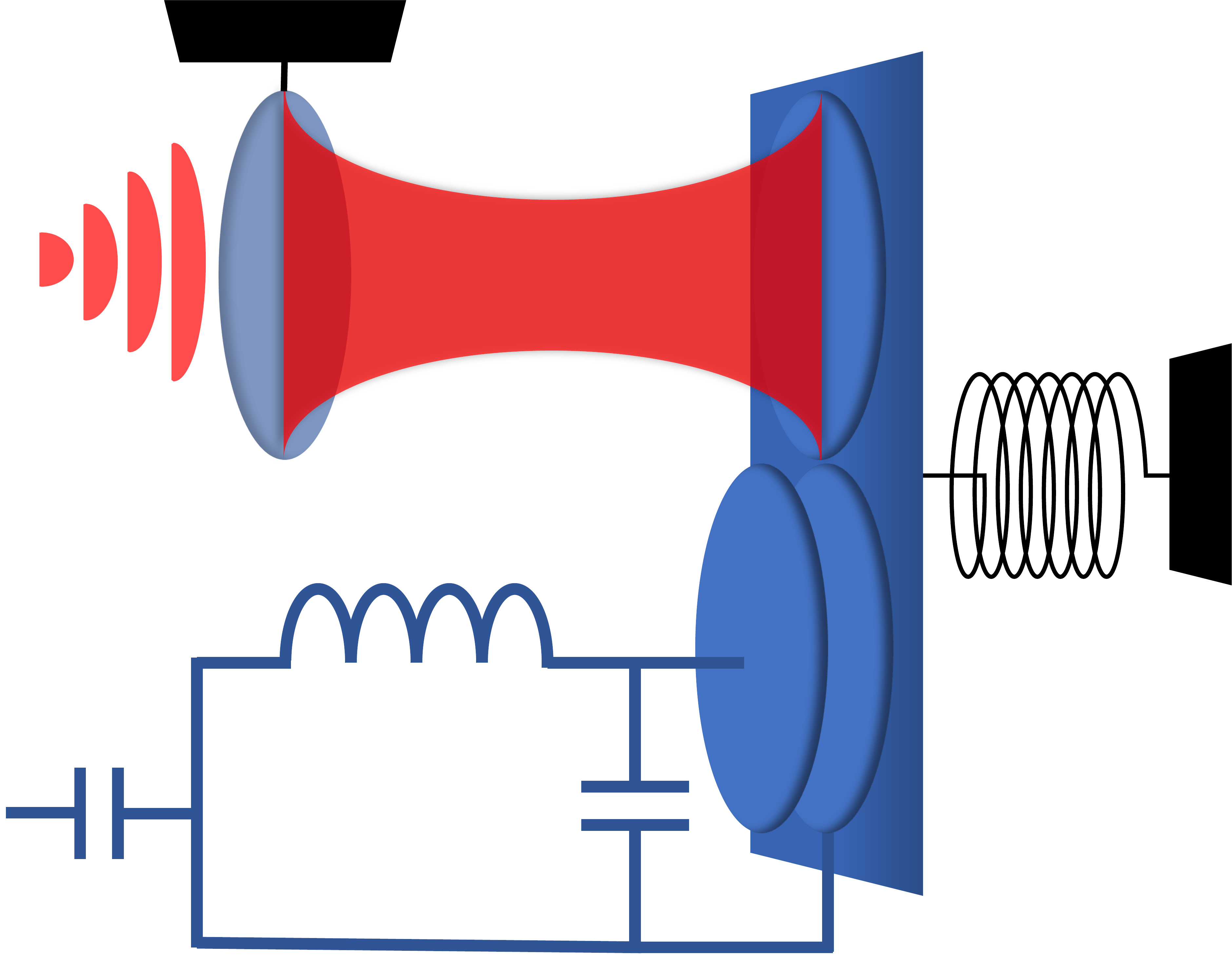}
    \caption[Opto-electromechanical sysetm]{An optical cavity can be coupled to a microwave resonator by a mechanical medium. Such a tripartite system can be used to generate continuous-variable microwave-optical entangled states or transduce quantum states between microwave and optical frequency regimes.}
    \label{fig:opto-electr-mechanical-system-scheme}
\end{figure}

In addition to quantum sensors, one can use the flexibility of mechanical systems in coupling to different subsystems to make hybrid quantum systems \cite{barzanjeh2017mechanical, rueda2019electro, parkins1999quantum, macquarrie2013mechanical, teissier2014strain, zhang2016cavity, li2018magnon, zoepfl2020single, schliesser2010cavity, shandilya2022diamond}. As an example of a hybrid quantum system, we can mention quantum transduction using opto-electro-mechanical systems \cite{stannigel2010optomechanical, arnold2020converting, higginbotham2018harnessing, bagci2014optical}. It has been demonstrated that microwave and optical cavities can be effectively coupled using a mechanical oscillator (see Figure \ref{fig:opto-electr-mechanical-system-scheme}). With such devices, it becomes possible to generate continuous-variable microwave-optical entangled states or perform bidirectional transduction of quantum states between microwave and optical frequency regimes. Microwave-to-optical transduction plays a crucial role in connecting future microwave quantum processors using optical communication methods (for example fiber or satellite), thus enabling the establishment of quantum networks and the realization of a quantum internet \cite{pirandola2016physics, wehner2018quantum, cacciapuoti2019quantum}.\\

\section{The state-of-the-art}
To date, a wide range of optomechanical, electromechanical, and opto-electro-mechanical devices have been developed and studied. In this section, we provide a concise overview of a few different systems and offer a comparison among them. You can also find a summary of the details of these systems in Table (\ref{tab:opto/electromechanical-experiments}).

In 2009, an experiment conducted by S. Gröblacher et al. \cite{groblacher2009observation} showed a remarkable demonstration of strong optomechanical coupling. The device they constructed was a Fabry-Pérot optical cavity, where one of the two mirrors served as a micromechanical oscillator. The measured single-photon optomechanical coupling rate was found to be $g_0 = 2\pi\times2.7$ Hz, which was smaller than both the optical cavity damping rate ($\kappa = 2\pi\times215$ kHz) and the mechanical damping rate ($\gamma = 2\pi\times140$ Hz). However, by effectively pumping the cavity, the optomechanical coupling was enhanced to reach a value of $g=2\pi\times325$ kHz, surpassing the damping rates of both the mechanical oscillator and the cavity.

In another optomechanical experiment in 2011, J. Chan et al. \cite{chan2011laser} fabricated a one-dimensional Silicon photonic crystal that localized optical fields at $\omega_c/2\pi=195$ THz and mechanical fields at $\Omega/2\pi=3.68$ GHz in the center of the structure. By utilizing the optomechanical interaction, they successfully cooled down the mechanical degree of freedom of the photonic crystal. The achieved average phonon number was $0.85\pm0.08$, even when the system was immersed in a heat bath at a temperature of 20 K, corresponding to approximately 113 thermal phonon quanta.

In a comparable experimental setup, a study conducted by R. Reidinger et al. in 2016 \cite{riedinger2016non} employed a Silicon photonic crystal structure that confined optical fields at a frequency of $\omega_c/2\pi=193$ THz, while simultaneously localizing mechanical fields at $\Omega/2\pi=5.3$ GHz. Through this configuration, they achieved a single-photon coupling rate of $g_0=2\pi\times825$ kHz, thereby harnessing the optomechanical interaction to generate entangled photon-phonon pairs.

In a breakthrough experiment in 2011, J. D. Teufel et al. \cite{teufel2011sideband} achieved the electromechanical cooling of a micrometer-scale mechanical oscillator to its ground state. Their experimental setup consisted of a suspended Aluminum membrane (with a mechanical frequency of $\Omega/2\pi=\times10.56$ MHz), serving as one of the plates of a capacitor. This capacitor was connected to an integrated superconducting coil, which acted as an LC circuit (with a frequency of $\omega_c/2\pi=\times7.54$ GHz) and was fabricated on a sapphire substrate. By exploiting a single-photon electromechanical coupling rate of approximately $g_0\simeq200$ Hz, they effectively cooled the membrane to an average phonon number of approximately $0.34$, within a heat bath environment at a temperature of around $15$ mK. It is worth noting that the average phonon number of the membrane at $15$ mK corresponds to approximately $30$.

\begin{table}[]
    \centering
    \begin{tabular}{c||c|c|c|c|c|c|}
         Research group & Type & $\omega_c/2\pi$ & $\kappa/2\pi$ & $\Omega/2\pi$ & $\gamma/2\pi$ & $g_0/2\pi$  \\\hline\hline
         Gr\"oblacher et al. \cite{groblacher2009observation} & OM & $282$THz & $215$kHz & $947$kHz & $140$Hz & $2.7$Hz \\\hline
         Chan et al. \cite{chan2011laser} & OM & $195$THz & $500$MHz  & $3.68$GHz & $35$kHz & $910$kHz \\\hline
         Reidinger et al. \cite{riedinger2016non} & OM & $293$THz & $1.3$GHz & $5.3$GHz & $4.8$kHz & $825$kHz\\\hline
         Teufel et al. \cite{teufel2011sideband} & EM & $7.54$GHz & $200$kHz & $10.56$MHz & $32$Hz & $200$Hz\\\hline
         \multirow{2}{*}{Higginbotham et al. \cite{higginbotham2018harnessing}} & OM & $281$THz & $3$MHz & \multirow{2}{*}{$1.47$MHz} & \multirow{2}{*}{$11$Hz} & $6.6$Hz \\\cline{2-4}\cline{7-7}
         & EM & $6.16$GHz & $2.5$MHz &  &  & $3.8$Hz \\\hline
         \multirow{2}{*}{Arnold et al. \cite{arnold2020converting}} & OM & $198$THz & $1.6$GHz & \multirow{2}{*}{$11.84$MHz} & \multirow{2}{*}{$15$Hz} & $662$kHz \\\cline{2-4}\cline{7-7}
         & EM & $10.50$GHz & $1.6$MHz &  &  & $67$Hz \\\hline
    \end{tabular}
    \caption[A summary of a few opto/electromechanical experiments]{\textbf{A summary of a few opto/electromechanical experiments} - OM and EM indicate optomechanical and electromechanical, respectively.}
    \label{tab:opto/electromechanical-experiments}
\end{table}

There are also reported experiments in which optical and microwave cavities have been coupled through a mechanical medium to form optical-microwave transduction \cite{Lauk_2020}. A. P. Higgeinbotham et al. \cite{higginbotham2018harnessing} in 2018 coupled a Fabry-P\'erot optical cavity to an integrated LC circuit through a suspended dielectric membrane which was placed in the middle of the optical cavity, and was partially coated with Aluminum to make an oscillating plate of a capacitor that modulates the microwave cavity. They reported $47\%$ optical-microwave conversion efficiency with 38 noise quanta added to the output signal. Although this experiment showed an outstanding bi-directional conversion efficiency, the added noise is still too high for quantum transduction.\\
As another example of the transduction project, G. Arnold et al. \cite{arnold2020converting} can be mentioned. In 2020, they reported the results of an experiment in which they made a Silicon zipper photonic crystal connected to a nano-string, while the nano-string holds a plate of a capacitor. The optical field localized on the zipper photonic crystal was coupled to the microwave field trapped in an integrated LC circuit through the mechanical mode of the structure (photonic crystal and the nano-string). The bi-directional optical-Microwave conversion efficiency was reported $1.2\%$ with 224(145) noise quanta added to the output optical(microwave) signal. Despite low efficiency and a large number of added noise quanta, this project is interesting due to the integrity of the structure: The entire system was fabricated on a Silicon chip.

Although there have been many great works done in optomechanics and electromechanics, ultra-strong coupling is still aspirational.

\section{Thesis aim and outline}
The primary aim of this thesis is to undertake a comprehensive theoretical investigation, design, simulation, and possibly fabrication and characterization of an optimized electromechanical device. Among all the possible candidate structures for an electromechanical device, we mainly decided to implement a nano-string coupled to a 3D microwave cavity. 3D microwave cavities are more robust to noise compared to integrated resonators, and nano-strings can be coupled to photonic crystals \cite{arnold2020converting}, for possible future research plans on opto-electro-mechanical systems.
Up to the date of writing this document, we have investigated the theoretical model and done the required simulations completely. The recipe for fabricating the electromechanical device has been developed, however, it is still in the optimization process. Consequently, the characterization of the device is planned for when we will have a functioning electromechanical device. Nevertheless, the measurement setup for characterizing an electromechanical device is already prepared and will be reviewed in the thesis.

The second chapter of this thesis describes the theoretical model of opto/electromechanics. It begins by constructing the Hamiltonian and making necessary approximations to render it more practical for analysis. Subsequently, the equations of motion are extracted and analyzed to show how the opto/electromechanical interaction gives rise to various functionalities.

Moving on to the third chapter, I provide an explanation of our design for an electromechanical device, which centers around nano-string structures. Furthermore, I detail the simulation procedures conducted using COMSOL and COMSOL LiveLink for MATLAB. COMSOL was utilized for simulating the structures, while LiveLink for MATLAB enabled automated post-processing and the implementation of geometric variations.

The fourth chapter focuses on the fabrication methods employed for electromechanical devices. Two primary designs are discussed, one involving coupling to 3D microwave cavities, and the other utilizing integrated LC circuits as the cavities. A step-by-step demonstration of the fabrication process for each design is provided, along with an explanation of the nano-fabrication tools utilized.
Ongoing efforts to optimize the fabrication process are being undertaken to address challenges and enable the observation of electromechanical interaction. 

In the fifth chapter, the characterization of the electromechanical device takes center stage. To measure the quantum characteristics of the microwave device, specific electronic equipment and a cryogenic system are employed. The chapter begins with an explanation of the equipment used and how the cryogenic system functions. Subsequently, the measurement setup is demonstrated, and details are provided on device preparation and the extraction of characteristics from the measurements. 

Finally, the last chapter presents a theoretical model based on electromechanics that enables entanglement between a microwave cavity and a magnonic system. While a pure electromagnonic interaction does not lead to entanglement generation, the incorporation of a mechanical subsystem through the electromechanical interaction ultimately gives rise to tripartite entanglement. This chapter serves as a foundation for future projects centered around the electromechanical device that is the focal point of our research endeavors.

  \chapter{Theoretical background}
\label{chp:theory}

Electromechanics is the field of studying the interaction of mechanical displacement with microwave (GHz frequency regime) electromagnetic fields in the form of AC electrical signals. An electromechanical system is typically an LC circuit that includes a capacitor with one oscillating plate. The oscillating plate plays the role of the mechanical oscillator in the system. The oscillations of the plate affect the capacitance, hence, the mode frequency of the LC circuit. On the other hand, the electromagnetic field in the LC circuit applies mechanical pressure on the oscillating plate through the potential difference between the plates.\\
In fact, electromechanics is the same as optomechanics in essence, and the theoretical models behind them are quite similar.
An optomechanical system can be described as a Fabry-P\'erot optical ($200$ THz frequency regime) cavity with one oscillating mirror. The oscillating mirror changes the length of the cavity and shifts the electromagnetic mode frequency, and the electromagnetic field pushes the mirror due to the radiation pressure.\\
The differences between electromechanics and optomechanics are the frequency regime of the electromagnetic field and the methods of practical implementation. Due to the similarities between the two types of systems, the phrase optomechanics often is used instead of electromechanics. In the following, we review the theory of opto/electromechanics.

\section{Hamiltonian}

An electromechanical system consists of one electromagnetic cavity and one mechanical oscillator. In the case of a linear electromagnetic resonator and small mechanical displacements, both subsystems can be considered harmonic oscillators. Therefore, according to the quantum harmonic oscillator model, the Hamiltonian reads,
\begin{equation}
    \hat{\mathcal{H}} = \hbar\tilde{\omega}_c \hat{a}^\dagger \hat{a} + \hbar\Omega\hat{b}^\dagger\hat{b} + \hat{\mathcal{H}}_\text{drive}
\end{equation}
where $\hbar$ is the reduced Planck's constant, $\hat{a}$ and $\hat{b}$ are the annihilation operators of the intracavity electromagnetic and mechanical fields, and $\tilde{\omega_c}$ and $\Omega$ are the natural frequencies of the electromagnetic resonator and the mechanical oscillator, respectively. $\hat{\mathcal{H}}_\text{drive}$ is the drive Hamiltonian which indicates the effect of the external drive field (pump) on the cavity. We will talk about this term in the next section. Note that we have neglected the constant terms in the Hamiltonian as they only shift the energy levels by a constant, and do not affect the equations of motion.\\
Since the oscillations of the mirror change the fundamental mode of the cavity (by changing the capacitance in electromechanics or changing the length of the cavity in optomechanics), we can consider $\tilde{\omega}_c$, the mode frequency, as a function of $\hat{x}$, the position operator of the mirror \cite{cavityOptomechanics};
\begin{equation}
    \tilde{\omega}_c(\hat{x}) = \omega_c +
    \frac{\partial\tilde{\omega}_c}{\partial\hat{x}}\hat{x} +
    \frac{1}{2}\frac{\partial^2\tilde{\omega}_c}{\partial\hat{x}^2}\hat{x}^2 + \cdots.
\end{equation}
where $\omega_c$ is the resonance frequency of the cavity mode when the mechanical oscillator is in its rest position. For small oscillations, we can neglect the non-linear terms, $O (\hat{x}^2)$, and keep the first two terms. Considering the relation between $\hat{x}$ and the mechanical ladder operators,
\begin{equation*}
    \hat{x} = x_\text{ZPF}\left( \hat{b}^\dagger + \hat{b} \right),
\end{equation*}
with $x_\text{ZPF} = \sqrt{\hbar/2m_\text{eff}\Omega}$ being the zero point fluctuations of the mechanical oscillator with the effective mass $m_\text{eff}$. The Hamiltonian then takes the following form,
\begin{equation}
    \hat{\mathcal{H}} \simeq \hbar\omega_c\hat{a}^\dagger\hat{a} + \hbar\Omega\hat{b}^\dagger\hat{b} + \hbar x_\text{ZPF}\frac{\partial\tilde{\omega}_c}{\partial\hat{x}} \hat{a}^\dagger\hat{a}\left(\hat{b}^\dagger+\hat{b}\right) + \hat{\mathcal{H}}_\text{drive}.
\end{equation}
The first two terms are free energy terms that govern the individual evolution of the cavity and mechanical modes. 
By defining the single photon coupling rate, $g_0$, as, 
\begin{equation}
g_0 = x_\text{ZPF}\frac{\partial\tilde{\omega}_c}{\partial\hat{x}},    
\end{equation}
the third term in the Hamiltonian,
\begin{equation}
    \hat{\mathcal{H}}_\text{int} = \hbar g_0 \hat{a}^\dagger\hat{a}\left(\hat{b}^\dagger+\hat{b}\right),
\end{equation}
describes the interaction between the cavity photons and phonons of the mechanical mode, leading to,
\begin{equation}
    \hat{\mathcal{H}} = \hbar\omega_c\hat{a}^\dagger\hat{a} + \hbar\Omega\hat{b}^\dagger\hat{b} + \hat{\mathcal{H}}_\text{int} = \hbar\omega_c\hat{a}^\dagger\hat{a} + \hbar\Omega\hat{b}^\dagger\hat{b} + \hbar g_0 \hat{a}^\dagger\hat{a}\left(\hat{b}^\dagger+\hat{b}\right) + \hat{\mathcal{H}}_\text{drive}.\label{eq:non-linearezied-Hamiltonian}
\end{equation}

\subsection{Hamiltonian linearization}
\label{chp:Hamiltonian-linearization}
The Hamiltonian in Equation (\ref{eq:non-linearezied-Hamiltonian}) leads to nonlinear equations of motion because the interaction Hamiltonian, $\hat{a}^\dagger\hat{a}\left(\hat{b}^\dagger+\hat{b}\right)$, contains third-degree terms of ladder operators. Since we are interested only in the linear responses of the system, we make an approximation on the interaction term to linearize the equations of motion. To simplify further, we first need to transform the Hamiltonian from the regular frame to a frame rotating at the frequency of the external drive pump, $\omega_p$, with respect to the lab. The Hamiltonian of the external drive in the lab frame reads,
\begin{equation}
    \hat{\mathcal{H}}_\text{drive} = i\hbar\sqrt{\kappa_\text{ex}}\left(\hat{a}\alpha^*_\text{in}e^{i\omega_pt}+\hat{a}^\dagger\alpha_\text{in}e^{-i\omega_pt}\right),
\end{equation}
in which $\kappa_\text{ex}$ is called the extrinsic damping rate of the cavity and indicates the rate of energy exchange between the cavity and the externally applied field, and $\alpha_\text{in}$ indicates the complex amplitude of the drive field. In the rotating frame, the time dependency of $\hat{\mathcal{H}}_\text{drive}$ should disappear.
The transformation required for finding the equations in the rotating frame can be accomplished using the unitary operator $\hat{U} = \exp\left(i\hat{a}^\dagger\hat{a}\omega_pt\right)$\footnote{Using the interaction picture formalism \cite{sakurai1995modern}.}. The new Hamiltonian in the rotating frame will be obtained as,
\begin{equation}
    \hat{\mathcal{H}} \rightarrow \hat{U}\hat{\mathcal{H}}\hat{U}^\dagger - i\hbar\hat{U}\frac{\partial\hat{U}^\dagger}{\partial t}.
\end{equation}
It can be shown\footnote{$e^{\hat{A}}\hat{B}e^{-\hat{A}} = B + \left[\hat{A}, \hat{B}\right] + \frac{1}{2!}\left[\hat{A}, \left[\hat{A}, \hat{B}\right]\right] + \frac{1}{3!}\left[\hat{A}, \left[\hat{A}, \left[\hat{A}, \hat{B}\right]\right]\right] + \cdots$\cite{louisell1973quantum}} that,
\begin{align}
    \hat{U}\hat{a}\hat{U}^\dagger =& \hat{a}e^{-i\omega_pt},\\
    \hat{U}\hat{a}^\dagger\hat{U}^\dagger =& \hat{a}^\dagger e^{i\omega_pt},\\
    \hat{U}\hat{a}^\dagger\hat{a}\hat{U}^\dagger =& \hat{a}^\dagger\hat{a},\\
    i\hbar\hat{U}\frac{\partial\hat{U}^\dagger}{\partial t} =& \hbar\omega_p\hat{a}^\dagger\hat{a},
\end{align}
therefore, the transformation carried out by $\hat{U}$ will eliminate the time dependency of $\hat{\mathcal{H}}_\text{drive}$, and the Hamiltonian in the rotating frame will be,
\begin{equation}
    \hat{\mathcal{H}} = \hbar\Delta\hat{a}^\dagger\hat{a} + \hbar\Omega\hat{b}^\dagger\hat{b} + \hbar g_0\hat{a}^\dagger\hat{a}\left(\hat{b}^\dagger+\hat{b}\right) + \hat{\mathcal{H}}_\text{drive},
\end{equation}
where $\Delta = \omega_c-\omega_p$ is the detuning of the pump with respect to the cavity. 

The presence of $\hat{\mathcal{H}}_\text{drive}$ and non-zero $\left\vert\alpha_\text{in}\right\vert$ gives rise to a steady non-zero mean photon number in the cavity (see Appendix \ref{appendix:input-output-theorem} for more information). Now, we assume that the mean value of the photon number in the cavity is much larger than its small fluctuations over time. Therefore, it is reasonable to approximate $\hat{a}$ and $\hat{a}^\dagger$ operators with the combination of a classical field with amplitude $\alpha$ (where $\left|\alpha\right|^2 = \langle\hat{a}^\dagger\hat{a}\rangle = \bar{n}_c$ with $\bar{n}_c$ being the steady mean photon number in the cavity) and fluctuation operators $\delta\hat{a}$ and $\delta\hat{a}^\dagger$.
\begin{align}
    \hat{a} &\simeq \alpha + \delta\hat{a},\\
    \hat{a}^\dagger &\simeq \alpha + \delta\hat{a}^\dagger.
\end{align}
Note that we assume $\alpha$ to be real. From now on, we do not write the Hamiltonian of the drive field in the total Hamiltonian explicitly, but as we will see, we will include it in the equations of motion through Langevin equations. By applying this approximation the interaction Hamiltonian will be expanded as,
\begin{align}
    \hat{\mathcal{H}}_\text{int} =& \hbar g_0 \hat{a}^\dagger\hat{a}\left(\hat{b}^\dagger + \hat{b}\right)  \nonumber\\
    =& \hbar g_0 \left(\alpha+\delta\hat{a}^\dagger\right)\left(\alpha+\delta\hat{a}\right)\left(\hat{b}^\dagger + \hat{b}\right) \nonumber\\
    = &\hbar g_0 \left|\alpha\right|^2\left(\hat{b}^\dagger + \hat{b}\right) + \hbar g_0\alpha\left(\delta\hat{a}^\dagger+\delta\hat{a}\right)\left(\hat{b}^\dagger + \hat{b}\right) + \hbar g_0 \delta\hat{a}^\dagger\delta\hat{a} \left(\hat{b}^\dagger + \hat{b}\right).\label{eq:interaction_hamiltonian_before_approximations}
\end{align}
As previously mentioned, the operator $\delta\hat{a}$ represents small fluctuations around the steady photon number $\left|\alpha\right|^2$, therefore, $\left|\langle\delta\hat{a}^\dagger\delta\hat{a}\rangle\right|\ll\left|\alpha\langle\delta\hat{a}\rangle\right|\ll\left|\alpha\right|^2$. Besides, for the case of a weak single-photon coupling we have, $\left|g_0\langle\hat{b}^\dagger+\hat{b}\rangle\right|\ll\omega_c$. As a result of inequalities, the contribution of the third degree term $\hbar g_0 \delta\hat{a}^\dagger\delta\hat{a} \left(\hat{b}^\dagger + \hat{b}\right)$ to the Hamiltonian can be considered negligible.\\
Furthermore, term $\hbar g_0 \left|\alpha\right|^2\left(\hat{b}^\dagger + \hat{b}\right)$ indicates an average steady force,
\begin{equation}
    \bar{F}_\text{rad} = \left|-\frac{\partial}{\partial\hat{x}}\hbar g_0\left|\alpha\right|^2\left(\hat{b}^\dagger+\hat{b}\right)\right| = \hbar G\left|\alpha\right|^2 = \hbar G \bar{n}_c,\label{eq:radiation_pressure_force}
\end{equation}
exerted on the mechanical oscillator due to the presence of the steady field in the cavity. This force is evidence of radiation pressure.
In Equation (\ref{eq:radiation_pressure_force}), $G = g_0/x_{ZPF}$ represents the coupling strength between the optical field and the mechanical oscillator.
It can be shown\footnote{$H = \frac{P}{2m} + \frac{1}{2}m\omega^2X^2 + FX = \frac{P}{2m} + \frac{1}{2}m\omega^2\left(X+\frac{F}{m\omega^2}\right)^2 - \frac{F^2}{2m\omega^2}.$} that a constant force applied on a harmonic oscillator shifts the rest position of the oscillator and adds a constant term to its Hamiltonian.
Similarly, $\hbar g_0 \left|\alpha\right|^2\left(\hat{b}^\dagger + \hat{b}\right)$ can be absorbed by the free mechanical Hamiltonian (The Hamiltonian describing an isolated mechanical oscillator in the case of no interaction) with the result of the shift
\begin{equation}
    \delta x_\alpha = \frac{\hbar G\left\vert\alpha\right\vert^2}{m_\text{eff}\Omega^2},
\end{equation}
in its rest position, and an added constant term in the Hamiltonian which we will neglect. Consequently, the mode frequency of the cavity shifts by
\begin{equation}
    \delta\omega_\alpha = G\delta x_\alpha = \frac{\hbar G^2\left\vert\alpha\right\vert^2}{m_\text{eff}\Omega^2} = \frac{g_0^2\left\vert\alpha\right\vert^2}{\Omega},
\end{equation}
as the length of the cavity has changed.
With re-defining the position operator and cavity mode frequency and including the shifts, $\hbar g_0 \left|\alpha\right|^2\left(\hat{b}^\dagger + \hat{b}\right)$ will be taken into account implicitly and will not appear in the interaction Hamiltonian anymore.

Ultimately, after applying the approximations and re-definitions, and by changing the notation, $\{\delta\hat{a}, \delta\hat{a}^\dagger\} \rightarrow \{\hat{a}, \hat{a}^\dagger\}$, for the sake of simplicity, the Hamiltonian reduces to,
\begin{align}
    \hat{\mathcal{H}} =& \hbar\omega_c\hat{a}^\dagger\hat{a} + \hbar\Omega\hat{b}^\dagger\hat{b} + \hat{\mathcal{H}}_\text{int,lin},\\
    \hat{\mathcal{H}}_\text{int,lin} =& \hbar g\left(\hat{a}^\dagger+\hat{a}\right)\left(\hat{b}^\dagger+\hat{b}\right),
\end{align}
where, $g=g_0\left|\alpha\right|=g_0\sqrt{\bar{n}_c}$, is known as the enhanced electromechanical coupling rate. The recent analysis shows that we can enhance the coupling by pumping the cavity. The relation of the intracavity number of photons, $\bar{n}_c$, and the external pump power is described in Appendix (\ref{appendix:input-output-theorem}).

\subsection{Rotating-wave approximation}

The linearized interaction Hamiltonian contains four terms in its general form:

\begin{equation}
    \hat{\mathcal{H}}_\text{int,lin} = \hbar g\left(\hat{a}^\dagger\hat{b}+\hat{a}\hat{b}^\dagger\right) + \hbar g\left(\hat{a}^\dagger\hat{b}^\dagger+\hat{a}\hat{b}\right).
\end{equation}
The initial two terms in the equation depict a beam-splitter interaction between photons and phonons, allowing for exchanging a single photon into a single phonon or vice versa. Furthermore, the subsequent two terms describe a parametric interaction between photons and phonons, resulting in the possibility of generation of entanglement between the cavity mode and the mechanical oscillator. We will explore in detail how one of these interactions can dominate the other under specific conditions. However, by employing the Rotating-Wave Approximation (RWA) \cite{fleming2010rotating}, we can investigate the effect of a properly tuned external applied field on the interaction Hamiltonian.

We previously transformed the Hamiltonian to a rotating frame which rotates at the frequency of the external pump. Therefore, as it is obvious from the cavity's free Hamiltonian (the Hamiltonian of the isolated cavity in case of no interaction), the frequency of $\hat{a}$ and $\hat{a}^\dagger$ is $\Delta$ in the rotating frame, leading to,
\begin{align}
    \mathcal{H} =& \hbar\Delta\hat{a}^\dagger\hat{a} + \hbar\Omega\hat{b}^\dagger\hat{b} +
    \hbar g \left(\hat{a}^\dagger e^{i\Delta t}\hat{b}e^{-i\Omega t}+\hat{a}e^{-i\Delta t}\hat{b}^\dagger e^{i\Omega t}\right) + 
    \hbar g\left(\hat{a}^\dagger e^{i\Delta t}\hat{b}^\dagger e^{i\Omega t}+\hat{a}e^{-i\Delta t}\hat{b}e^{-i\Omega t}\right) \nonumber\\
    =& \hbar\Delta\hat{a}^\dagger\hat{a} + \hbar\Omega\hat{b}^\dagger\hat{b} +
    \hbar g \left(\hat{a}^\dagger\hat{b}e^{i(\Delta-\Omega)t}+\hat{a}\hat{b}^\dagger e^{-i(\Delta-\Omega)t}\right) + 
    \hbar g\left(\hat{a}^\dagger\hat{b}^\dagger e^{i(\Delta+\Omega)t}+\hat{a}\hat{b}e^{-i(\Delta+\Omega)t}\right).
\end{align}

\begin{figure}[!h]
    \centering
    \begin{subfigure}[b]{0.8\textwidth}
        \includegraphics[width=\textwidth]{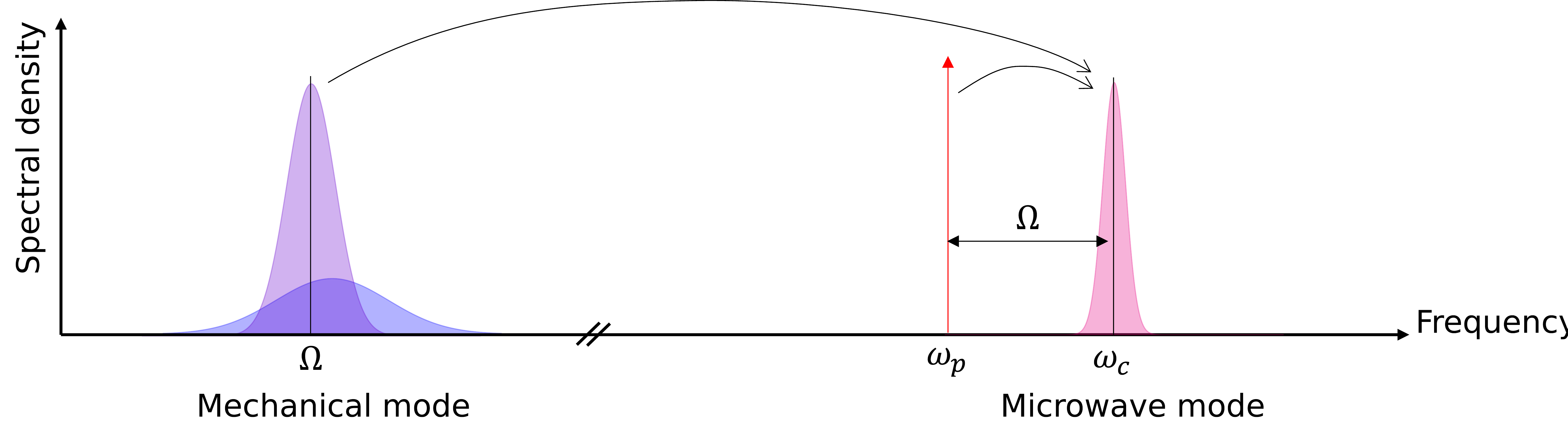}
        \subcaption{}
        \label{fig:energy-diagrams-red-detuned}
    \end{subfigure}
    \begin{subfigure}[b]{0.8\textwidth}
        \includegraphics[width=\textwidth]{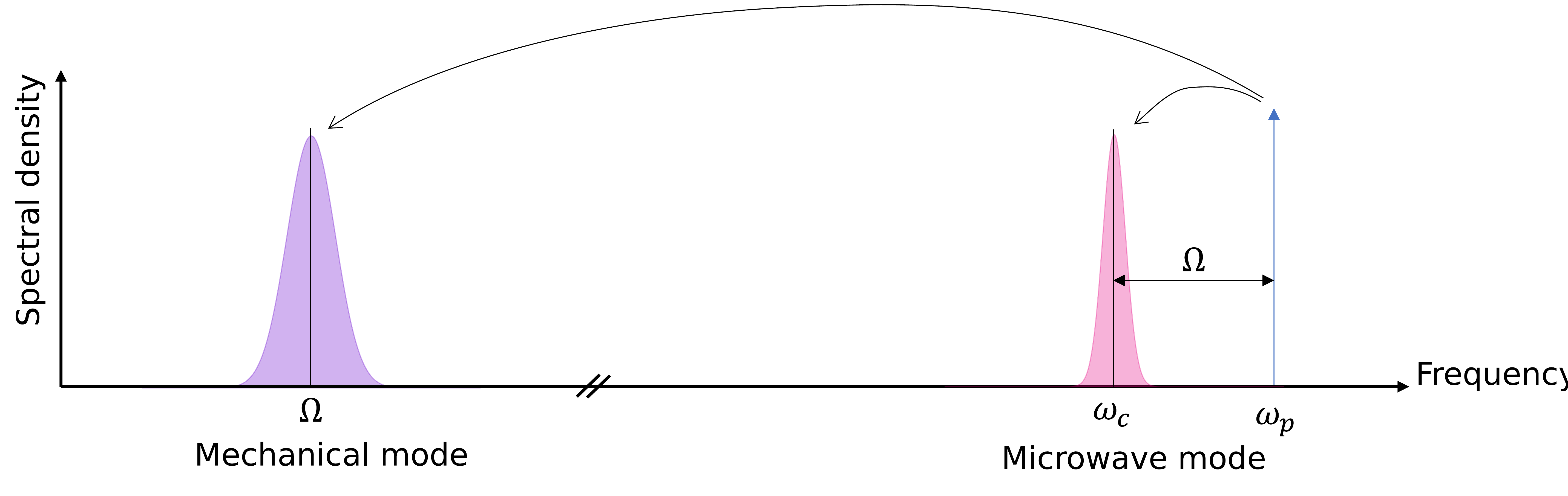}
        \subcaption{}
        \label{fig:energy-diagrams-blue_detuned}
    \end{subfigure}
    \caption[Optomechanical interactions]{The pump makes the cavity and the mechanical oscillator interact. The interaction depends on the frequency of the pump relative to the frequency of the cavity's mode. (a) For a red-detuned pump, the beam-splitter Hamiltonian becomes dominant and opens an energy exchange gate between the cavity and the mechanical oscillator. As a result of the interaction, an extra damping rate is introduced to the mechanical oscillator and broadens and deflates its spectral density. (b) For a blue-detuned pump, the parametric interaction dominates and the interaction can lead to photon-phonon entangled pair generation.}
    \label{fig:energy-diagrams}
\end{figure}

When we tune the applied field in the \textbf{red-detuning} regime ($\omega_p<\omega_c$) such that $\Delta \sim \Omega$, we encounter a situation where $\left|\Delta+\Omega\right|\gg\left|\Delta-\Omega\right|$. In this scenario, terms oscillating at the frequency $\Delta+\Omega$ can be considered fast-oscillating terms, and applying the RWA allows us to neglect them while retaining the other terms. Therefore, the Hamiltonian can be simplified to,
\begin{equation}
    \mathcal{H}_\text{red} \simeq \hbar\Delta\hat{a}^\dagger\hat{a} + \hbar\Omega\hat{b}^\dagger\hat{b} +
    \hbar g \left(\hat{a}^\dagger\hat{b}+\hat{a}\hat{b}^\dagger\right).
\end{equation}
As we see, the red-detuned pump makes the beam-splitter terms survive and we can observe energy exchange between photons and phonons (see Figure \ref{fig:energy-diagrams-red-detuned}). On the other hand, if we tune the applied field in the \textbf{blue-detuning} regime ($\omega_p>\omega_c$) such that $\Delta \sim -\Omega$ then we will have $\left|\Delta+\Omega\right|\ll\left|\Delta-\Omega\right|$. In this case, the fast-oscillating terms are the ones with $\Delta-\Omega$ frequency and the Hamiltonian reduces to,
\begin{equation}
    \mathcal{H}_\text{blue} \simeq \hbar\Delta\hat{a}^\dagger\hat{a} + \hbar\Omega\hat{b}^\dagger\hat{b} +
    \hbar g \left(\hat{a}^\dagger\hat{b}^\dagger+\hat{a}\hat{b}\right),
\end{equation}
after the RWA. It is possible to show that a blue-detuned pump leads to two-mode squeezing or generates entanglement and amplification \cite{barzanjeh2017mechanical, PhysRevA.84.042342, PhysRevLett.114.080503} (see Figure \ref{fig:energy-diagrams-blue_detuned}).

\section{Equations of motion}
Having the linearized Hamiltonian in the rotating frame, we can derive the equations of motion for the creation and annihilation operators using Langevin equations that are introduced in Appendix (\ref{appendix:input-output-theorem}) \cite{scully1999quantum}.

\begin{align}
\dot{\hat{a}} =& \frac{1}{i\hbar} \left[a,\mathcal{H}\right] - \frac{\kappa}{2}\hat{a} + \sqrt{\kappa_\text{in}} \hat{a}_\text{in} + \sqrt{\kappa_\text{ex}} \hat{a}_\text{ex}\nonumber\\
=& -i\Delta\hat{a} - \frac{\kappa}{2}\hat{a} - ig\left(\hat{b}^\dagger+\hat{b}\right) + \sqrt{\kappa_\text{in}} \hat{a}_\text{in} + \sqrt{\kappa_\text{ex}} \hat{a}_\text{ex},\label{eq:time-dom-EM-equation-of-motion}\\
\dot{\hat{b}} =& \frac{1}{i\hbar}\left[b,\mathcal{H}\right] - \frac{\gamma}{2}\hat{b} + \sqrt{\gamma}\hat{b}_\text{in} \nonumber\\
=& -i\Omega\hat{b} - \frac{\gamma}{2}\hat{b} -ig\left(\hat{a}^\dagger+\hat{a}\right) + \sqrt{\gamma}\hat{b}_\text{in}\label{eq:time-dom-Mech-equation-of-motion},
\end{align}
where the parameters $\kappa_\text{in}$ and $\kappa_\text{ex}$ represent the intrinsic (the rate of energy exchange between the cavity and the environment) and extrinsic (the rate of energy exchange between the cavity and the communication line) damping rates of the cavity, respectively. The total damping rate of the cavity mode is denoted by $\kappa$, which is the sum of the intrinsic and extrinsic damping rates, $\kappa = \kappa_\text{in} + \kappa_\text{ex}$. Furthermore, the parameter $\gamma$ corresponds to the mechanical damping rate (the rate of energy exchange between the mehcanical oscillator and the environment). The operators $\hat{a}_\text{in}$ and $\hat{a}_\text{ex}$ describe the input noises arising from intrinsic and extrinsic losses, respectively, while the operator $\hat{b}_\text{in}$ represents the mechanical input noise.

Equations (\ref{eq:time-dom-EM-equation-of-motion}) and (\ref{eq:time-dom-Mech-equation-of-motion}) describe the dynamics of the intracavity fields and mechanical mode, meaning that the operators are defined in the time domain. We can transform the equations of motion from the time domain to the frequency domain by taking the Fourier transform of the equations in the time domain (see Appendix \ref{chp:frequency_domain_analysis});
\begin{align}
    -i\omega\hat{a} =& \left(-i\Delta-\frac{\kappa}{2}\right)\hat{a} - ig\left(\hat{b}^\dagger+\hat{b}\right) + \sqrt{\kappa_\text{in}}\hat{a}_\text{in} + \sqrt{\kappa_\text{ex}}\hat{a}_\text{ex} \Rightarrow\nonumber\\
    \chi_a^{-1}(\omega)\hat{a} =& - ig\left(\hat{b}^\dagger+\hat{b}\right) + \sqrt{\kappa_\text{in}}\hat{a}_\text{in} + \sqrt{\kappa_\text{ex}}\hat{a}_\text{ex},\label{eq:freq-dom-Mech-equation-of-motion_a}\\
    -i\omega\hat{b} =& \left(-i\Omega-\frac{\gamma}{2}\right)\hat{b} - ig\left(\hat{a}^\dagger+\hat{a}\right) + \sqrt{\gamma}\hat{b}_\text{in}\Rightarrow\nonumber\\
    \chi_b^{-1}(\omega)\hat{b} =& -ig\left(\hat{a}^\dagger+\hat{a}\right) + \sqrt{\gamma}\hat{b}_\text{in}\label{eq:freq-dom-Mech-equation-of-motion_b},
\end{align}
where $\chi_a(\omega) = \left(-i\left(\omega-\Delta\right)+\kappa/2\right)^{-1}$ and $\chi_b(\omega)= \left(-i\left(\omega-\Omega\right)+\gamma/2\right)^{-1}$ are the electromagnetic and the mechanical free susceptibilities (susceptibilities of individual subsystems in the case of no cross interaction). Having $\mathcal{F}\left\{\hat{a}^\dagger(t)\right\} = \hat{a}^\dagger(\omega)$ (see Appendix \ref{chp:frequency_domain_analysis}), 
one can similarly find the equations of motion of the creation operators,
\begin{align}
    \tilde{\chi}_a^{-1}(\omega)\hat{a}^\dagger =& ig\left(\hat{b}+\hat{b}^\dagger\right) + \sqrt{\kappa_\text{in}}\hat{a}_\text{in}^\dagger  + \sqrt{\kappa_\text{ex}}\hat{a}_\text{ex}^\dagger,\label{eq:freq-dom-Mech-equation-of-motion_a_dagger}\\
    \tilde{\chi}_b^{-1}(\omega)\hat{b}^\dagger =& ig\left(\hat{a}+\hat{a}^\dagger\right) + \sqrt{\gamma}\hat{b}_\text{in}^\dagger,
\end{align}
where $\tilde{\chi}_a(\omega) = \chi_a^*(-\omega) = \left(-i\left(\omega+\Delta\right)+\kappa/2\right)^{-1}$ and $\tilde{\chi}_b(\omega) = \chi_b^*(-\omega) = \left(-i\left(\omega+\Omega\right)+\gamma/2\right)^{-1}$.

\section{Optomechanical interaction}
\begin{figure}
     \centering
     \includegraphics[width=0.7\textwidth]{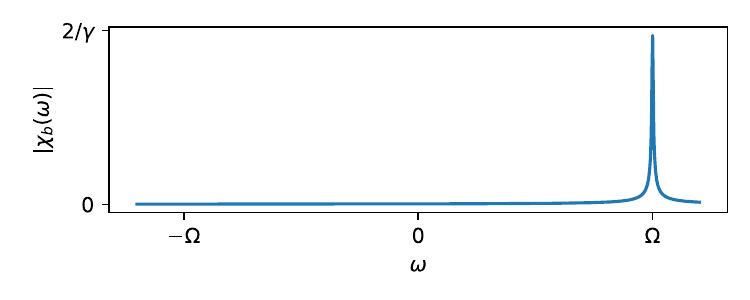}
     \caption[Mechanical susceptibility]{\textbf{Mechanical susceptibility} - The mechanical susceptibility has only one peak around $\omega=\Omega$, and is negligible at other frequencies.}
     \label{fig:Mech_susceptibiliy}
 \end{figure}

A good approach to understanding the effects of the opto/electromechanical interaction is analyzing the susceptibilities. The expression $\chi_b^{-1}(\omega) = -i\left(\omega-\Omega\right)+\gamma/2$ is the inverse of the mechanical free susceptibility. The free susceptibility describes how the mechanical oscillator responds to the input fields (environmental noise in this case) when there is no opto/electromechanical interaction.
To see how the opto/electromechanical interaction affects the mechanical susceptibility, we need to decouple the equations of motion and find an equation for $\hat{b}$ in terms of the input fields only. We start with Equation (\ref{eq:freq-dom-Mech-equation-of-motion_b}), and replace $\hat{a}$ and $\hat{a}^\dagger$ with Equations (\ref{eq:freq-dom-Mech-equation-of-motion_a}) and (\ref{eq:freq-dom-Mech-equation-of-motion_a_dagger}).
\begin{gather}
    \chi_b^{-1}(\omega)\hat{b} = -ig\left(\frac{-ig}{\chi_a^{-1}(\omega)}\left(\hat{b}^\dagger+\hat{b}\right) + \frac{ig}{\tilde{\chi}_a^{-1}(\omega)}\left(\hat{b}+\hat{b}^\dagger\right)\right) + \sqrt{\gamma}\hat{b}_\text{in} + \hat{A}_\text{in} \Rightarrow\nonumber\\
    \chi_b^{-1}(\omega)\hat{b} = -g^2\left(\frac{1}{\chi_a^{-1}(\omega)}-\frac{1}{\tilde{\chi}_a^{-1}(\omega)}\right)\left(\hat{b}+\hat{b}^\dagger\right) + \hat{B}_\text{in}\Rightarrow\nonumber\\
    \chi_b^{-1}(\omega)\left(\hat{b}+g^2\left(\frac{1}{\chi_a^{-1}(\omega)}-\frac{1}{\tilde{\chi}_a^{-1}(\omega)}\right)\left(\frac{\hat{b}+\hat{b}^\dagger}{\chi_b^{-1}(\omega)}\right)\right) = \hat{B}_\text{in} \Rightarrow\nonumber\\
    \chi_b^{-1}(\omega)\left(\hat{b}+g^2\left(\frac{1}{\chi_a^{-1}(\omega)}-\frac{1}{\tilde{\chi}_a^{-1}(\omega)}\right)\left(\frac{\hat{b}}{\chi_b^{-1}(\omega)}\right)\right) \simeq \hat{B}_\text{in} \Rightarrow\nonumber\\
    \left(\chi_b^{-1}(\omega)+g^2\left(\frac{1}{\chi_a^{-1}(\omega)}-\frac{1}{\tilde{\chi}_a^{-1}(\omega)}\right)\right)\hat{b} = \hat{B}_\text{in},
\end{gather}
where $\hat{A}_\text{in} = \left(\sqrt{\kappa_\text{in}}\hat{a}_\text{in}+\sqrt{\kappa_\text{ex}}\hat{a}_\text{ex}\right)/\chi_a^{-1} + \left(\sqrt{\kappa_\text{in}}\hat{a}_\text{in}^\dagger+\sqrt{\kappa_\text{ex}}\hat{a}_\text{ex}^\dagger\right)/\tilde{\chi}_a^{-1}$ and $\hat{B}_\text{in} = \sqrt{\gamma}\hat{b}_\text{in}+\hat{A}_\text{in}$. In the fourth line, we can neglect $\hat{b}^\dagger$ as $\langle\hat{b}^\dagger(\omega)\rangle$ has only one peak around $\omega=-\Omega$ and is negligible at other regions, while $\left|\chi_b(\omega)\right|$ has one peak around $\omega=\Omega$ and is negligible at other regions 
 (Figure \ref{fig:Mech_susceptibiliy}).\\
Now, we can rewrite the equation of motion for $\hat{b}$,
\begin{equation}
    \left(\chi_b^{-1}(\omega) + \chi_{b, \text{opt}}^{-1}(\omega)\right)\hat{b} = \hat{B}_\text{in},
\end{equation}
in which $\chi_{b, \text{opt}}(\omega) = \left(g^2\left(1/\chi_a^{-1}(\omega) - 1/\tilde{\chi}_a^{-1}(\omega)\right)\right)^{-1}$, is the correction to the mechanical susceptibility due to the optomechanical interaction. As it is obvious, in case of no optomechanical interaction (i.e. $g=0$), this correction vanishes. The terms $\chi_a(\omega)$  and $\tilde{\chi}_a(\omega)$ in $\chi_{b, \text{opt}}$ correspond to $\hat{a}$ and $\hat{a}^\dagger$, respectively, in Equation (\ref{eq:freq-dom-Mech-equation-of-motion_b}), which is obtained from the linearized interaction Hamiltonian. In the case of a red-detuned pump ($\Delta\simeq\Omega$), the correction susceptibility at the frequency $\omega=\Omega$ reads,
\begin{align*}
    \chi_{b, \text{opt}}^{-1}(\Omega) =& g^2\left(\frac{1}{-i(\Omega-\Delta)+\kappa/2}-\frac{1}{-i(\Omega+\Delta)+\kappa/2}\right)\simeq\nonumber \\
    &g^2\left(\frac{1}{-i(\Omega-\Omega)+\kappa/2}-\frac{1}{-i(\Omega+\Omega)+\kappa/2}\right) = \nonumber\\
    &g^2\left(\frac{1}{\kappa/2}-\frac{1}{-2i\Omega+\kappa/2}\right).
\end{align*}
and for $4\Omega\gg\kappa$, we get,
\begin{equation}
     \chi_{b, \text{opt}}^{-1}(\Omega) \simeq g^2\left(\frac{1}{\kappa/2}-0\right),
\end{equation}
which means the contribution of $1/\tilde{\chi}_a(\omega)$ can be neglected. It is equivalent to removing $\hat{a}^\dagger$ from the equation of motion of $\hat{b}$ (Equation \ref{eq:freq-dom-Mech-equation-of-motion_b}), which could be done through RWA from the beginning. Such condition ($4\Omega\gg\kappa$) is called \textbf{resolved side-band condition}. One can do the same analysis for the case of blue-detuned pump
($\Delta\simeq-\Omega$).
\begin{align*}
    \chi_{b, \text{opt}}^{-1}(\Omega) =& g^2\left(\frac{1}{-i(\Omega-\Delta)+\kappa/2}-\frac{1}{-i(\Omega+\Delta)+\kappa/2}\right)\simeq\nonumber \\
    &g^2\left(\frac{1}{-i(\Omega+\Omega)+\kappa/2}-\frac{1}{-i(\Omega-\Omega)+\kappa/2}\right) = \nonumber\\
    &g^2\left(\frac{1}{-2i\Omega+\kappa/2}-\frac{1}{\kappa/2}\right),
\end{align*}
which means that under the resolved side-band condition and for the blue-detuned pump, the contribution of $\chi_a(\omega)$ in $\chi_{b, \text{opt}}^{-1}(\Omega)$ is negligible, and validates the RWA for the mentioned conditions.

\subsection{Optomechanical damping rate}
As we will see in detail, the optomechanical interaction introduces an extra source of damping to the mechanical oscillator. The extra damping rate is called the optomechanical damping rate ($\gamma_\text{opt}$) and modifies the total damping rate of the mechanical oscillator ($\gamma_\text{total}=\gamma+\gamma_\text{opt}$). 
This damping rate is responsible for optomechanical cooling or heating of the mechanical oscillator.
Using the correction susceptibility, we can understand how cooling or gain occurs. The real part of the inverse of the free mechanical susceptibility, $\Re\left[\chi_b^{-1}(\Omega)\right]=\gamma/2$, describes the intrinsic damping rate of the mechanical mode. Similarly, we can extract the effect of the optomechanical interaction on mechanical damping rate by analyzing the real part of the inverse of the correction susceptibility:
\begin{align}
    \gamma_\text{opt}(\Delta) = \Re\left[\chi_{b, \text{opt}}^{-1}(\Omega)\right] =& g^2\Re\left[\frac{1}{-i(\Omega-\Delta)+\kappa/2}-\frac{1}{-i(\Omega+\Delta)+\kappa/2}\right] \nonumber\\
    =& \frac{g^2\kappa}{2}\left(\frac{1}{\left(\Omega-\Delta\right)^2+\kappa^2/4}-\frac{1}{\left(\Omega+\Delta\right)^2+\kappa^2/4}\right).
\end{align}
As defined earlier, $\gamma_\text{opt}$ is the damping rate introduced to the mechanical oscillator by the optomechanical interaction, and consequently, the cavity mode. In the resolved side-band regime, we can simplify the optomechanical damping rate
\begin{align}
    \gamma_\text{opt}(\Omega) =& \frac{g^2\kappa}{2}\left(\frac{1}{\kappa^2/4}-\frac{1}{4\Omega^2 +\kappa^2/4}\right)\simeq\frac{g^2\kappa}{2}\left(\frac{4}{\kappa^2}\right) = \frac{2g^2}{\kappa},\\
    \gamma_\text{opt}(-\Omega) =& \frac{g^2\kappa}{2}\left(\frac{1}{4\Omega^2+\kappa^2/4}-\frac{1}{\kappa^2/4}\right)\simeq\frac{g^2\kappa}{2}\left(-\frac{4}{\kappa^2}\right) = -\frac{2g^2}{\kappa}.
\end{align}

In the case of red-detuning ($\Delta=\omega_c-\omega_p=\Omega$), a positive damping rate corresponds to optomechanical cooling. Intuitively, when pump photons with a frequency $\omega_p = \omega_c-\Delta = \omega_c-\Omega$ are present, they bridge the energy gap between cavity photons and phonons ($\hbar\omega_p+\hbar\Omega=\hbar\omega_c$), allowing for energy exchange between photons and phonons. Specifically, the annihilation of one pump photon and one phonon results in the creation of one cavity photon. During this process, the cavity extracts energy from the mechanical oscillator, inducing an effective broadening that leads to mechanical cooling (shown in Figure \ref{fig:energy-diagrams-red-detuned}).

In the case of blue-detuning ($\Delta=\omega_c-\omega_p=-\Omega$), a negative damping rate indicates a mechanical gain.
In this scenario, the cavity field bridges the energy gap between the phonons and pump photons ($\hbar\omega_c+\hbar\Omega=\hbar\omega_p$), causing one pump photon with frequency $\omega_p=\omega_c-\Delta=\omega_c+\Omega$ to split into one cavity photon and one phonon. As a consequence, the excitation of mechanical phonons increases, leading to phonon gain. As long as the total mechanical damping rate, $\gamma+\gamma_\text{opt}$, remains positive, the system remains stable. However, if $-\gamma_\text{opt}>\gamma$ and the total mechanical damping rate becomes negative, the system enters an instability regime \cite{ludwig2008optomechanical}.

In order to demonstrate the coherent interaction between photons and phonons, it is necessary to introduce the concept of interaction cooperativity. This parameter is defined as the ratio between the optomechanical damping rate and the natural mechanical damping rate, denoted as $\mathcal{C} = \frac{\gamma_\text{opt}}{\gamma}$. In the resolved sideband condition, specifically, we have $\mathcal{C} = \frac{2g^2}{\kappa \gamma}$. The cooperativity serves as an important indicator of the efficiency of state transfer between the phononic and photonic modes. When $\mathcal{C}>1$, it implies that the energy exchange between photons and phonons surpasses the damping rates of each subsystem. This indicates a coherent interaction between the two modes, demonstrating the ability to transfer energy coherently between them.

\subsection{OptoMechanically Induced Transparency}
\label{chp:theory-OMIT}
Electromagnetically Induced Transparency (EIT) is a well-known quantum effect in atomic systems, in which an atomic ensemble couples to a probe and a pump optical fields. EIT makes a transparent window appear within the absorption profile of the atomic ensemble due to destructive interference of the atomic excitation paths \cite{EIT1, EIT2}. The optomechanical analogous of EIT is called OptoMechanically Induced Transparency  (OMIT), in which the excitation paths of the cavity, (i) directly due to the absorption of the probe field, and (ii) indirectly due to the creation of a cavity photon followed by the annihilation of a pump photon and a phonon, destructively interfere and lead to a non-absorption behavior from the cavity \cite{OMIT2}.\\
To understand how this effect can be investigated in experiments, we can start by finding the reflection of a cavity (see Appendix \ref{appendix:input-output-theorem}) in the presence of a mechanical oscillator. To couple the cavity and the mechanical resonator, we assume the red-detuned pump, under the resolved side-band condition,
\begin{align}
    &\begin{cases}
        \chi_a\hat{a} = -ig\hat{b} + \sqrt{\kappa_\text{in}}\hat{a}_\text{in} + \sqrt{\kappa_\text{ex}}\hat{a}_\text{ex}\\
        \chi_b\hat{b} = -ig\hat{a} + \sqrt{\gamma}\hat{b}_\text{in}
    \end{cases} \Rightarrow \nonumber \\
    &\chi_a\hat{a} = 
    \frac{-ig}{\chi_b} \left(-ig\hat{a}+\sqrt{\gamma}\hat{b}_\text{in}\right) + \sqrt{\kappa_\text{in}}\hat{a}_\text{in} + \sqrt{\kappa_\text{ex}}\hat{a}_\text{ex}\Rightarrow\nonumber\\
    &\left(\chi_a + \frac{g^2}{\chi_b}\right)\hat{a} = -\frac{ig\sqrt{\gamma}}{\chi_b}\hat{b}_\text{in} + \sqrt{\kappa_\text{in}}\hat{a}_\text{in} + \sqrt{\kappa_\text{ex}}\hat{a}_\text{ex} \Rightarrow\nonumber\\
    &\left(\chi_a + \frac{g^2}{\chi_b}\right)\left(\hat{a}_\text{out}+\hat{a}_\text{ex}\right) = -\frac{ig\sqrt{\gamma\kappa_\text{ex}}}{\chi_b}\hat{b}_\text{in} + \sqrt{\kappa_\text{in}\kappa_\text{ex}}\hat{a}_\text{in} + \kappa_\text{ex}\hat{a}_\text{ex} \Rightarrow\nonumber\\
    & \mathcal{R} = \frac{\langle\hat{a}_\text{out}\rangle}{\langle\hat{a}_\text{ex}\rangle} = -\frac{\chi_a + \frac{g^2}{\chi_b}-\kappa_\text{ex}}{\chi_a+\frac{g^2}{\chi_b}} = -\frac{-i(\omega-\Delta)+\frac{\kappa_\text{in}-\kappa_\text{ex}}{2}+\frac{g^2}{-i(\omega-\Omega)+\gamma/2}}{-i(\omega-\Delta)+\frac{\kappa_\text{in}+\kappa_\text{ex}}{2}+\frac{g^2}{-i(\omega-\Omega)+\gamma/2}},
    \label{eq:OMIT}
\end{align}
in which $\hat{a}_\text{out}$ is the output field of the cavity. In the derivation of the recent equation, we used the second Langevin equation (Appendix \ref{appendix:input-output-theorem}), and we assumed $\langle\hat{a}_\text{in}\rangle = \langle\hat{b}_\text{in}\rangle = 0$, as they represent thermal noise.\\
It should be noted that the effective coupling rate, $g$, depends on both the single-photon coupling rate ($g_0$) and the number of photons in the cavity ($\bar{n}_c$). The single-photon coupling rate is an inherent property of the system and cannot be controlled once the system is fabricated. However, we have the ability to manipulate the number of photons, $\bar{n}_c$, by adjusting the power of the applied field. By sweeping the power of the applied field, we can control the optomechanical coupling rate ($g$), allowing us to modulate the strength of the interaction. \\
\begin{figure}[ht]
    \centering
    \includegraphics[width=0.7\textwidth]{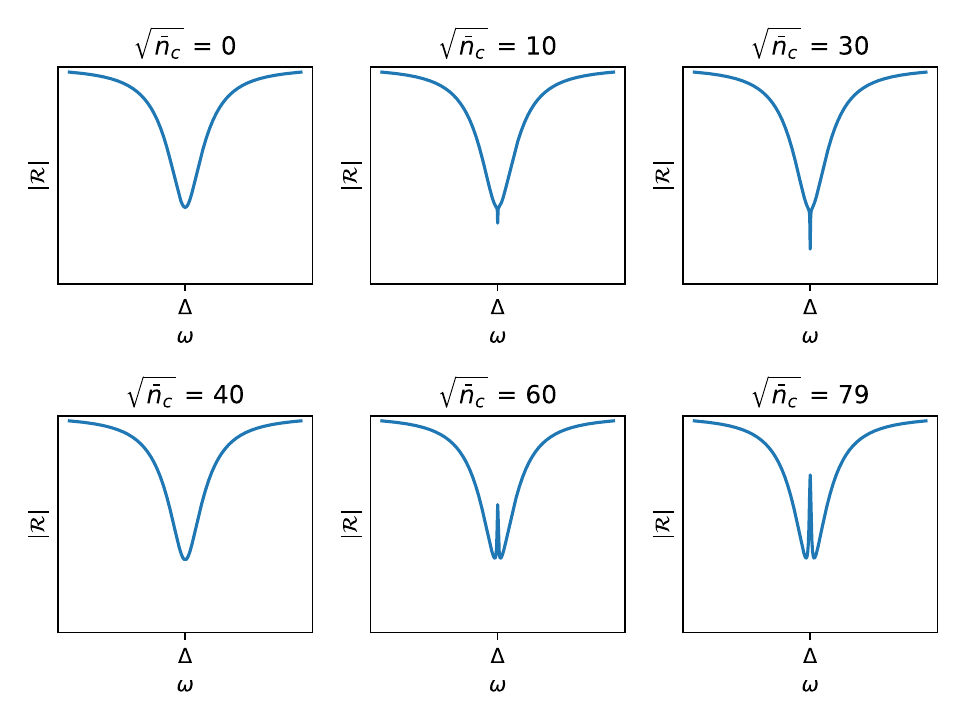}
    \caption[OMIT in the reflection of a cavity]{For an arbitrary set of physical parameters, the reflection is a simple Lorentzian when $\bar{n}_c = 0$. As $\bar{n}_c$, and consequently, the coupling rate, $g$ increases, a narrow dip appears in the Lorentzian and turns into a narrow peak.}
    \label{fig:OMIT}
\end{figure}

In the limit where the number of photons in the cavity, $\bar{n}_c$, approaches zero ($\bar{n}_c\rightarrow0$) corresponding to the absence of pump power, the equation resembles that of a simple cavity reflection. In this case, the modulus of the reflection coefficient, $\left|\mathcal{R}(\omega)\right|$, takes the form of a Lorentzian function centered at $\omega=\Delta$. Figure (\ref{fig:OMIT}) illustrates this behavior, where the Lorentzian function appears as a dip around the cavity frequency, with the width of the dip determined by the total damping rate of the cavity, $\kappa = \kappa_\text{in} + \kappa_\text{ex}$.

By setting $\Delta=\Omega$, we establish coupling between the cavity and the mechanical oscillator. Increasing the pump power subsequently amplifies the coupling rate, $g$, resulting in the formation of a small dip in the cavity response precisely at its center. The width of this dip is proportional to the natural mechanical damping rate, $\gamma$. Given that $\gamma\ll\kappa$ in typical optomechanical systems, the dip becomes exceptionally narrow compared to the cavity profile. From the perspective of absorption and transparency, this ultra-narrow dip corresponds to additional absorption at the cavity frequency. This phenomenon is known as OptoMechanically (or Electromagnetically, in the case of an atomic ensemble) Induced Absorption (OMIA) \cite{lezama1999electromagnetically, hocke2012electromechanically, qu2013phonon, naweed2005induced}.

However, the most interesting behavior arises when we further increase the coupling rate. As the pump power and coupling rate rise, the small dip gradually shortens, disappears, and eventually transforms into a small peak. This central peak in the cavity profile signifies partial transparency in the system, where the absorption (reflection) of the signal is slightly reduced (increased) precisely at the cavity frequency compared to neighboring frequencies. As evident from these observations, we refer to this effect as OptoMechanically Induced Transparency (OMIT). Continuing to increase the coupling rate causes the peak to grow, resulting in enhanced induced transparency.

One of the applications of OMIT is in making slow-light and quantum memory \cite{OMIT1}. But at this stage, we are only interested in the modulation of the reflection itself. The height and the width of the peak depend on the mechanical damping rate and optomechanical coupling rate. Therefore, all the optomechanical characteristics can be extracted from the reflection of the cavity, by fitting the Equation (\ref{eq:OMIT}) to the measured data points.

  \chapter{Designs and simulations}

In this chapter, I provide a detailed explanation of the sample design, offering comprehensive insights into its various aspects. Furthermore, I introduce numerical methods for simulating the electromechanical system, which allow us to thoroughly analyze and explore its behavior. For the simulations, I employ the finite element method and specifically utilize COMSOL along with LiveLink for MATLAB. This combination enables us to simulate the design and investigate the impact of geometric variations on its performance.
To extract the single-photon coupling rates from the simulations, I use equations that establish the relationship between these factors and the parameters obtained directly from the simulations. 
In the first section of this chapter, I present the mathematical derivations leading to the necessary equations. In the second section, I provide a detailed description of the structure we have designed for the electromechanical system. Additionally, I explain the simulation process and discuss the creation of MATLAB scripts for modulating the simulation and performing optimizations.

\section{Coupling rate}

The primary objective of the simulations is to determine the electromechanical coupling rate and investigate the impact of fabrication inaccuracies on the coupling rate. Prior to conducting the simulations, it is necessary to manipulate the expression of the electromechanical coupling rate so that it is formulated in terms of parameters that can be obtained from the simulations. This modified equation will then be utilized to establish a similar relationship for the electromechanical coupling rate.
It is important to note that the arguments and mathematical principles presented in this chapter are applicable to both optomechanical and electromechanical systems, unless specifically stated otherwise.

We have seen in the previous chapter that the single-photon electromechanical coupling rate is defined as,
\begin{equation*}
    g_0 = x_\text{ZPF}\frac{\partial \Tilde{\omega}_c}{\partial \hat{x}},
\end{equation*}
where $x_\text{ZPF} = \sqrt{\hbar/2m_\text{eff}\Omega}$ with $m_\text{eff}$ and $\Omega$ to be the effective motional mass and the mechanical mode frequency, respectively. The mechanical frequency of any arbitrary structure can be determined through COMSOL simulations without much complication. However, the concept of effective mass for an oscillator can be slightly complicated. This concept arises when we draw an analogy between two harmonically oscillating systems that share the same frequency: a rigid body oscillating in space and a flexible structure that is anchored at one or more points while allowing the remaining parts to oscillate. Each infinitesimal segment of the flexible oscillator oscillates with the same frequency and phase as the other segments, but their amplitudes, masses, and contributions to the overall kinetic and potential energies of the system can vary.\\
We define the deformation of the structure while it oscillates as, $\vec{Q}(\vec{r}, t) = \vec{Q}(\vec{r})\exp(-i\Omega t)$, where $\vec{Q}(\vec{r})$ describe the deformation vector of the infinitesimal piece of the structure located at $\vec{r}$. To find the potential energy of such a system, we break the entire body into infinitely many infinitesimal pieces. The potential energy of the entire system is $U = \sum_n U_n$, with $U_n$ being the potential energy of the $n$th piece located at $\vec{r}_n$.
\begin{equation}
    U = \sum_n U_n = \sum_n \frac{1}{2}\delta M_n \Omega^2 \left|\vec{Q}(\vec{r}_n)\right|^2 =
    \frac{1}{2}\Omega^2\sum_n\rho(\vec{r}_n)\delta\nu\left|\vec{Q}(\vec{r}_n)\right|^2,
\end{equation}
where $\delta M_n = \rho(\vec{r}_n)\delta\nu$ is the mass of the $n$th infinitesimal piece, $\rho(\vec{r}_n)$ is the mass density of the structure at the position $\vec{r}_n$, and $\delta\nu$ is the volume of each infinitesimal piece. We assumed the potential energy of the $n$th infinitesimal piece, $U_n = \frac{1}{2}\delta M_n\Omega^2\left|\vec{Q}(\vec{r}_n)\right|^2$, according to Hooke's law. By transforming the discrete distribution to a continuous distribution (i.e. $\sum_n\delta\nu \rightarrow \int d^3\vec{r}$ ), the potential energy of the entire system reads,
\begin{equation}
    U = \frac{1}{2}\Omega^2\int\rho(\vec{r})\left|\vec{Q}(\vec{r})\right|^2 d^3\vec{r}.
\end{equation}
Now, instead of dealing with infinitely many amplitudes and masses (the continuous vector field, $\vec{Q}(\vec{r})$, and the mass density of the structure, $\rho(\vec{r})$), we consider a single rigid oscillator with mass $m_\text{eff}$ and amplitude $\alpha$ (which we assume to be equal to the maximum displacement amplitude of the oscillating continuum), and the same frequency and potential energy as the flexible oscillator. 
Therefore, making the analogy we have,
\begin{equation}
    U = \frac{1}{2}\Omega^2\int\rho(\vec{r})\left|\vec{Q}(\vec{r})\right|^2 d^3\vec{r} = \frac{1}{2}m_\text{eff}\Omega^2\alpha^2 \Rightarrow m_\text{eff} = \int\rho(\vec{r})\left|\frac{\vec{Q}(\vec{r})}{\alpha}\right|^2d^3\vec{r}. \label{eq:effective_mass}
\end{equation}
Note that $\alpha$ normalizes the displacement field, $\vec{Q}$, so that $\left|\vec{Q}(\vec{r})/\alpha\right|$ will be bounded between 0 and 1. Going forward, we can employ the effective mass, denoted as $m_\text{eff}$, and the amplitude $\alpha$ as representatives of the entire structure. Thus,

\begin{equation}
    g_0 = x_\text{ZPF}\frac{\partial\tilde{\omega}_c}{\partial\hat{x}} \rightarrow g_0 = x_\text{ZPF}\frac{d\tilde{\omega}_c}{d\alpha}.
\end{equation}
The next step is to reform the ratio $d\tilde{\omega}_c/d\alpha$ and relate it to the parameters that can be extracted conveniently from the simulations. To do so, we first need to understand how mechanical deformation affects the electromagnetic field. There are different materials in the structures that we study (e.g. silicon, air, aluminum) each of which has a specific electric permittivity. Mechanical deformations shift the boundaries between the materials, consequently, change the electric permittivity at the fixed points close to the boundaries. This shift in the geometry of the electric permittivity distribution is equivalent to a shift in the geometry of the system and leads to a shift in the mode frequency. Since the deformations are typically much smaller than the scale of the structures, we can understand the interaction between the mechanical oscillations and the electromagnetic modes as a perturbation in the electric permittivity distribution, $\varepsilon(\vec{r})$. Only for optomechanical systems, in addition to the moving boundaries between the materials, mechanical deformations apply tension to the dielectrics, and in the optical frequency regime, the tension affects the electric permittivity due to the photoelastic effect \cite{nelson1971theory, biegelsen1974photoelastic}. Both moving boundaries and photoelastic effects contribute to the optomechanical coupling rate. Since the photoelasticity effect does not appear in the microwave frequency regime, we only consider the moving boundary effect.

Following Ref. \cite{johnson2002perturbation}, we can use the first-order perturbation and find an expression for $d\tilde{\omega}_c/d\alpha$.
\begin{equation}
    \frac{d\tilde{\omega}_c}{d\alpha} = -\frac{\omega_c}{2}\frac{\left\langle\vec{E}(\vec{r})\right|\frac{d\varepsilon(\vec{r})}{d\alpha}\left|\vec{E}(\vec{r})\right\rangle}{\left\langle\vec{E}(\vec{r})\right|\varepsilon(\vec{r})\left|\vec{E}(\vec{r})\right\rangle} = -\frac{\omega_c}{2}\frac{\displaystyle\int\frac{d\varepsilon(\vec{r})}{d\alpha}\left|\vec{E}(\vec{r})\right|^2d^3\vec{r}}{\displaystyle\int\varepsilon(\vec{r})\left|\vec{E}(\vec{r})\right|^2d^3\vec{r}},\label{eq:dw/da_1}
\end{equation}
where $\vec{E}(\vec{r})$ is the unperturbed electric field of the electromagnetic mode, at the location $\vec{r}$.
It is worth mentioning that the denominator of the last fraction, $\int\varepsilon(\vec{r})\left|\vec{E}(\vec{r})\right|^2d^3\vec{r}$, is twice the total electric energy stored in the cavity in the case of no deformation, and the numerator, $\int\frac{d\varepsilon(\vec{r})}{d\alpha}\left|\vec{E}(\vec{r})\right|^2d^3\vec{r}$, is twice the rate of change in the electric energy with respect to $\alpha$ when the mechanical deformation is introduced to the structure.

The integrand of the numerator in Equation (\ref{eq:dw/da_1}) vanishes everywhere except in the proximity of the boundaries of the materials: The mechanical deformations cause the material filling a specific fixed point in the space to change to the neighbouring material only close to the boundaries. Therefore, we can approximate the volume integral in the numerator with a surface integral, over all the interface surfaces between different materials in the structure.\\ 
We consider an infinitesimal surface with area $dA$ on the interface of two materials with electric permittivities $\varepsilon_1$ and $\varepsilon_2$. The displacement of the element of the surface is described by $\vec{Q}(\vec{r})$ where $\vec{r}$ is the location of the surface element. We can show that the contribution of the displacement of the surface element to the integral in the numerator approximately reads \cite{johnson2002perturbation},
\begin{equation}
    \left\langle\vec{E}(\vec{r})\right|\frac{d\varepsilon}{d\alpha}\left|\vec{E}(\vec{r})\right\rangle\Biggr|_{dA} = \left(\frac{\vec{Q}(\vec{r})\cdot\hat{n}(\vec{r})}{\alpha}\right)\left(\Delta\varepsilon\left|\vec{E}_\parallel(\vec{r})\right|^2 - \Delta\varepsilon^{-1}\left|\vec{D}_\perp(\vec{r})\right|^2\right)dA,\label{eq:numerator_OM_coupling_factor}
\end{equation}
where $\hat{n}(\vec{r})$ is the normal vector of the surface element, $\vec{E}_\parallel(\vec{r})$ and $\vec{D}_\perp(\vec{r})$ are the parallel component of the electric field and the perpendicular component of the electric displacement field with respect to the surface element, $\Delta\varepsilon = \varepsilon_1 - \varepsilon_2$, and $\Delta\varepsilon^{-1} = \varepsilon_1^{-1} - \varepsilon_2^{-1}$. If the system is made of only one dielectric material surrounded by air (or vacuum), we have,
\begin{equation}
    g_0 = -x_\text{ZPF}\frac{\omega_c}{2}\frac{\displaystyle\oint\left(\frac{\vec{Q}(\vec{r})\cdot\hat{n}(\vec{r})}{\alpha}\right)\left(\Delta\varepsilon\left|\vec{E}_\parallel(\vec{r})\right|^2 - \Delta\varepsilon^{-1}\left|\vec{D}_\perp(\vec{r})\right|^2\right)dA}{\displaystyle\int\varepsilon(\vec{r})\left|\vec{E}(\vec{r})\right|^2d^3\vec{r}},\label{eq:g_mb_optical}
\end{equation}
in which the surface integral in the numerator is taken over all the interface surfaces between the dielectric and air, and the volume integral in the denominator is taken over all the space.
The above equation is highly valuable for our objectives. By conducting a mechanical simulation, we can determine the mechanical frequency of the mode $\Omega$, and the deformation field of the mechanical mode, represented by $\vec{Q}(\vec{r})$. Additionally, through an electromagnetic simulation (optical or microwave), we can obtain $\vec{E}(\vec{r})$ and $\vec{D}(\vec{r})$. Subsequently, we can perform integrals as a post-processing step in the simulation to calculate the value of $g_0$.

As mentioned before, all the statements are valid for both electromechanical and optomechanical systems, as we have not made any assumption specific to any of the systems. However, because of a critical difference between simulating optomechanical and electromechanical systems, the analysis should be pushed one step further for electromechanical systems.
For the case of an optical cavity (a 1D photonic crystal for instance), we simulate the entire cavity and find $\vec{E}(\vec{r})$ for the whole system. The reason for that is the mechanical oscillator in such systems cannot be considered a separate component from the optical cavity. For electromechanical systems, in contrast, the capacitor with a mechanical oscillator and the microwave cavity can be considered separate components in the system: We can design and fabricate a capacitor with a mechanical oscillator, and then couple it to either a 3D microwave cavity or an integrated on-chip LC circuit. Therefore, for extracting the electromechanical coupling rate, we only simulate the capacitor and then include the microwave cavity in the equation implicitly, by adding extra factors to the equations. Let us see how we can do it. The frequency of an LC circuit reads,
\begin{equation}
    \omega_c = 2\pi\sqrt{\frac{1}{L\left(C_s+C_m\right)}},
\end{equation}
where $L$ is the total self-inductance of the circuit, $C_s$ is the total stray capacitance (summation of the self and stray capacitances) of the circuit, and $C_m$ is the capacitance of the capacitor with oscillating plates. Using the same definition for $\alpha$ from the previous section, the electromechanical coupling will be,
\begin{align}
    \frac{d\omega_c}{d\alpha} =& \frac{d}{d\alpha}2\pi\sqrt{\frac{1}{L\left(C_s+C_m\right)}} = -2\pi\frac{1}{2}\sqrt{\left(\frac{1}{L\left(C_s+C_m\right)}\right)^3}L\frac{dC_m}{d\alpha} \nonumber\\
    =&-\frac{\omega_c}{2}\frac{C_m}{C_s+C_m}\frac{1}{C_m}\frac{dC_m}{d\alpha} = -\frac{\omega_c}{2}\eta\frac{1}{C_m}\frac{dC_m}{d\alpha},
\end{align}
the quantity $\eta = C_m/\left(C_s+C_m\right)$ is referred to as the participation ratio which characterizes the influence of stray capacitance on the electromechanical coupling rate. A higher stray capacitance leads to a decrement in the electromechanical coupling rate. Therefore, for electromechanical applications, microwave cavities with smaller stray capacitance are generally preferred.

Now, let us review Equation (\ref{eq:dw/da_1}) once more. The integral in the numerator depicts twice the rate of the change in the electric energy stored in the cavity with respect to $\alpha$ due to the movements of the boundaries, and the integral in the denominator describes twice the total electric energy stored in the cavity. One can show that,
\begin{equation}
    \frac{1}{C_m}\frac{dC_m}{d\alpha} = \frac{1}{\frac{1}{2}C_mV_C^2}\frac{d\left(\frac{1}{2}C_mV_C^2\right)}{d\alpha} = \frac{1}{\mathcal{E}_E}\frac{d\mathcal{E}_E}{d\alpha} = \frac{d\mathcal{E}_E/d\alpha}{\mathcal{E}_E},
\end{equation}
in which $V_C$ is the electric voltage applied on the plates of the capacitor, and
\begin{equation}
    \mathcal{E}_E = \frac{1}{2} C_m V_C^2,
    \label{eq:electric_energy_of_a_capacitor}
\end{equation}
is the electric energy stored in the capacitor. The recent relation along with the physical justification allows us to use Equation (\ref{eq:dw/da_1}) to make a similar equation for the electromechanical coupling rate
\begin{equation}
    \frac{1}{C_m}\frac{dC_m}{d\alpha} = \frac{\displaystyle\int\frac{d\varepsilon(\vec{r})}{d\alpha}\left|\vec{E}(\vec{r})\right|^2d^3\vec{r}}{\displaystyle\int\varepsilon(\vec{r})\left|\vec{E}(\vec{r})\right|^2d^3\vec{r}}.
\end{equation}

By making the same approximation as in Equation (\ref{eq:numerator_OM_coupling_factor}), we arrive at a similar equation as (\ref{eq:g_mb_optical}) for the electromechanical system. However, there is a notable distinction that needs to be addressed. Unlike typical optical systems, electromechanical systems involve multiple materials, resulting in more than one interface surface. In our specific design, for instance, we have silicon as the substrate, aluminum for the capacitor plates, and air surrounding the structure. This gives rise to three sets of interface surfaces: silicon-aluminum, silicon-air, and aluminum-air interfaces. Consequently, a separate surface integral needs to be performed for each interface surface. Thus, the electromechanical coupling rate can be expressed as:
\begin{align}
    g_0 = -\frac{x_\text{ZPF}\eta\frac{\omega_c}{2}}{\displaystyle\int\varepsilon(\vec{r})\left|\vec{E}(\vec{r})\right|^2d^3\vec{r}}\Biggl(&\displaystyle\int_\text{Si-Al}\left(\frac{\vec{Q}(\vec{r})\cdot\hat{n}(\vec{r})}{\alpha}\right)\left(\Delta\varepsilon\left|\vec{E}_\parallel(\vec{r})\right|^2 - \Delta\varepsilon^{-1}\left|\vec{D}_\perp(\vec{r})\right|^2\right)dA+\nonumber\\
    &\displaystyle\int_\text{Si-Air}\left(\frac{\vec{Q}(\vec{r})\cdot\hat{n}(\vec{r})}{\alpha}\right)\left(\Delta\varepsilon\left|\vec{E}_\parallel(\vec{r})\right|^2 - \Delta\varepsilon^{-1}\left|\vec{D}_\perp(\vec{r})\right|^2\right)dA+\nonumber\\
    &\displaystyle\int_\text{Al-Air}\left(\frac{\vec{Q}(\vec{r})\cdot\hat{n}(\vec{r})}{\alpha}\right)\left(\Delta\varepsilon\left|\vec{E}_\parallel(\vec{r})\right|^2 - \Delta\varepsilon^{-1}\left|\vec{D}_\perp(\vec{r})\right|^2\right)dA\Biggr)\label{eq:g_mb_electrical}.
\end{align}

\section{Electromechanical system}
There are several different designs for implementing electromechanical devices and many of them have proven to be quite efficient. However, each design has its own advantages and disadvantages. Our design is based on a specific structure known as a nano-string. Nano-strings are the only proposed electromechanical structures that can be coupled to photonic crystals. Using nano-strings opens the way to further research on opto-electro-mechanical systems based on photonic crystals, as well as other tripartite systems.
\begin{figure}[!h]
    \centering
     \begin{subfigure}[b]{0.6\textwidth}
         \centering
         \includegraphics[width=\textwidth]{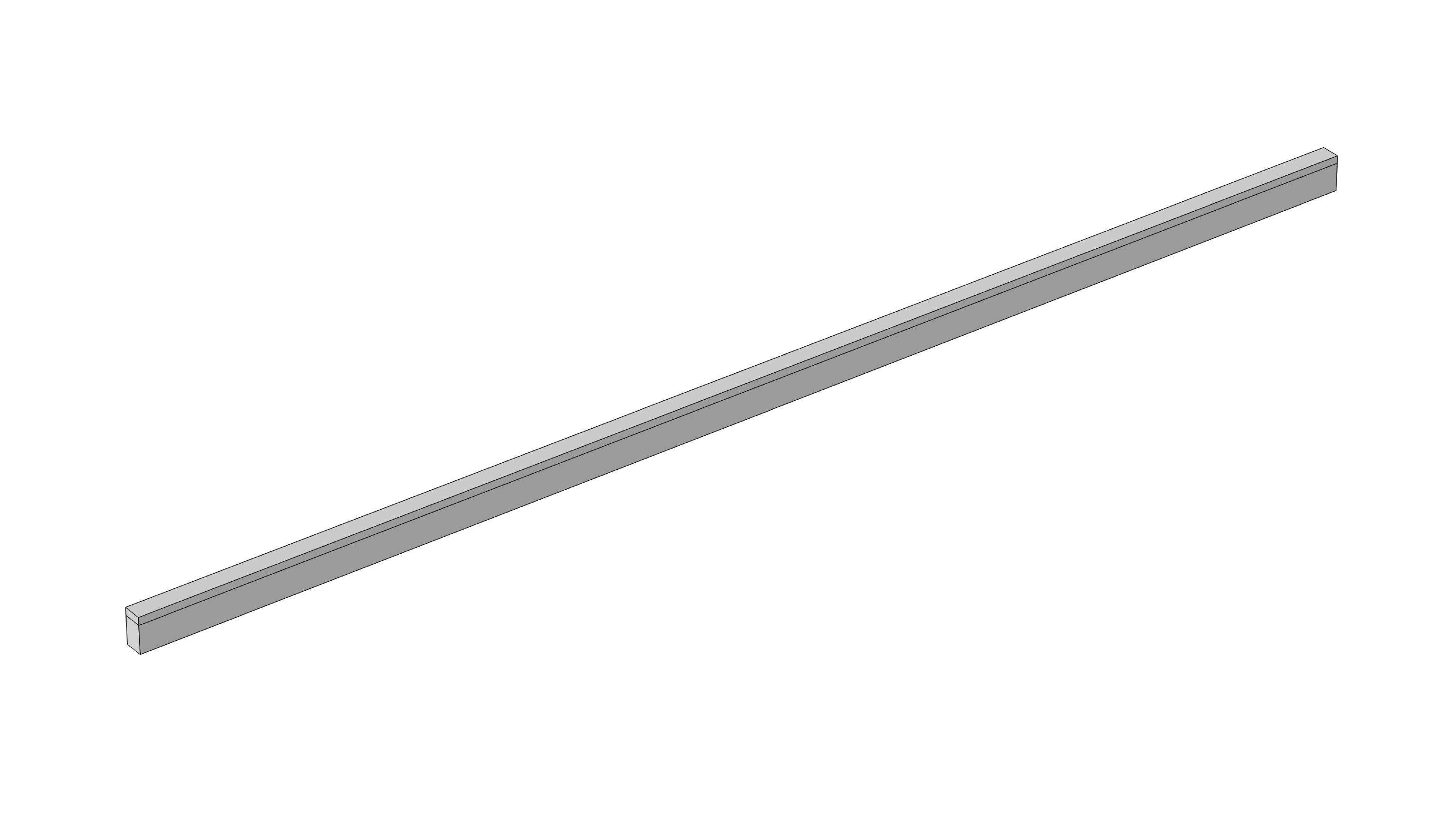}
         \caption{}
         \label{fig:nanostring-concept}
     \end{subfigure}
     \hfill
     \begin{subfigure}[b]{0.47\textwidth}
         \centering
         \includegraphics[width=\textwidth]{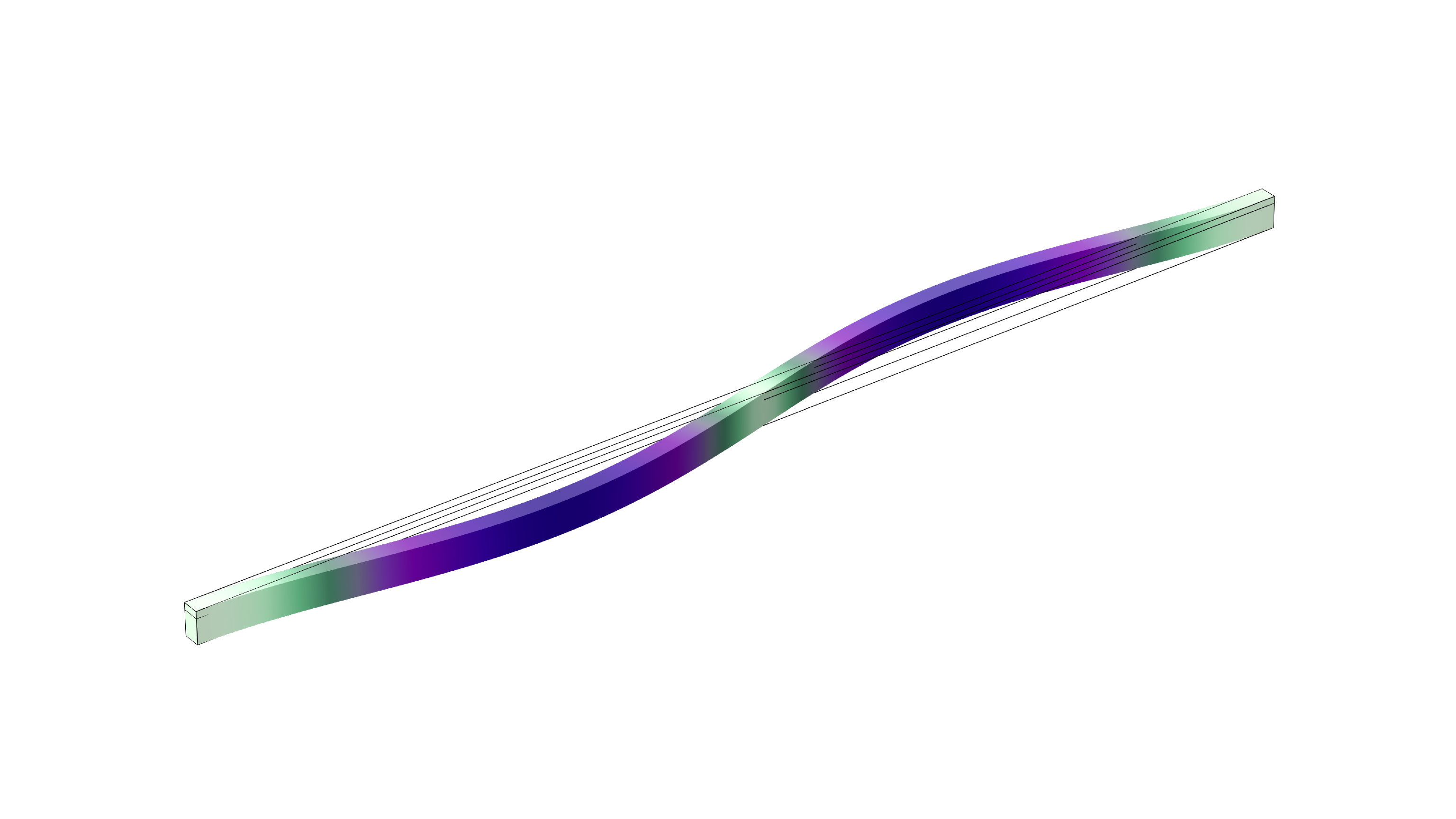}
         \caption{}
         \label{fig:nanostring_first_mode}
     \end{subfigure}
     \begin{subfigure}[b]{0.47\textwidth}
         \centering
         \includegraphics[width=\textwidth]{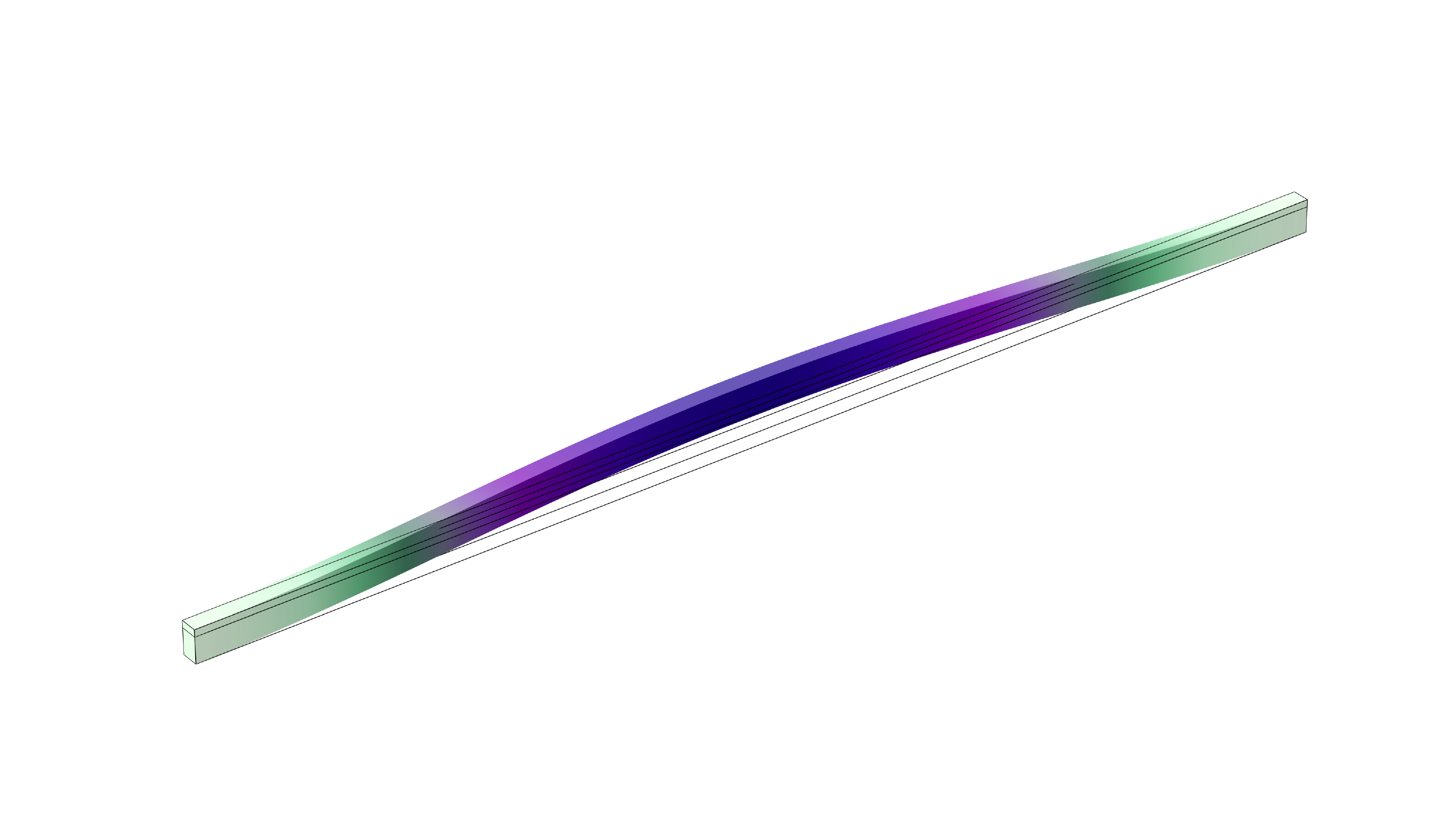}
         \caption{}
         \label{fig:nanostring_second_mode}
     \end{subfigure}
     \caption[Nano-string basic concept]{(a) The nano-string is made out of a dielectric, with a thin layer of metal covering its surface. (b) The first in-plane mechanical mode. (c) The second in-plane mechanical mode.}
     \label{fig:nano-string_basic_concept_modes}
\end{figure}
A nano-string refers to a micrometer-scale string composed of a dielectric material, characterized by its narrow width and thickness in the nanometer range (Figure \ref{fig:nanostring-concept}). To facilitate its functionality, the nano-string is coated with a thin layer of metal, effectively forming a long and slender conductive element. With the two ends of the nano-string securely fixed, it becomes possible to generate standing mechanical waves along both the nano-string itself and the metal layer that is attached to its surface. Figures (\ref{fig:nanostring_first_mode}) and (\ref{fig:nanostring_second_mode}) demonstrate the first and the second in-plane mechanical mode of the nano-string.

\begin{figure}[!h]
    \centering
     \begin{subfigure}[b]{0.7\textwidth}
         \centering
         \includegraphics[width=\textwidth]{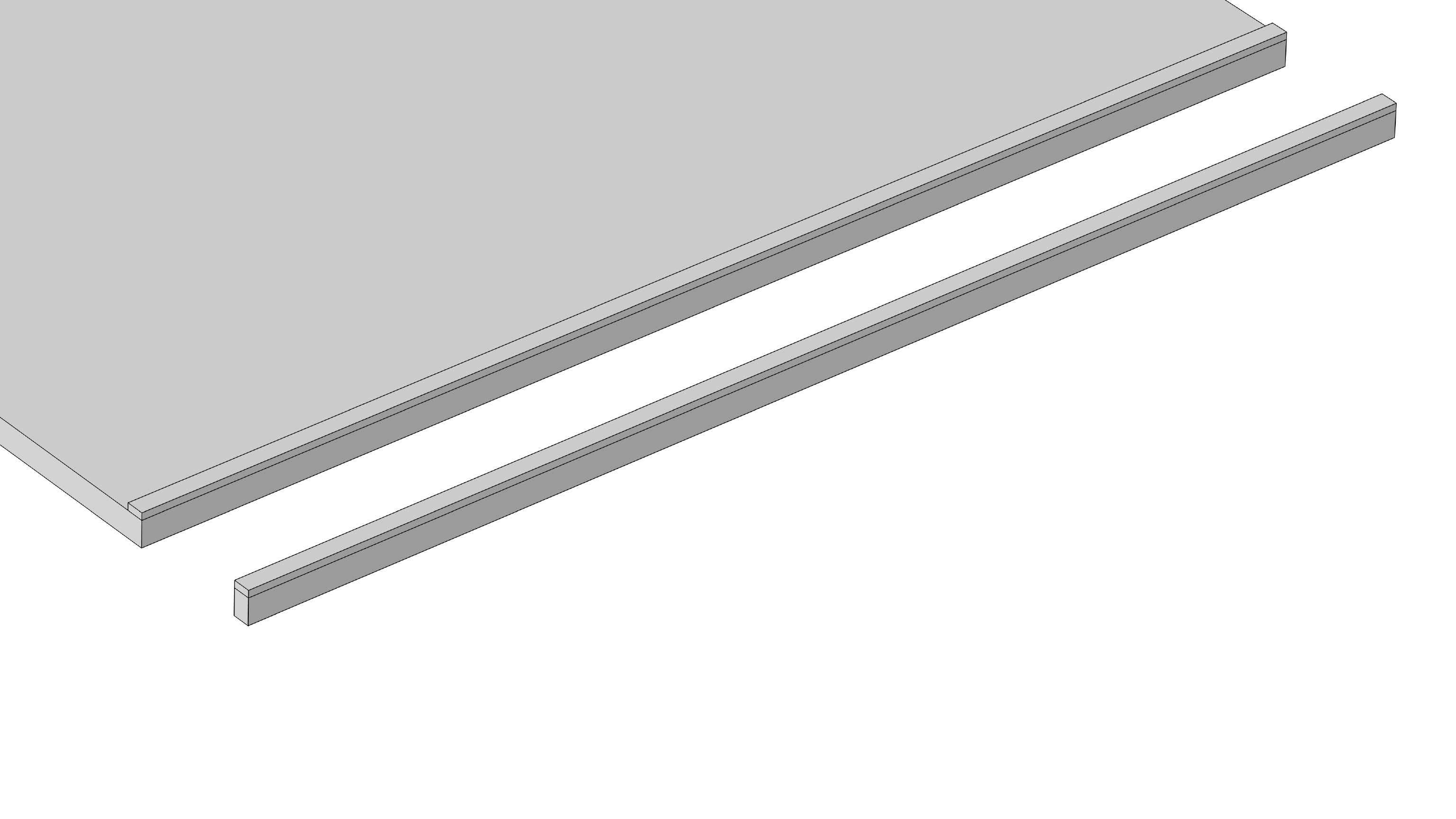}
         \caption{}
         \label{fig:nanostring-as-a-capacitor}
     \end{subfigure}
     \hfill
     \begin{subfigure}[b]{0.47\textwidth}
         \centering
         \includegraphics[width=\textwidth]{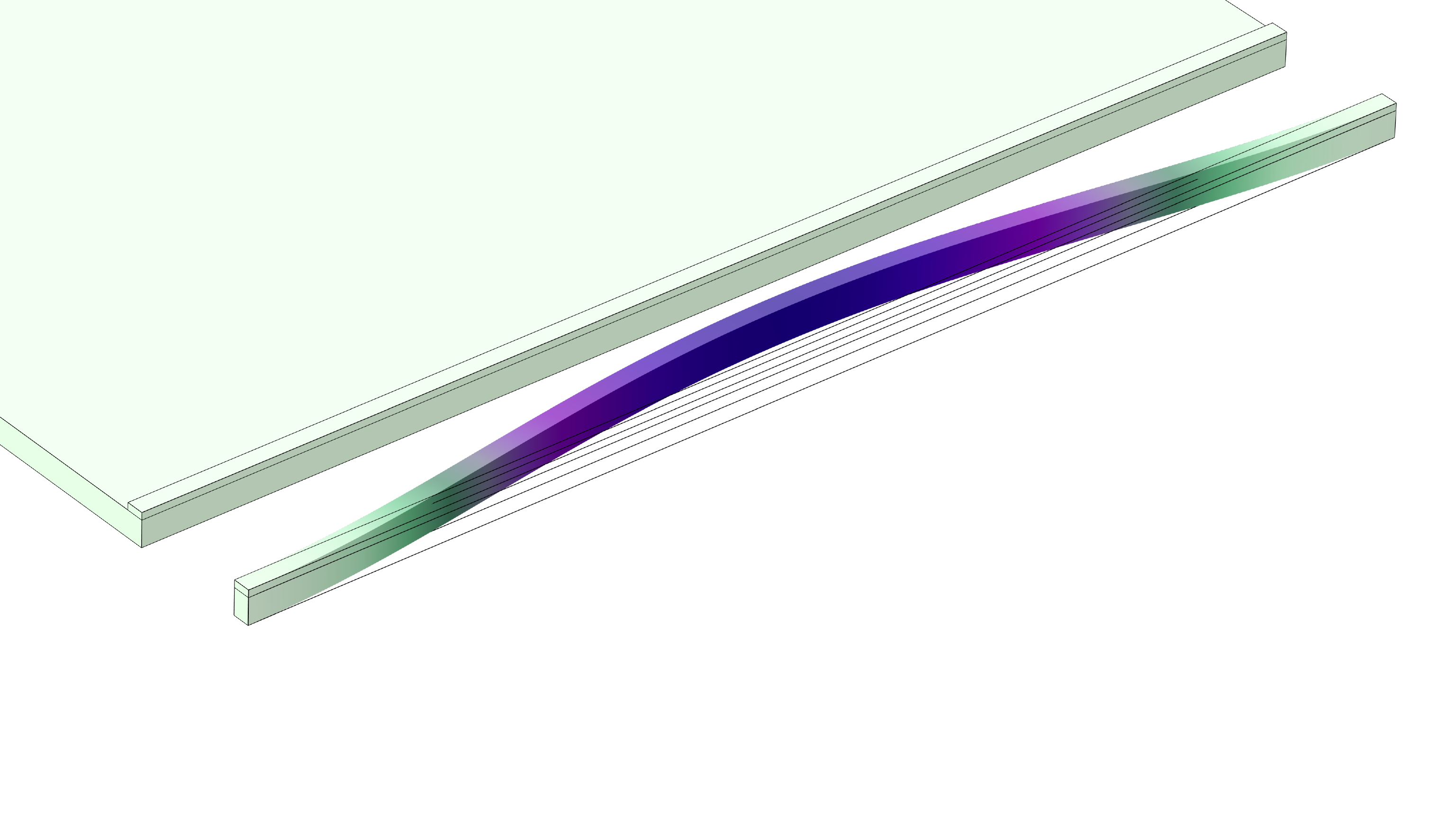}
         \caption{}
         \label{fig:capacitor_first_mode}
     \end{subfigure}
     \begin{subfigure}[b]{0.47\textwidth}
         \centering
         \includegraphics[width=\textwidth]{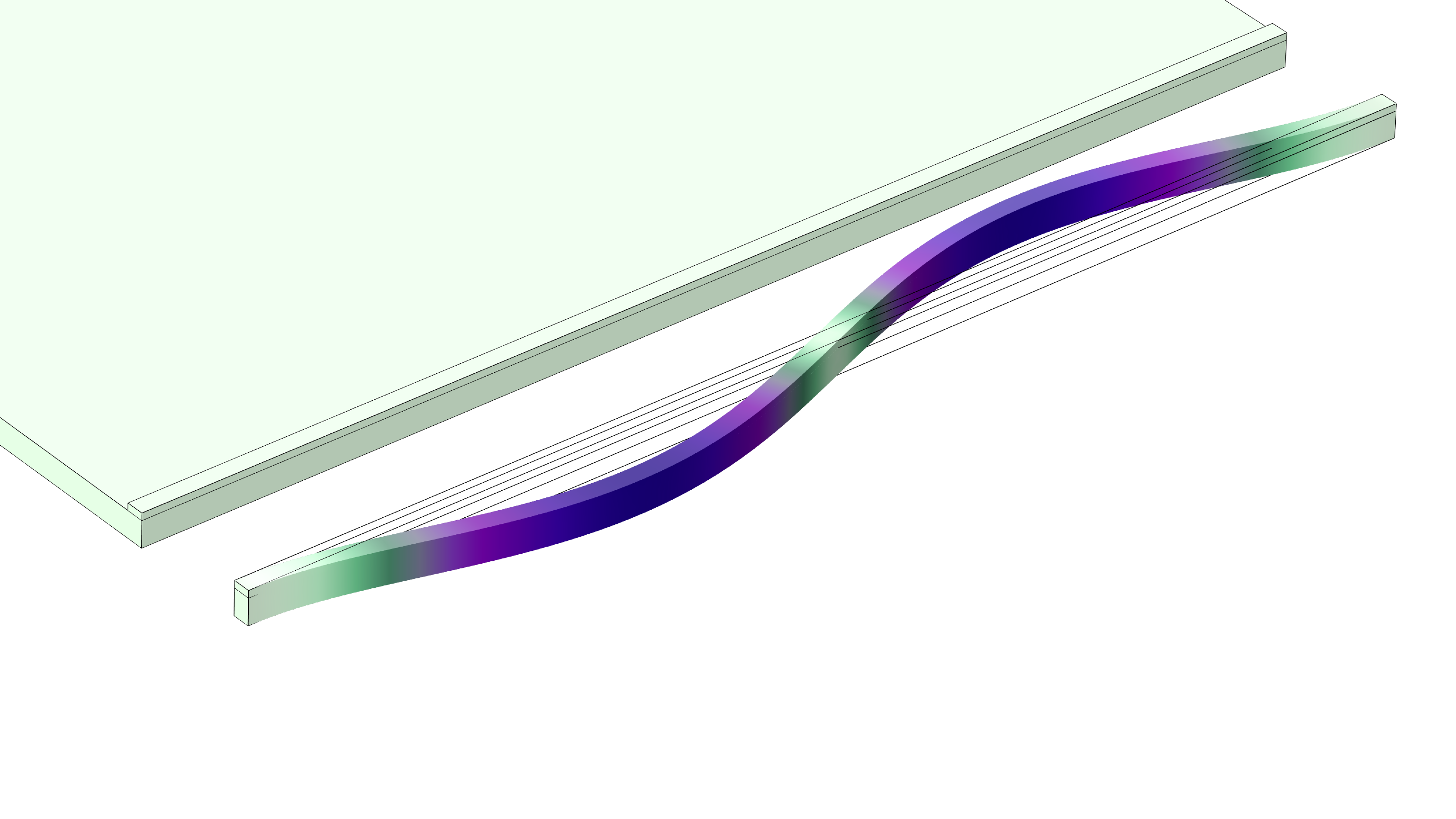}
         \caption{}
         \label{fig:capacitor_second_mode}
     \end{subfigure}
     \caption[The electromechanical capacitor concept]{(a) The nano-string and the fixed (larger) piece of dielectric, and the two metal electrodes on top of the dielectrics. The structure is a nano-capacitor and can be used for electromechanical purposes. For quantum electromechanics, the gap between the nano-string and the fixed part should typically be in $<100$ nm range so the interaction would be measurable. However, in the figure, the gap is exaggerated to express the concept better. (b) The first in-plane mechanical mode, (c) The second in-plane mechanical mode.}
     \label{fig:nano-string-capacotor_modes}
\end{figure}
Now, if we implement a long but narrow piece of metal on top of a fixed piece of the dielectric, in the proximity of the nano-string, we will have a capacitor with one fixed and one oscillating plate (see Figure \ref{fig:nanostring-as-a-capacitor}). Then we can connect the plates of the capacitor to an arbitrary type of microwave cavity (e.g. 3D cavity or an integrated LC circuit). Ultimately, the mechanical modes of the nano-string (Figures \ref{fig:capacitor_first_mode} and \ref{fig:capacitor_second_mode}) can be coupled to the microwave mode of the resonator and form an electromechanical device.

In terms of material selection, we have opted for aluminum as the metal component due to its superconducting properties. Superconductivity causes less loss in the system resulting in higher quality factors. As for the dielectric material, silicon emerges as our primary choice given its high refractive index ($n_\text{silicon} = 3.48$). However, we have also explored the potential of diamonds in our simulations. Despite diamond's lower refractive index ($n_\text{diamond} = 2.3$) compared to silicon, it is worth noting that diamond can accommodate NV-Centers \cite{schirhagl2014nvcenters}, which opens up possibilities for future investigations on tripartite atomic-electro-mechanical systems. Although the electromechanical coupling rate may be smaller in diamond, its unique properties make it a compelling avenue for further studies.

\subsection{Design}
\begin{figure}[!h]
    \centering
    \begin{subfigure}[b]{0.9\textwidth}
        \centering
        \includegraphics[width=\textwidth]{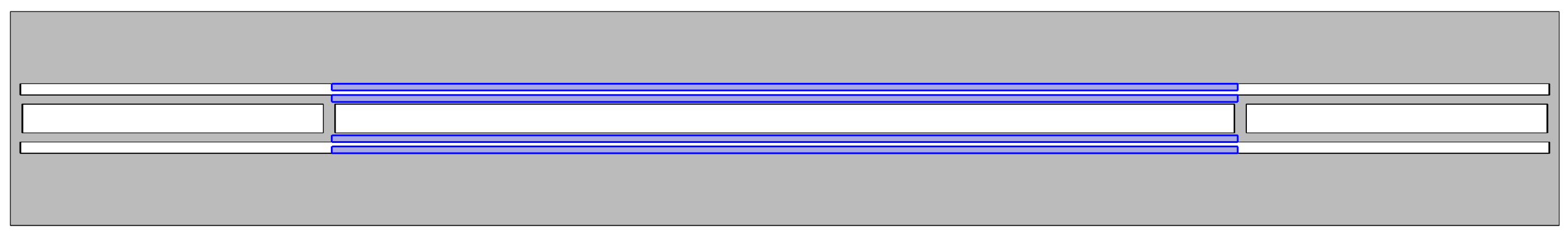}
        \caption{}
        \label{fig:EM_topview1}
    \end{subfigure}
    \hfill
    \begin{subfigure}[b]{0.9\textwidth}
        \centering
        \includegraphics[width=\textwidth]{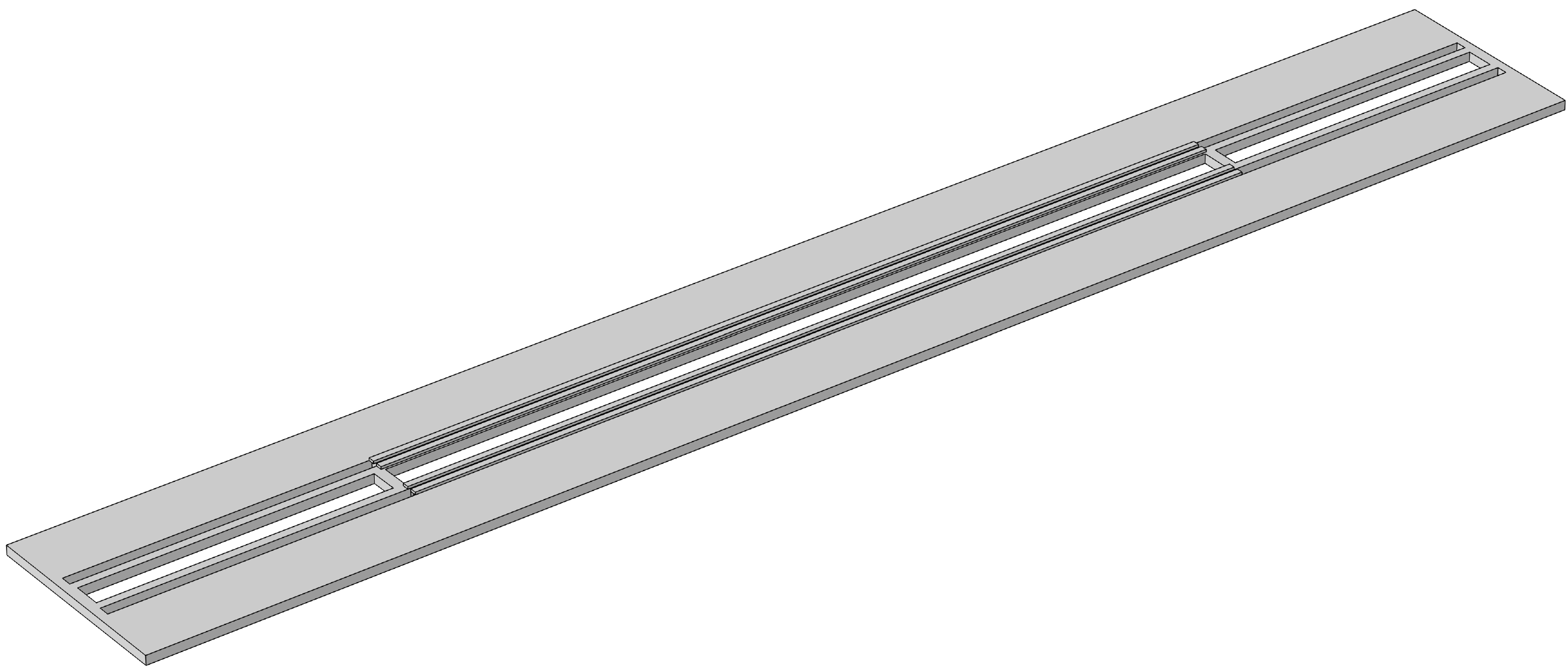}
        \caption{}
        \label{fig:EM_angleview1}
    \end{subfigure}
    \caption[The nano-string design]{The structure is formed on a $220$ nm thick silicon slab. Some parts of the silicon slab are removed to make the oscillating parts, and four aluminum electrodes are placed on the surface of the slab to form two identical capacitors. (a) The top view of the design. The gray parts are silicon and the blue parts indicate aluminum electrodes. (b) An oblique view of the design.}
    \label{fig:EM1}
\end{figure}
In the subsequent sections, we will provide an overview of the design and simulations conducted for our nano-string structure. Figures (\ref{fig:EM_topview1}) and (\ref{fig:EM_angleview1}) demonstrate the specific design of our nano-string from different angles, which is implemented on a silicon slab with a thickness of $220$ nm. The central region of the structure exhibits a $\doubleH$-shaped pattern, achieved by creating voids within the slab. The two elongated sections within the central $\doubleH$ shape correspond to the nano-strings previously mentioned. These nano-strings are designed to undergo oscillatory motion in harmonic modes. Positioned on the structure are four elongated aluminum pieces with a thickness of $60$ nm, which serve as two capacitors. The two outer aluminum electrodes are fixed and located on the larger and heavier section of the structure, thereby remaining stationary. On the other hand, the inner electrodes are placed on the nano-strings and can experience oscillatory motion in the harmonic modes as shown previously.

\begin{figure}
    \centering
    \includegraphics[width=0.75\textwidth]{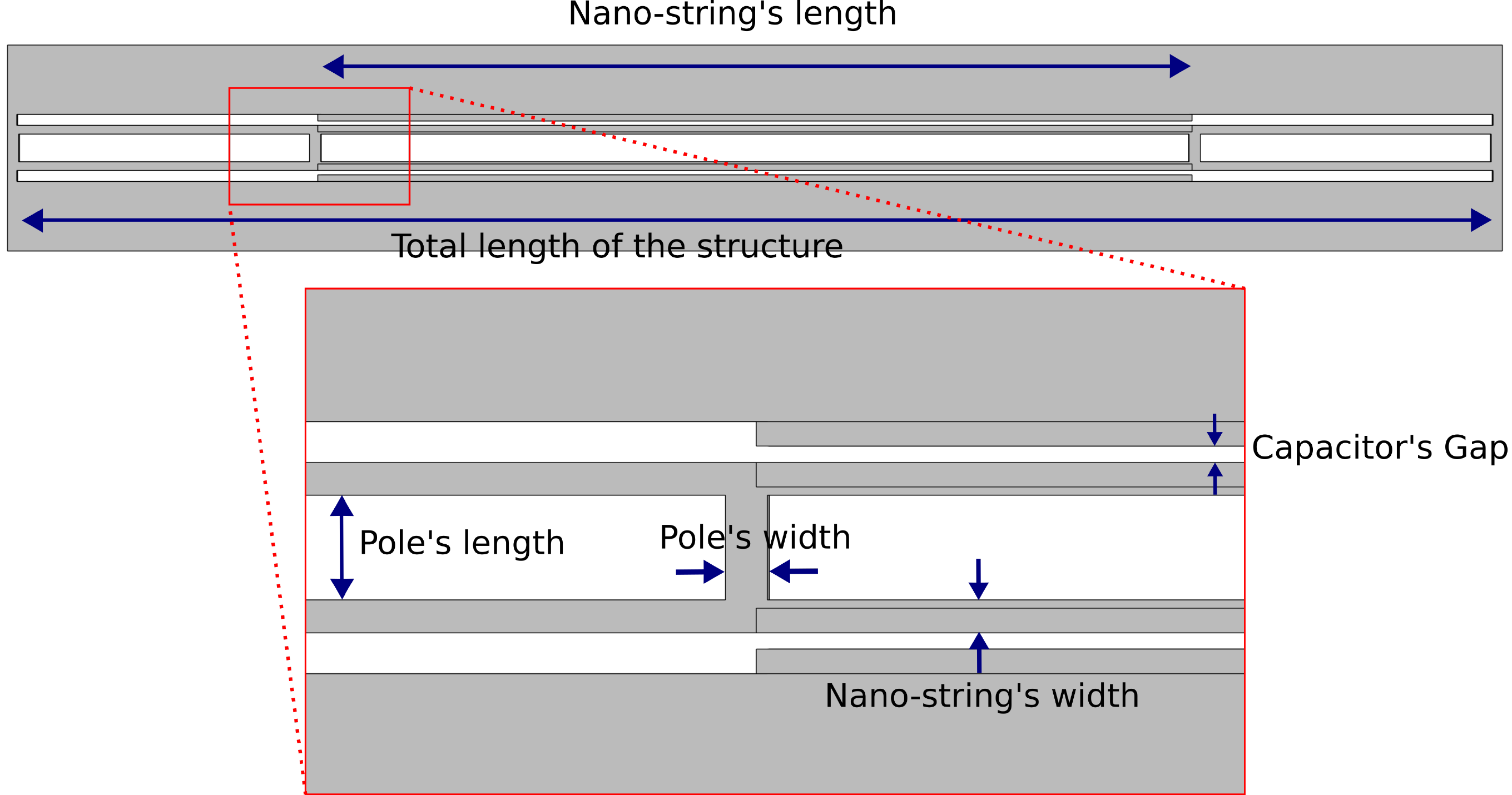}
    \caption[Key geometrical parameters]{Some of them will be set fixed through the whole simulation process, and some of them will be swept to see how they affect the behavior of the system.}
    \label{fig:EM_geometrical_parameters1}
\end{figure}

The key geometrical parameters of the system are shown in Figure (\ref{fig:EM_geometrical_parameters1}).
The most affecting parameter is the Capacitor's gap. The capacitance and the electromechanical coupling rate are highly dependent on the gap.
The smaller the gap is, the higher the capacitance and the coupling rate become, but there is a lower limit to the gap size that is set by the accuracy of the fabrication techniques.
In general, the gap can be different for the upper and the lower capacitors to make different capacities for some purposes, but we keep the structure symmetric and set both gaps equal.
The nominal value for the gap in our design is $55$ nm, however, since the fabrication inaccuracy is inevitable at this scale and the gap can be different, we will sweep the gap up to $100$ nm through the simulations.
Another important geometrical parameter of the nano-string structure is its length. It is worth noting that a longer nano-string leads to a larger capacitance, resulting in a higher electromechanical coupling rate. However, the frequencies of the mechanical modes are inversely proportional to the length of the nano-string. This means that increasing the length of the nano-string leads to a decrement in frequency and an increment in the occupation number of noise at a fixed temperature (see Table \ref{tab:thermal-noise-quanta}). Therefore, there exists a trade-off between the electromechanical coupling rate and the level of mechanical noise for the chosen length of the nano-string.\\
In our specific design, we have fixed the length of the nano-string to $20$ $\mu$m. Furthermore, the total length of the structure is set to $33.9$ $\mu$m, the width and length of the poles are $250$ nm and $640$ nm respectively, and the width of the nano-string is $200$ nm. All of the aluminum electrodes are identical in size, with a width and thickness of $150$ nm and $60$ nm respectively. It is important to note that the length of the electrodes always matches the length of the nano-strings in our design.

\begin{figure}[!h]
    \centering
    \begin{subfigure}[b]{0.9\textwidth}
        \centering
        \includegraphics[width=\textwidth]{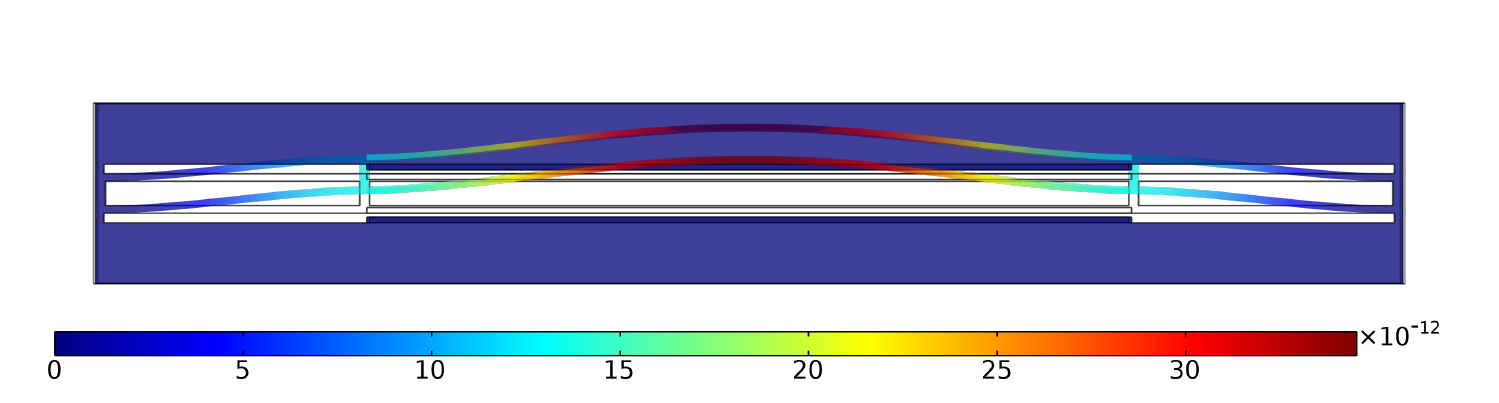}
        \caption{}
        \label{fig:EM_first_mechanical_mode}
    \end{subfigure}
    \begin{subfigure}[b]{0.9\textwidth}
        \centering
        \includegraphics[width=\textwidth]{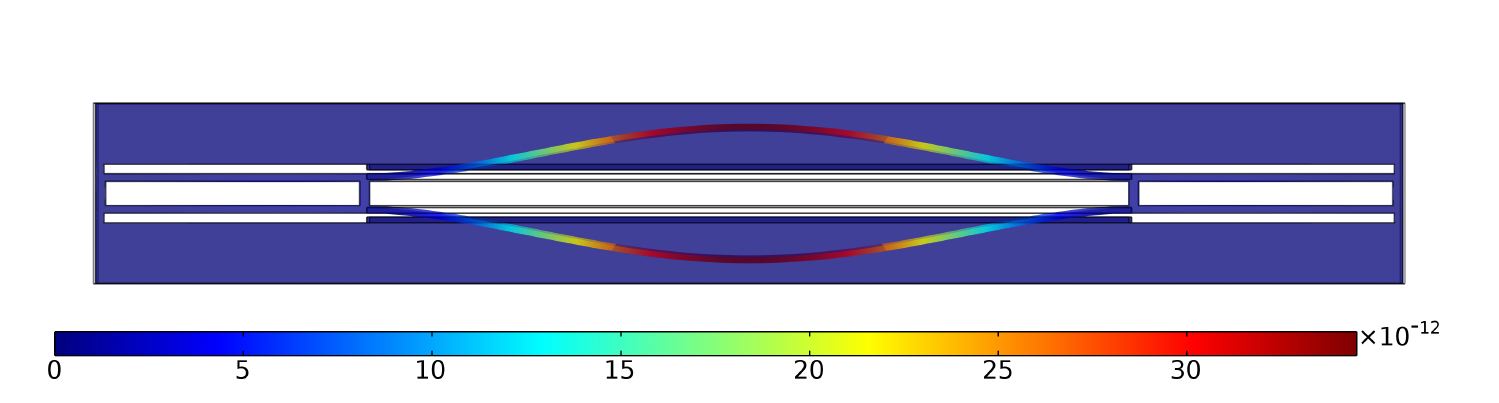}
        \caption{}
        \label{fig:EM_second_mechanical_mode}
    \end{subfigure}
    \caption[First two in-plane mechanical modes]{(a) The first in-plane mechanical mode. (b) The second in-plane mechanical mode. Note that the magnitude of the deformation is exaggerated so the shape of the mode becomes obvious. By looking at the color legend, one can see that the maximum displacement is $\sim 35$ pm which is much less than the gap size, $55$ nm.}
    \label{fig:EM_in_plane_mechanical_modes}
\end{figure}

As we will see in the next section, one can simulate the mechanical behavior of the structure and find the modes. 
Figures (\ref{fig:EM_first_mechanical_mode}) and (\ref{fig:EM_second_mechanical_mode}) show the first and the second in-plane mechanical modes of the structure. The first mechanical mode has a greater impact on the capacitance of each capacitor, compared to the second mode, due to the larger cumulative displacement of the nano-strings. However, one should note that in the first mode, with increasing the gap of one of the capacitors, the gap of the other one decreases, so increasing one capacitance happens along with decreasing the other. Therefore, if we connect the two capacitors and connect the system of capacitors to one microwave cavity, the ratio $\frac{d C_m}{d\alpha}$, and consequently, the electromechanical coupling rate will vanish. In conclusion, the first mechanical mode is suitable in the case we connect each capacitor to a separate microwave cavity.

On the other hand, the second mechanical mode has a smaller cumulative displacement of the nano-strings but the capacitance of the two capacitors oscillates in phase. Therefore, in the case that both capacitors are connected to one microwave cavity, the electromechanical coupling rate will be non-vanishing. Note that we still can use the second mechanical mode for the case of two separate microwave cavities but as the cumulative displacement of the nano-strings and the electromechanical coupling rate will be less than the first mode, the first mode is a better option for this case. To conclude, the second mechanical mode is suitable in case we connect the capacitors to each other and to one microwave cavity.

There are two primary options for implementing microwave cavities in our system. The first option involves integrating an LC circuit on the same chip as the capacitors and connecting the capacitors to the LC circuit. This approach offers advantages such as higher typical quality factors and scalability, as the entire system is fully integrated on a single silicon chip.

The second option is to utilize a 3D microwave cavity. A 3D microwave cavity consists of an empty space carved within a highly-conductive or superconductive material. The walls surrounding this empty space trap electromagnetic fields, and the frequency of the cavity modes depends on its inner dimensions. To couple the capacitors to a 3D microwave cavity, we implement two pads on the sides of the silicon chip and connect the capacitors to them. The pads can be coupled to the cavity either capacitively or by galvanic connection to the cavity walls. The advantages of 3D microwave cavities include easier fabrication and higher robustness to noise, due to their higher volume-to-surface ratio when compared to integrated circuits.

\subsection{Electromechanical simulation}

With a clear understanding of the design, we can now proceed to simulate and extract the mechanical and electromechanical characteristics using COMSOL Multiphysics Software. In this section, I will outline the main steps involved. For more detailed instructions, please refer to Appendix (\ref{appendix:simulation-electromechanical}).

The first step is to create the geometry, which encompasses the silicon slab with the desired patterns. This can be achieved by initially generating a plain slab with the appropriate thickness. Subsequently, smaller slabs overlapping with the main plain slab are created, and the smaller slabs are subtracted from the main one. This process results in the desired patterned structure.
Next, we generate four long and narrow electrodes within the geometry. These electrodes play a crucial role in the functionality of the capacitors.
As the final step, we create an air box that surrounds the entire structure. This ensures an appropriate simulation environment.

\begin{figure}[!h]
    \centering
    \includegraphics[width=0.9\textwidth]{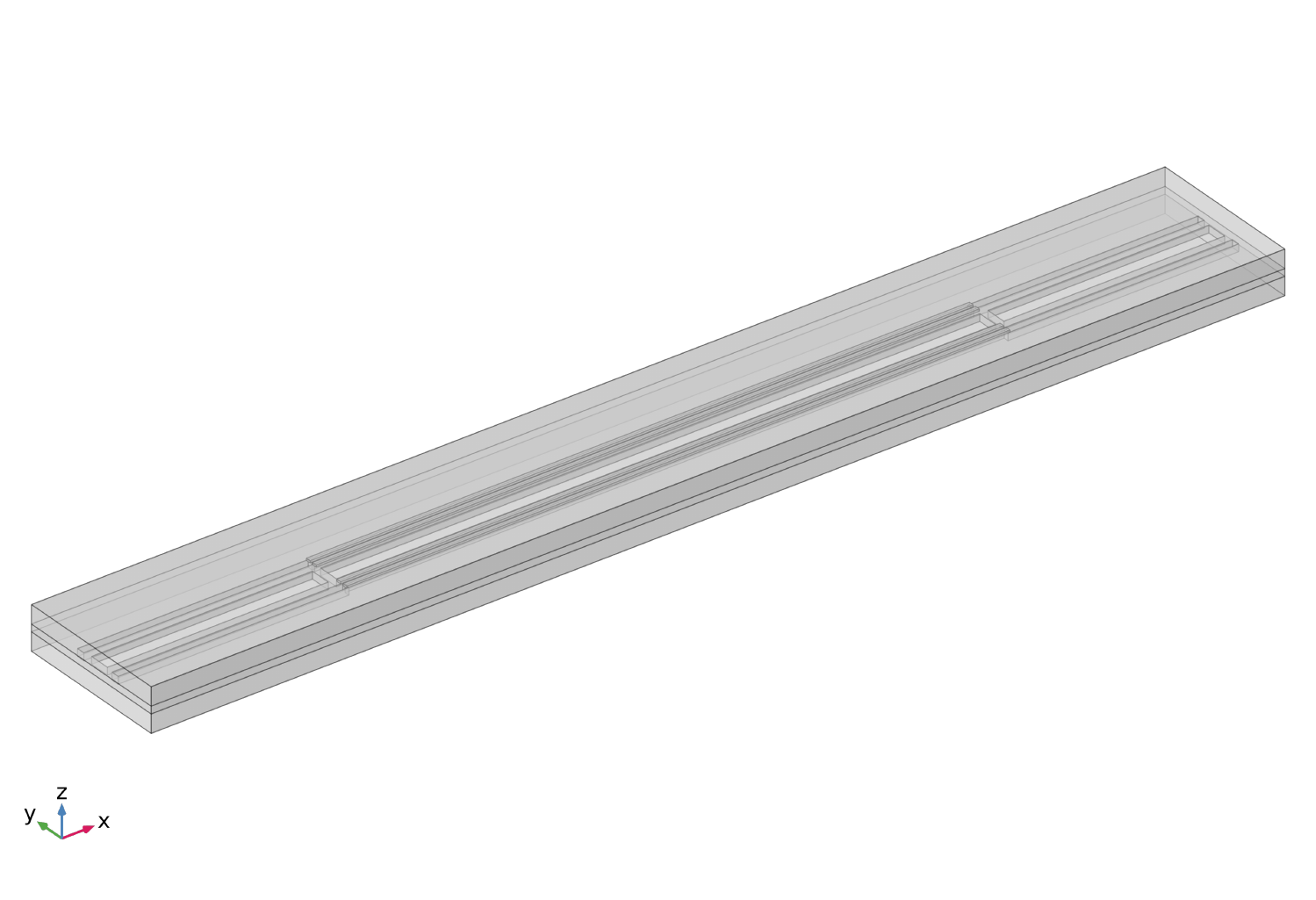}
        \caption[The final geometry of a nano-string]{The final state of the geometry contains the substrate with the cuts on it, four electrodes, and an air box surrounding the structure.}
    \label{fig:EM_with_airbox2}
\end{figure}

Figure (\ref{fig:EM_with_airbox2}) provides a visual representation of the expected final geometry, illustrating all the components and their relative positions within the design. 

The next step involves setting the materials within the simulation. This can be accomplished by adding the materials `Silicon', `Aluminum', and `Air' to the model and assigning them to the respective domains. It's important to note that the software requires the `Relative permittivity' parameter of all materials involved in the electrostatic simulation to be defined numerically.

In the case of `Aluminum', the relative permittivity is not defined by default in the COMSOL material library. To address this, we manually define the relative permittivity of `Aluminum' in our model. While the relative permittivity of perfect conductors is theoretically infinite for static electric fields, the software requires a finite number to be entered. Therefore, a large number, such as $1 \times 10^{12}$, is sufficient for our purposes.

In the third step, we need to add the required `Physics' to the model and set them properly. We will include `Solid Mechanics' to simulate the mechanical properties of the structure. To set up the mechanical simulation, we first exclude the air box as we are specifically interested in the mechanical properties of the structure itself. Then, we apply a fixed boundary condition to all four surfaces on the sides of the silicon slab. This ensures that we focus on the mechanical modes of the nano-strings rather than the slab itself. Optionally, a `Symmetry' boundary condition can be applied to the bottom surface of the slab if we are only interested in the in-plane mechanical modes. However, for our purposes, we want to find all the mechanical modes, so we do not apply this condition.

Once the `Solid Mechanics' is set, we add the `Electrostatics' physics. This physics should be applied to all the domains and the entire structure. We also need to apply a `Ground' boundary condition to all the surfaces of the inner electrodes as well as the `Terminal' boundary condition to all the surfaces of the outer electrodes. The `Terminal' boundary condition should be set to `Voltage' where the voltage value is $1$ V.

After setting up the physics, we need to define the mesh for the simulation. We can start with the `Physics-controlled mesh' option, but for optimal results, manual management of the mesh is required. The mesh elements in the vicinity of the electrodes and in the gap region should be sufficiently small to ensure reliable precision in the electrostatic simulation. However, using excessively small mesh elements can lead to an extremely large number of mesh elements, which may render the simulation impractical with limited computational resources. As a general guideline, having two mesh elements along the gap is typically enough to achieve precise results.

The last step before beginning the post-processing is to set studies, to perform the simulations. Using an `Eigenfrequency' study we simulate the mechanical modes, and a `Stationary' to simulate the electric field distribution. Therefore, the only physics used in the `Eigenfrequency' study is `Solid Mechanics', and in the `Stationary' study, we only use `Electrostatics' physics. Using a simpler separate mechanical simulation of a nano-string with the same geometrical parameters, we found that the second mechanical mode should occur close to $3.8$ MHz, so, it is a good initial guess for finding the eigenmodes.

\begin{figure}
    \centering
    \begin{subfigure}[b]{0.48\textwidth}
        \centering
        \includegraphics[width=\textwidth]{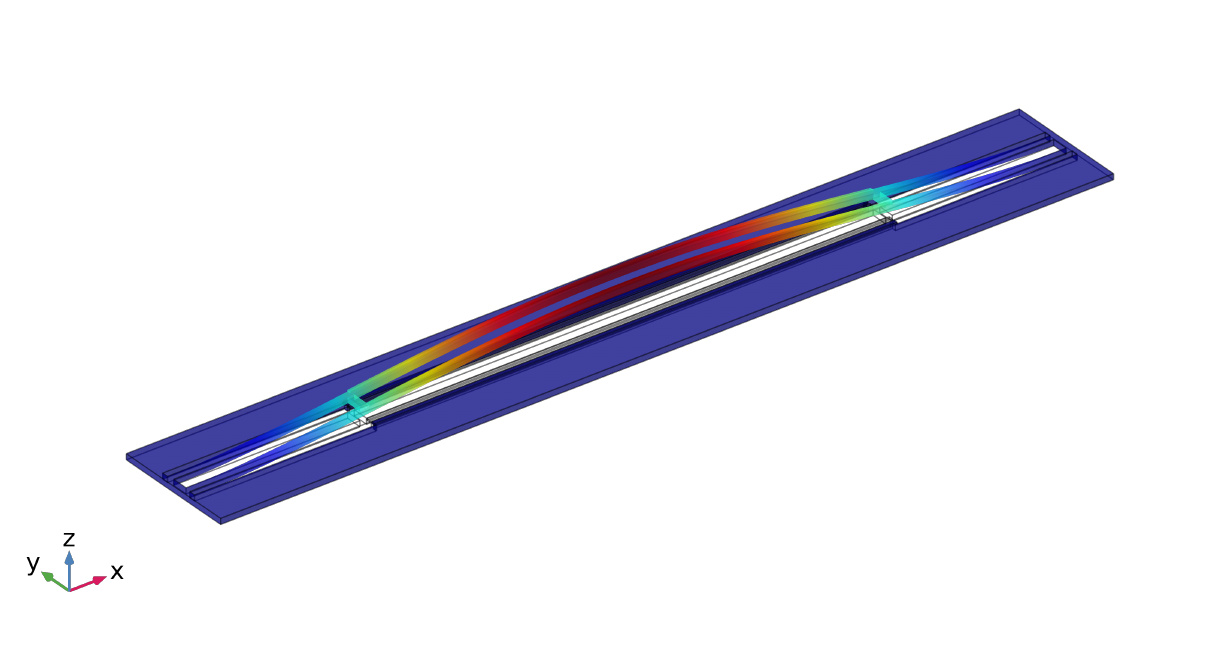}
        \caption{}
        \label{fig:EM-first-out-of-plane-mode}
    \end{subfigure}
    \begin{subfigure}[b]{0.48\textwidth}
        \centering
        \includegraphics[width=\textwidth]{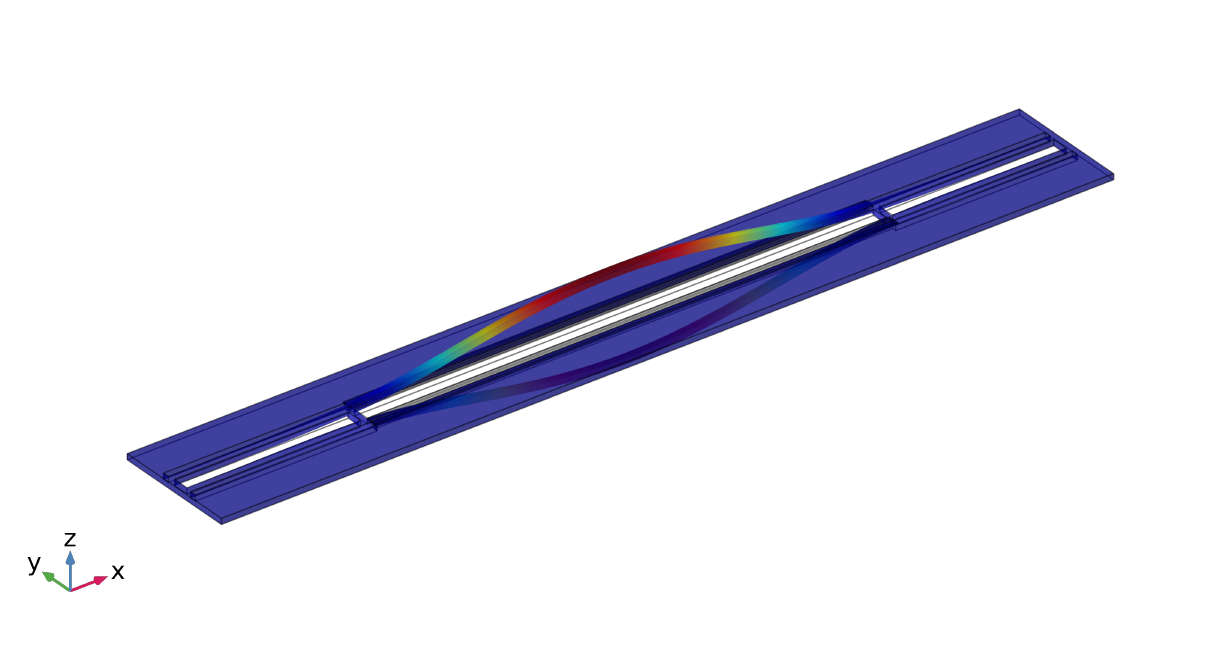}
        \caption{}
        \label{fig:EM-second-out-of-plane-mode}
    \end{subfigure}
    \caption[Out-of-plane mechanical modes]{(a) First out-of-plane mode. (b) Second out-of-plane mode.}
    \label{fig:out-of-plane-modes}
\end{figure}

You can see the quantitative results of the mechanical simulation for the first four modes in Table (\ref{tab:mech_modes1}). The in-plane modes are shown previously in Figure (\ref{fig:EM_in_plane_mechanical_modes}), and first and second out-of-plane modes are demonstrated in Figures (\ref{fig:EM-first-out-of-plane-mode}) and (\ref{fig:EM-second-out-of-plane-mode}), respectively.
\begin{table}[!h]
    \begin{center}
        \begin{tabular}{|c||c|c|c|c|}
            \hline
            Mode &  1st in-plane & 2nd in-plane & 1st out-of-plane & 2nd out-of-plane \\
            \hline\hline
            $\Omega$ [MHz] &  $2\pi\times2.81$ & $2\pi\times3.86$ & $2\pi\times1.60$ & $2\pi\times4.47$ \\
            \hline
            $m_\text{eff}$ [pg] & $3.03$ & $2.06$ & $3.51$ & $2.14$\\
            \hline
        \end{tabular}
        \caption[Mechanical simulation results]{\textbf{Mechanical simulation results} - The eigenfrequencies and effective masses of the mechanical modes are summarized in the table.}
        \label{tab:mech_modes1}
    \end{center}
\end{table}

\begin{figure}
    \centering
    \begin{subfigure}[b]{0.8\textwidth}
        \centering
        \includegraphics[width=\textwidth]{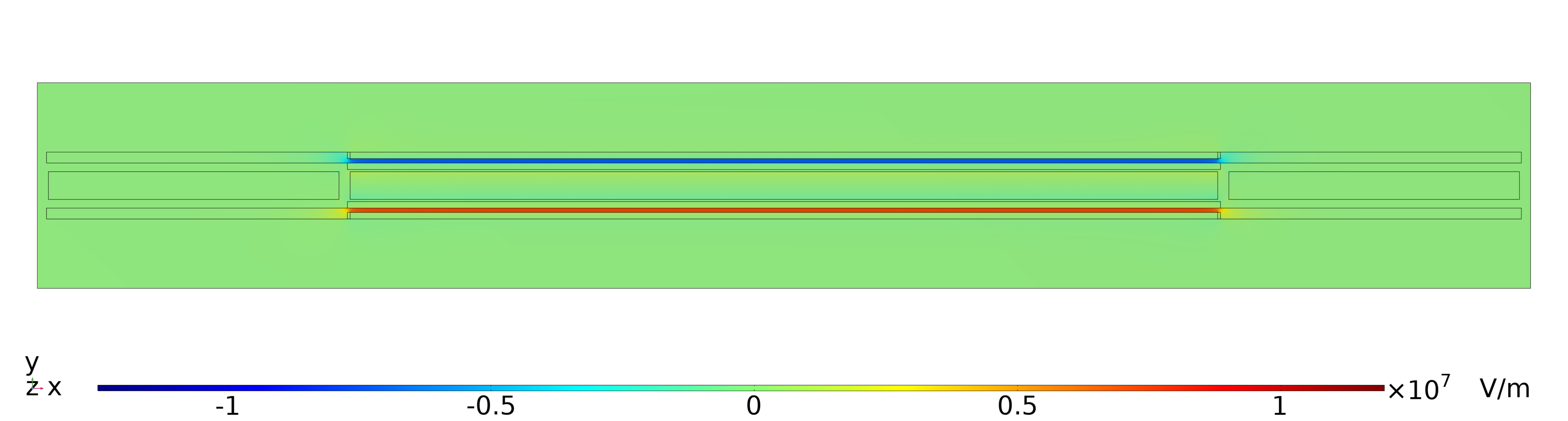}
        \caption{}
        \label{fig:electrostatic-simulation-Ey}
    \end{subfigure}
    \begin{subfigure}[b]{0.8\textwidth}
        \centering
        \includegraphics[width=\textwidth]{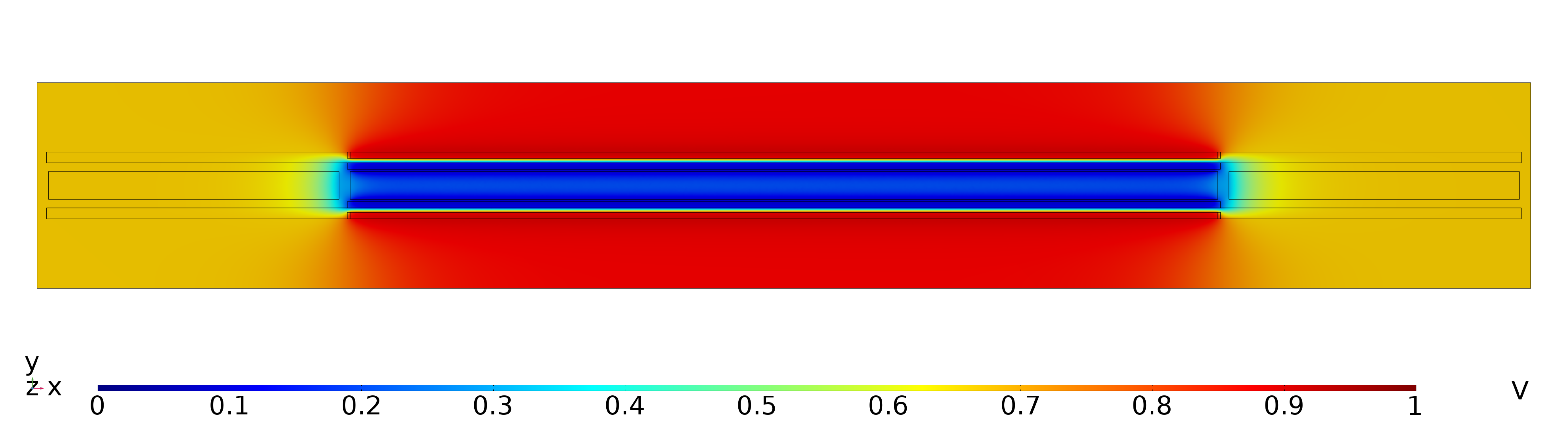}
        \caption{}
        \label{fig:electrostatic-simulation-V}
    \end{subfigure}
    \caption[Electrostatic simulation results]{The figures show (a) the y-component of the electric field, (b) the electric potential}
    \label{fig:electrostatic-simulation}
\end{figure}

The electrostatic simulation results in the electric field and the electric potential distribution (Figure \ref{fig:electrostatic-simulation-V}). As shown in Figure (\ref{fig:electrostatic-simulation-Ey}), the y-component of the electric field is concentrated in the gap regions, and as we will see, the mechanical modes that modulate the gap in the y-direction (in-plane modes) will have the largest electromechanical coupling rates.

\subsubsection{Post processing}
So far, we have successfully determined the mechanical modes, including their shape and frequency, as well as the distribution of the electric field. However, to extract the electromechanical coupling rate, we need to perform post-processing.\\
According to Equation (\ref{eq:effective_mass}), the effective mass of each mechanical mode can be calculated by a volume integral. Having the effective mass, one can find the zero-point fluctuation of each mechanical mode. The volume integral in the denominator of Equation (\ref{eq:g_mb_electrical}) can be also taken as we have the electric field distribution. Note that the integral not only is a part of the equation of the electromechanical coupling rate but also is equal to twice the total electric energy stored in the system, $\mathcal{E}_E$, due to its capacitance. According to the equation $\mathcal{E}_E=\frac{1}{2}C_mV_C^2$, and recalling that we set the electric potential difference between the plates of the capacitor $V_C=1$V, the capacitance of the capacitor, $C_m$, is equal to,
\begin{equation}
    C_m = 2\mathcal{E}_E = \displaystyle\int\varepsilon(\vec{r})\left|\vec{E}(\vec{r})\right|^2d^3\vec{r}.
\end{equation}
Therefore, we also obtain $C_m$, and having the stray capacitance, $C_s$, and the inductance, $L$, of the cavity of our interest (from relatively simple simulations either in COMSOL or Sonnet), the participation function, $\eta = C_m/(C_s+C_m)$, and the final microwave frequency, $\omega_c = 1/\sqrt{L(C_s+C_m)}$, will be calculated.\\
Finally, as the simulation leads to the electric field, $\vec{E}(\vec{r})$, the electric displacement field, $\vec{D}(\vec{r})$, and the mechanical deformation, $\vec{Q}(\vec{r})$, the surface integrals in Equation (\ref{eq:g_mb_electrical}) can be solved, and we eventually calculate the electromechanical coupling rate.
The post-processed parameters for the second in-plane mechanical mode and $70$ nm gap are presented in Table (\ref{tab:2nd-in-plane-mode-simulation_results}). The cavity specifications are $C_s = 10.97$ fF, and $\omega_c = 9.9$GHz, based on a separate simulation.

\begin{table}[!h]
    \centering
    \begin{tabular}{|c|c|c|c|c|c|c|}
        \hline
         $\Omega/2\pi$ & $m_\text{eff}$ & $x_\text{ZPF}$ & $C_m$ & $\eta$ & $g_0$  \\
         \hline\hline
         $3.85$ MHz & $2.06$ pg & $32.49$ fm & $1.78$ fF & $0.140$ & $65.63$ Hz\\
         \hline
    \end{tabular}
    \caption[Electromechanical properties of the second in-plane mode with silicon substrate]{\textbf{Electromechanical properties of the second in-plane mode with silicon substrate} - The numbers describe a nano-string structure based on a silicon substrate and for $70$ nm gap between the electrodes. The cavity parameters are assumed $C_S = 10.97$ fF and $\omega_c = 2\pi\times9.9$ GHz.}
    \label{tab:2nd-in-plane-mode-simulation_results}
\end{table}

\subsection{MATLAB scripts}
LiveLink for MATLAB is an interface between the COMSOL Multiphysics simulation platform and the MATLAB environment. The interface provides MATLAB syntax for preparing and running COMSOL simulations and lets the user write a script in MATLAB programming language that automates complicated post-processing.\\
Using LiveLink for MATLAB, one can make an entire COMSOL simulation in the form of a MATLAB function that might receive inputs as the parameters in the simulation, and return the model after the simulation is done. Then, another script can be written that calls the simulation function, passes the required inputs, receives the model, extracts the parameters out of the model, and finally applies the post-processing. Based on the case, the script can also call the simulation function in an iteration to sweep a parameter and investigate how it affects the characteristics, or even to optimize the behaviour of the simulated system.

The following boxes are pseudo-codes that describe the three MATLAB scripts that I made to investigate how the characteristics of the system change while we sweep the gap.\\
The first pseudo-code defines the \texttt{model\_builder} function that generates the simulation model based on the inputs, runs the simulations, and performs the primitive post-processing.
\begin{center}
\noindent\fbox{
    \begin{minipage}{0.95\textwidth}
        \texttt{\textcolor{blue}{function} model = model\_builder(material, gap\_size)}\\
        \null\qquad 1- Make the \texttt{model} object.\\
        \null\qquad 2- Set the \texttt{gap\_size}.\\
        \null\qquad 3- Make the geometry in \texttt{model}.\\
        \null\qquad 4- Set the \texttt{material} of the substrate.\\
        \null\qquad 5- Assign the materials to domains in \texttt{model}.\\
        \null\qquad 6- Make Physics and do settings in \texttt{model}.\\
        \null\qquad 7- Make the mesh in \texttt{model}.\\
        \null\qquad 8- Make Studies and do settings in \texttt{model}.\\
        \null\qquad 9- Run the simulation in \texttt{model}.\\
        \null\qquad 10- Take the integrals and save them in tables in \texttt{model}.\\
        \texttt{\textcolor{blue}{end}}
    \end{minipage}
}
\end{center}
The second pseudo-code defines the function \texttt{processor} which calls the \texttt{model\_builder} function and passes the inputs, extracts the results of the simulation from the output of the function \texttt{model\_builder}, performs the post-processing, and returns the final results of the post-processing. It also can save the simulation model if the user needs it.
\begin{center}
\noindent\fbox{
    \begin{minipage}{0.95\textwidth}
        \texttt{\textcolor{blue}{function} parameters = \texttt{processor}(material, gap\_size, save)}\\
        \null\qquad 1- Make the empty structure \texttt{parameters}.\\
        \null\qquad 2- \texttt{model = model\_builder(material, gap\_size)};\\
        \null\qquad 3- Extract the required parameters from the tables in object \texttt{model}.\\
        \null\qquad 4- Record $\Omega$ in \texttt{parameters}.\\
        \null\qquad 5- Calculate $x_\text{ZPF}$, $C_m$, $\eta$ and $g_0$, and record them in \texttt{parameters}.\\
        \null\qquad 6- If \texttt{save == true}, save the object \texttt{model} in the computer storage.\\
        \texttt{\textcolor{blue}{end}}
    \end{minipage}
}
\end{center}
And the final pseudo-code is a simple script that sweeps the gap, calls the \texttt{processor} function, and plots and saves the results of the simulation.
\begin{center}
\noindent\fbox{
    \begin{minipage}{0.95\textwidth}
        1- Make the array \texttt{gaps} containing the desired values for the gap size.\\
        2- Make empty arrays for saving the desired parameters.\\
        \texttt{\textcolor{blue}{for} i=1:length(gaps)}\\
        \null\qquad 3- \texttt{parameters = processor("Silicon", gaps(i))};\\
        \null\qquad 4- Extract values from \texttt{parameters} and save them in the corresponding array.\\
        \texttt{\textcolor{blue}{end}}\\
        5- Save the arrays.\\
        6- Plot the arrays versus \texttt{gaps}.
    \end{minipage}
}
\end{center}

\begin{figure}[!h]
    \centering
    \includegraphics[width=0.98\textwidth]{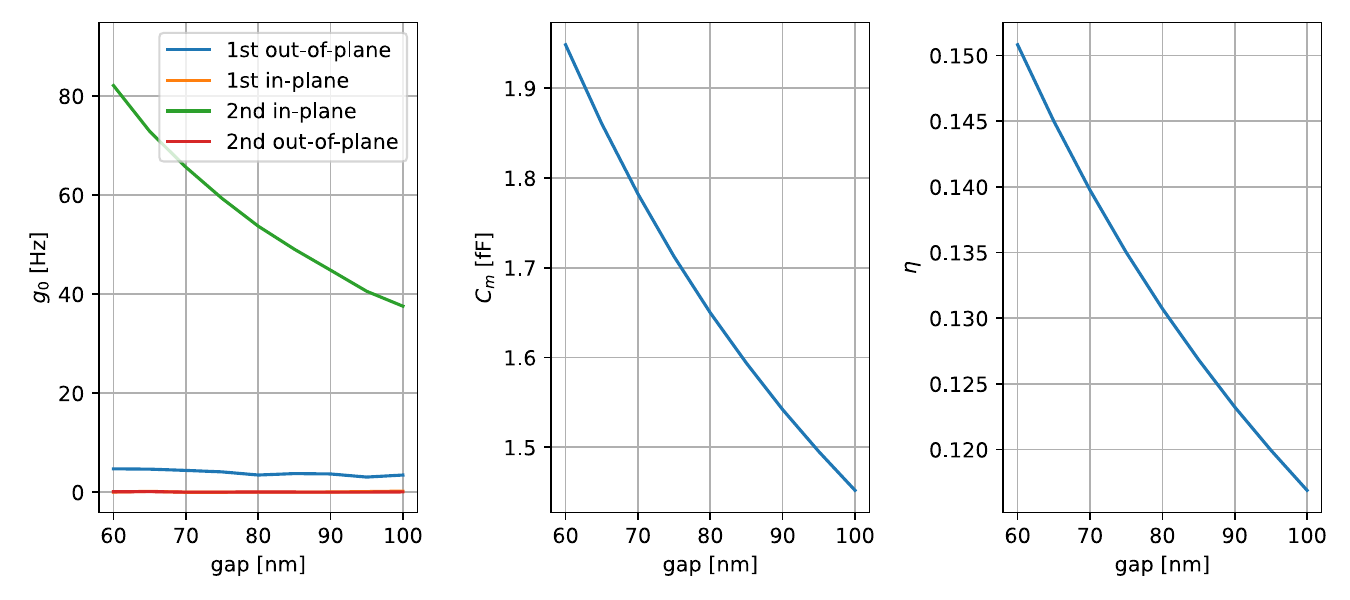}
    \caption[The result of gap sweep for silicon substrate]{The figure shows the single-photon coupling rate, $g_0$ for two in-plane and two out-of-plane mechanical modes, the  capacitance of the structure, $C_m$, and the participation ratio, $\eta$ versus the gap size, for \textbf{silicon substrate}. It is evident that the coupling rate for the first in-plane and the second out-of-plane mechanical modes is zero, as they are asymmetric modes and the nano-strings cancel out each other's effect on the capacitance. The first out-of-plane mode provides a non-zero but small coupling rate as the out-of-plane displacements do affect the capacitance weakly. The second in-plane mechanical mode leads to the strongest coupling rate as it maximizes the effect of the oscillations on the total capacitance.}
    \label{fig:Silicon-gap-sweep}
\end{figure}

You can see a more detailed explanation and instructions about the scripts in Appendix (\ref{chp:LiveLink_instruction}). The code has been used to demonstrate the relation between characteristic parameters and the gap size. You can see the result of the gap sweep in Figure (\ref{fig:Silicon-gap-sweep}).

\begin{figure}[!h]
    \centering
    \includegraphics[width=0.98\textwidth]{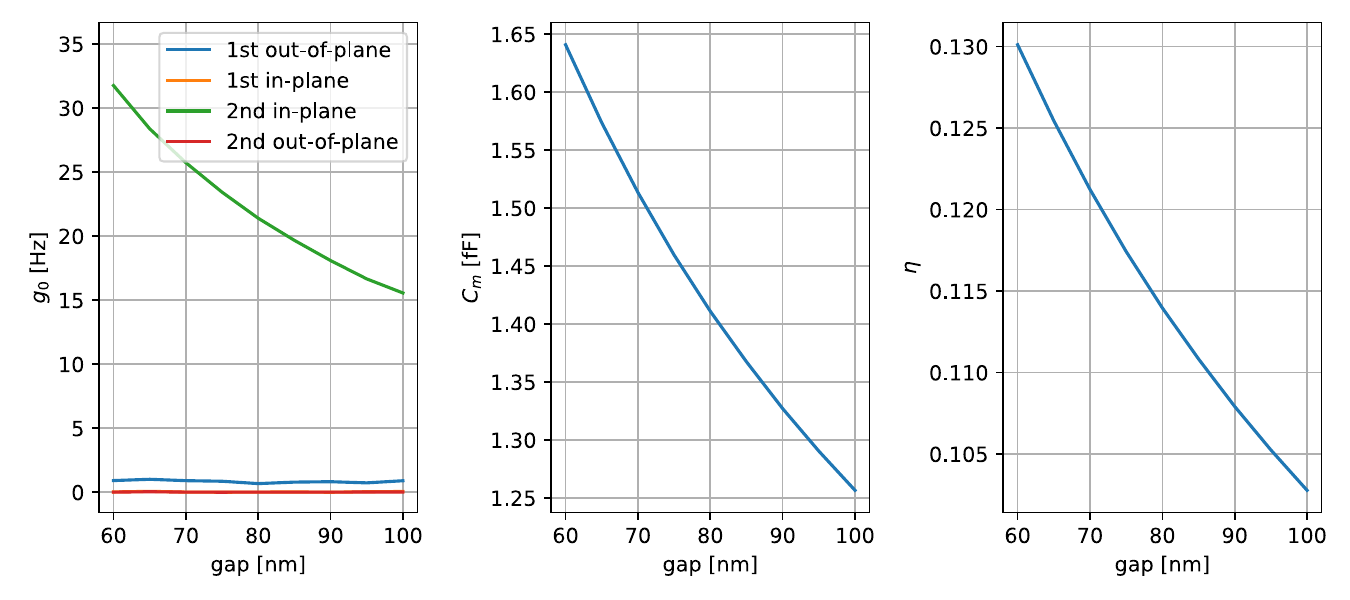}
    \caption[The result of gap sweep for diamond substrate]{The figure shows the single-photon coupling rate, $g_0$ for two in-plane and two out-of-plane mechanical modes, the  capacitance of the structure, $C_m$, and the participation ratio, $\eta$ versus the gap size, for \textbf{diamond substrate}. The effect of different mechanical modes on the coupling rate is similar to the silicon substrate case, however, the magnitude of the coupling rate for the second in-plane and first out-of-plane modes is less than silicon-based structure. The smaller refractive of diamond compared to silicon leads to a smaller capacitance, and consequently, a smaller participation ratio, and smaller participation ratio and zero-point fluctuations cause a smaller coupling rate.}
    \label{fig:Diamond-gap-sweep}
\end{figure}

Using the same scripts and functions, and only by adding the material properties of diamond in \texttt{model\_builder} function, we investigated the electromechanical properties of the system in the case that we use diamond instead of silicon. Figure (\ref{fig:Diamond-gap-sweep}) shows the result of the electromechanical simulation based on diamond.

Obviously, the capacitance (and consequently, the participation function) and the coupling rate are smaller for the diamond substrate, compared to silicon. The capacitance is smaller only because the relative electric permittivity of diamond ($\varepsilon_\text{diamond}=5.5$) is smaller than silicon's ($\varepsilon_\text{silicon}=12.1$), but for the coupling rate, there are more parameters involved. The higher mass density of diamond ($\rho_\text{diamond}=3150$kg/m$^3$) causes the frequencies and the effective mass of each mechanical mode to be larger compared to silicon ($\rho_\text{silicon}=2329$kg/m$^3$), and consequently, $x_\text{ZPF}$ will be smaller, as well as $C_m$. A summary of the electromechanical parameters for a diamond substrate and $70$nm gap is reported in Table (\ref{tab:2nd-in-plane-mode-simulation_results_Diamond}).

\begin{table}[!h]
    \centering
    \begin{tabular}{|c|c|c|c|c|c|c|}
        \hline
         $\Omega/2\pi$ & $m_\text{eff}$ & $x_\text{ZPF}$ & $C_m$ & $\eta$ & $g_0$  \\
         \hline\hline
         $8.28$ MHz & $2.64$ pg & $19.58$ fm & $1.51$ fF & $0.121$ & $25.73$ Hz\\
         \hline
    \end{tabular}
    \caption[Electromechanical properties of the second in-plane mode with diamond substrate]{\textbf{Electromechanical properties of the second in-plane mode with diamond substrate} - The numbers describe a nano-string structure based on a silicon substrate and for $70$ nm gap between the electrodes. The cavity parameters are assumed $C_S = 10.97$ fF and $\omega_c = 2\pi\times9.9$ GHz.}
    \label{tab:2nd-in-plane-mode-simulation_results_Diamond}
\end{table}

\section{Discussion}

Comparing the numbers presented in Table (\ref{tab:2nd-in-plane-mode-simulation_results}) with the electromechanical systems' specifications made and characterized before (see the summary of a few of them in Table \ref{tab:opto/electromechanical-experiments}), we realize the nano-string structure made on silicon provides an electromechanical coupling rate in the typical frequency regime and should be practically observable. Although making the nano-string structure on a diamond substrate results in a smaller electromechanical coupling rate (see Table \ref{tab:2nd-in-plane-mode-simulation_results_Diamond}), the coupling rate will remain in the typical range.\\
In conclusion, the structure investigated in this chapter is a suitable candidate for a cavity electromechanical system, and we can expect to observe electromechanical interaction out of it practically.

  \chapter{Fabrication}
\begin{figure}[!h]
    \centering
    \includegraphics[width=0.95\textwidth]{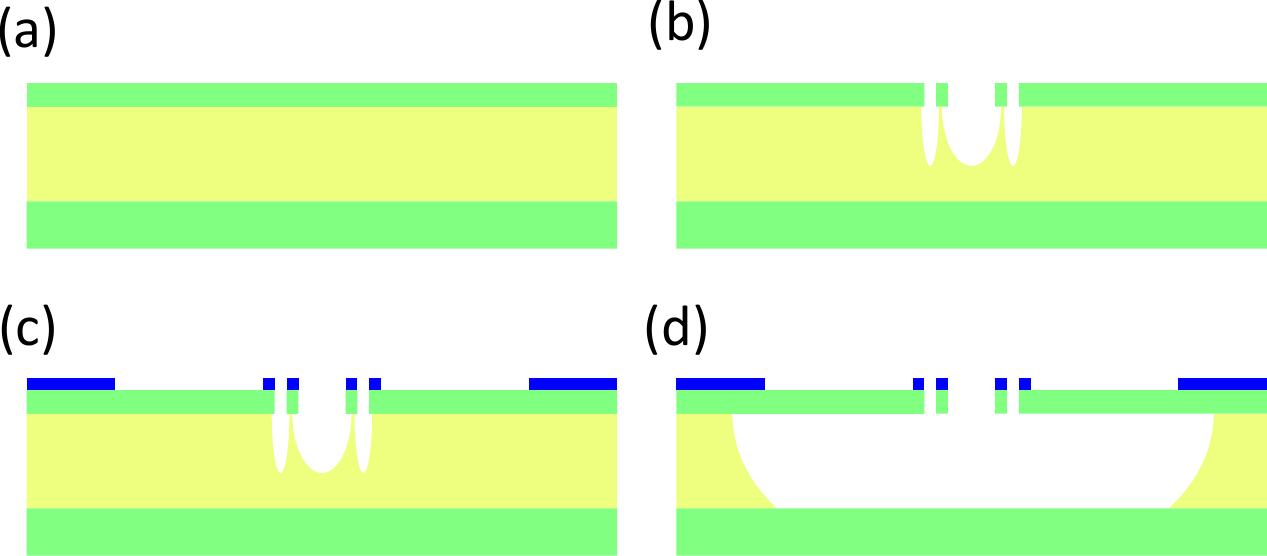}
    \caption[The major steps of fabrication]{The scheme shows a part of the cross-section of the chip. Green and yellow colors indicate silicon and silicon dioxide, and blue is aluminum. The fabrication starts with (a) a bare SOI chip. (b) Then we etch the desired pattern on the top silicon layer. After etching, (c) we coat the desired areas with aluminum, and at the end, (d) we release the top silicon layer.}
    \label{fig:fabrication-steps}
\end{figure}

In this section, we provide a concise overview of the steps taken to fabricate our samples. Given the presence of nanometer-scale features in our electromechanical system, the implementation of such a system necessitates the use of nano-fabrication techniques. To achieve this, we fabricate the structure on Silicon-On-Insulator (SOI) chips. These SOI chips consist of a thin layer of silicon (220 nm in our case) positioned on top of a $3$ $\mu$m-thick layer of silicon dioxide, which, in turn, rests on a silicon substrate. Generally, the fabrication process consists of three main parts: (i) etching the desired pattern on the top silicon layer, (ii) implementing the plates of the capacitor and on-chip circuits by a metal deposition method, and (iii) releasing the mechanical oscillator by removing the silicon dioxide under the structure. (Figure \ref{fig:fabrication-steps})\\
Based on the microwave cavity that we use, the design of the device, and consequently, the steps of the fabrication can be slightly different. To integrate an LC circuit on the SOI chip along with the mechanical capacitor, we also need to implement a ground plane all over the chip. In an ideal situation, the plates of the capacitor, wires, and the LC circuit can be made in the same lithography layer as the ground plane. But as the Electron-Beam Lithography (EBL) \cite{altissimo2010EBL} tool that we have access to does not exceed $30$ kV, and is slow relative to more advanced EBL tools, we separate the ground plane implementation and the circuit and do the ground plane lithography using an Optical Lithography \cite{menon2005masklessOpticalLithography, tilli2020handbook} tool. Therefore, for an integrated LC circuit, we will have one more layer of lithography.
In the following sections, we go through the fabrication steps in detail.

\section{The two-layer design}
\begin{figure}[!h]
    \centering
    \includegraphics[width=0.9\textwidth]{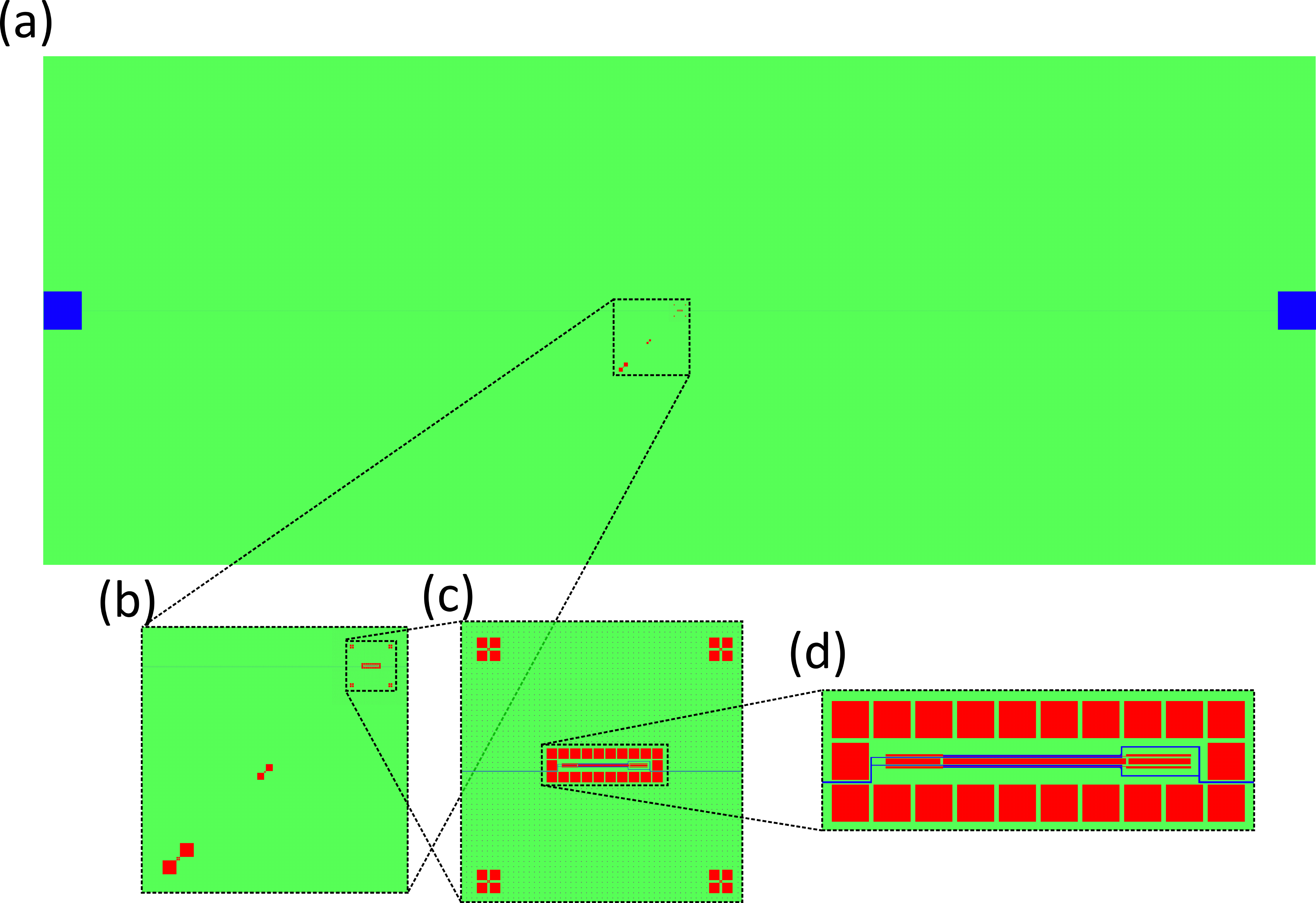}
    \caption[Two-layer design scheme]{The figure demonstrates the two-layer design. The green area indicates where the silicon layer neither is etched nor covered with aluminum, the red areas indicate where the silicon layer is etched, and the blue area indicates the regions that are covered with aluminum. (a) shows the entire design. The two blue squares on the sides are the aluminum pads implemented to connect the cavity to the electromechanical device in the middle of the chip. (b) two patterns in the lower-left corner and middle are the global markers which will be used to locate the device easier while doing electron microscopy and partially align the second layer lithography. The electromechanical device is in the upper-right corner. (c) Four patterns in the corners are the local markers that will be used to align the second layer precisely. (d) The electromechanical is in the middle and is surrounded by twelve etched squares.}
    \label{fig:two_layer_design}
\end{figure}

If we prefer to use a 3D microwave cavity, for the electromechanical part, we only need to implement a mechanical capacitor connected to two large pads on the SOI chip. Figure (\ref{fig:two_layer_design}) shows the design for such a device. Since in the case of more than one device on one chip we cannot choose which device to be coupled to the cavity, we only implement one device on each chip. Therefore, because one electromechanical device does not occupy much area on the surface of the chip, we use $4$ mm$\times10$ mm chips.
Two $300$ $\mu$m$\times300$ $\mu$m pads are implemented on the sides of the chip which are connected to the electromechanical device in the middle, with long wires of aluminum of width $1$ $\mu$m. There are two global markers close to the middle point of the chip. Global markers are meant to be features much larger than the electromechanical device so when we are looking for the device through a microscope they can be found easily and lead us to the device. Around the electromechanical device, there are four small markers that we call local markers. As we will see, etching the mechanical structure and placing the aluminum pieces will be done in two separate steps. The local markers are for the precise alignment of the aluminum electrodes and wires with the etched pattern.\\
One point to note is that the area around the electromechanical device and aluminum wires is filled with a periodic array of etched holes with a diameter of $200$ nm and center-to-center distance of $2$ $\mu$m. As will be explained in the last section of this chapter, the reason for etching holes is to remove the silicon dioxide anywhere close to wires. Since silicon dioxide is a lossy medium for the electric field, minimizing the overlap of the electric field of the circuit and the oxide improves the quality factor of the microwave cavity significantly.\\
Finally, we have the electromechanical device in the middle of the chip, surrounded by twelve squares of side $4$ $\mu$m and spacing $500$ nm. Implementing aluminum electrodes and wires takes place at an extremely high temperature ($\sim1300$ K), while we are going to characterize the device at an extremely low temperature ($<10$ mK). Due to the extreme temperature difference between fabrication and characterization, there will be thermal stress introduced to the silicon layer. The square cuts make room for the silicon layer to release the stress partially and also make partial mechanical isolation from the rest of the chip and decrease the effects of the vibrations of the measurement setup on the electromechanical device.

The fabrication of this design splits into two lithography layers. In the first layer, the pattern of nano-strings, the holes around the structure, and the markers will be written on the chip and etched. In the second lithography layer, the electrodes on top of the nano-strings, the wires, and two large pads will be written and covered by a $60$ nm-thick layer of aluminum.

\subsection{First layer}
\begin{figure}
    \centering
    \includegraphics[width=0.95\textwidth]{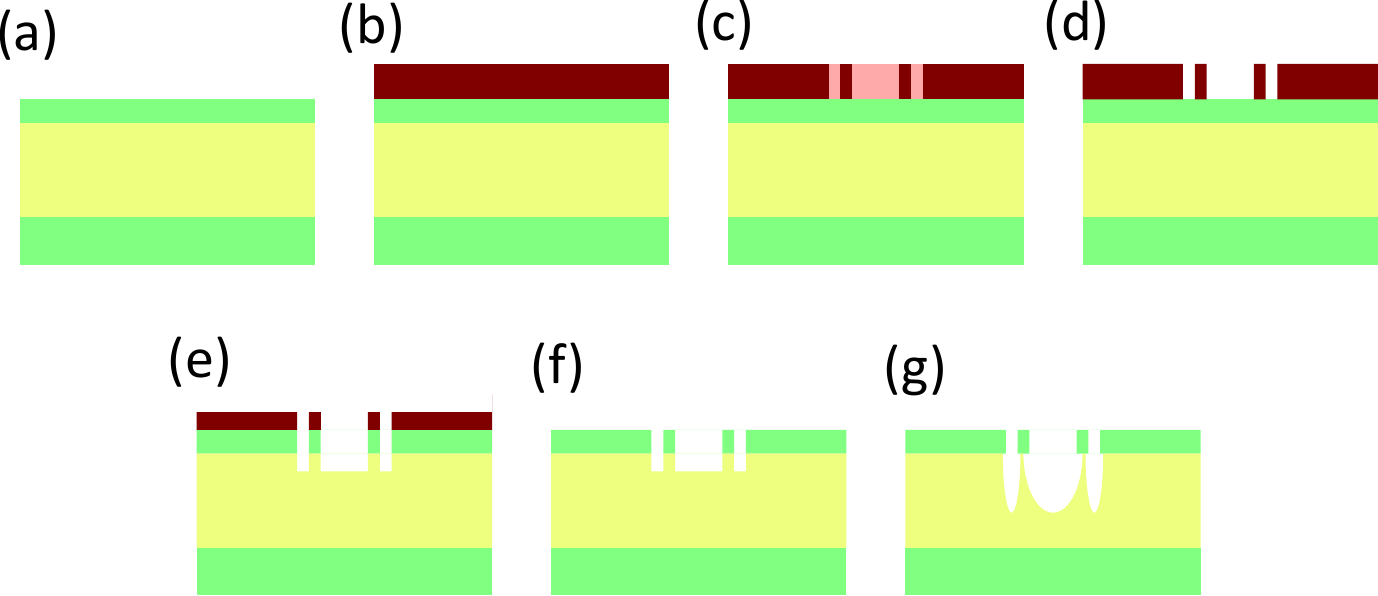}
    \caption[Steps of the first layer for the two-layer design]{The scheme shows a part of the cross-section of the chip. Green and yellow colors indicate silicon and silicon dioxide, crimson is the E-Beam unexposed resist, and pink is the exposed resist. (a) We start with a bare SOI chip. Then, (b) spin-coat the chip with the resist. (c) Using the EBL tool, we write the pattern of the first layer and (d) develop the chip. (e) Using ICP-RIE, we etch the pattern on the Silicon layer, (f) clean the resist, and finally (g) do the undercut using HF.}
    \label{fig:steps-of-first-layer}
\end{figure}

The aim of the first layer is to make the pattern of the nano-strings. We start with a bare SOI chip. Using a spinner tool, we coat the chip with CSAR 62 E-Beam resist \cite{thoms2014CSAR62, AllresistCSAR62}. Then, using the EBL tool, we write the pattern of the first layer (Figure \ref{fig:steps-of-first-layer}) on the chip. What happens at this step is that a highly focused beam of electrons targets a spot on the surface of the chip, and moves to glide over all the areas that are specified in the CAD file. The molecular structure of the resist at the regions that are exposed to high-energy electrons, and consequently, the solubility will be changed \cite{vutova2011nonlinear}. In the next step, we remove the exposed regions of the resist by submerging the chip in AR 600-549 developer \cite{AllresistAR600549}. The developer solves the exposed resist at a higher rate than the unexposed resist. To control the time of solving precisely, we submerge the chip in IPA to stop the solving process immediately. After the development step, the regions of the chip that are indicated in the CAD file should be exposed to air, and the rest of the surface should still be covered with the resist.

At this step, the chip is ready to be etched. Using an Inductively Coupled Plasma-Reactive Ion Etching (ICP-RIE) \cite{tilli2020handbook} tool, we etch the $220$ nm-thick silicon layer at the regions that are not covered with the resist. The ICP-RIE tool bombards the chip with high-energy ions. The ions are chosen specifically to react with silicon at a higher rate compared to the resist \cite{tilli2020handbook}. As the result, after the process, the unprotected areas of the silicon layer (and even a part of the silicon dioxide layer) are etched through, and the resist on top of the chip is slightly thinner. Now, the pattern is made on the chip, but there is still resist remaining on the surface. By submerging the chip in Remover PG \cite{KayakuRemoverPG} which is a strong solvent, we clear the resist off the chip.

As the last step of the second layer, we submerge the chip in HF solution. HF solution removes all the possible remaining resist on the chip or any other contaminants and also penetrates into the silicon dioxide layer and undercuts the chip \cite{tilli2020handbook, monk1993review}. Undercutting is to etch the silicon dioxide partially to make bubble-like empty spaces under the etched areas. These bubble-like vacancies increase the speed and efficiency of the release step. After the undercut, the first layer is completed and the chip is ready for the second layer.

\subsection{Second layer}
At the beginning of the second layer, we have the patterns etched on the chip. Similar to the first layer, we spin-coat the surface of the chip with CSAR 62 and write the pattern of the second layer on the resist with EBL tool. At the E-Beam exposure step, we use global and local markers to align the E-Beam coordinate with the already etched patterns on the chip. If the alignment is not accurate, the aluminum electrodes or wires might be placed in incorrect locations and the circuit will not be completed. Therefore, the second layer alignment is crucial to have a successful fabrication. After EBL, we develop the chip using AR 600-549 developer. Now we have the second layer's pattern removed from the resist. Using a Physical Vapour Deposition (PVD) \cite{mattox2010handbook} tool, a thin layer of aluminum will be deposited on the surface of the chip. The aluminum layer will stick to the resist where there is resist, and sticks to the silicon layer, where the resist is removed.\\
In the PVD method, a container of a specific metal is heated up to the point that the metal is vaporized. Then, the vaporized atoms of the metal deposit on the surface of the chip and form a thin layer of metal. The thickness can be precisely controlled by setting the rate of vaporization and the duration of the process.\\
After PVD, we submerge the chip in heated remover PG to lift-off the aluminum. Remover PG removes resist, and consequently, the aluminum pieces deposited on the resist will be removed while the aluminum pieces deposited on the silicon layer will remain on the chip.

\begin{figure}
    \centering
    \includegraphics[width=0.95\textwidth]{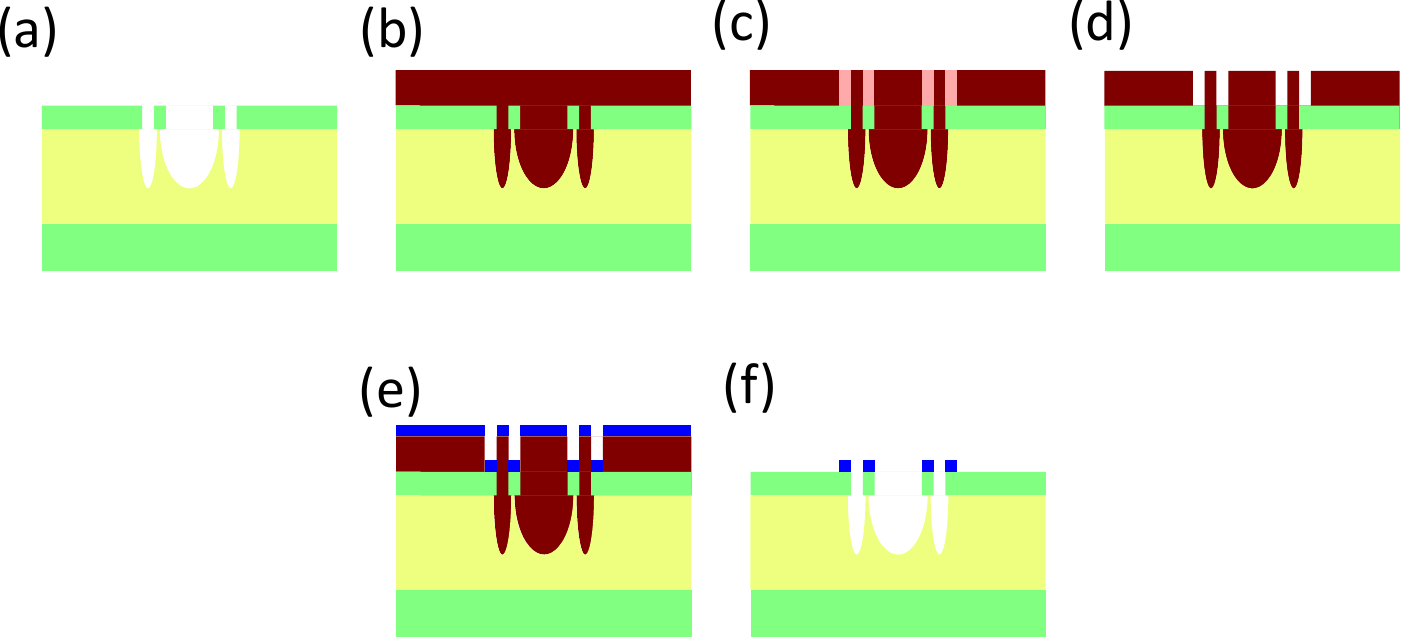}
    \caption[Steps of the second layer for the two-layer design]{The scheme shows a part of the cross-section of the chip. Green and yellow colors indicate silicon and silicon dioxide, crimson is the E-Beam unexposed resist, pink is the exposed resist, and blue is aluminum. (a) The second layer starts with the etched SOI chip. (b) We spin-coat the chip with the resist, and (c) using the EBL tool, we write the pattern of the second layer, and (d) develop the chip. (e) Using the PVD tool we coat the surface of the chip with aluminum and (f) remove the resist to lift-off the aluminum from unwanted areas.}
    \label{fig:steps-of-second-layer}
\end{figure}

\section{The three-layer design}
\begin{figure}[h!]
    \centering
    \includegraphics[height=0.6\textheight]{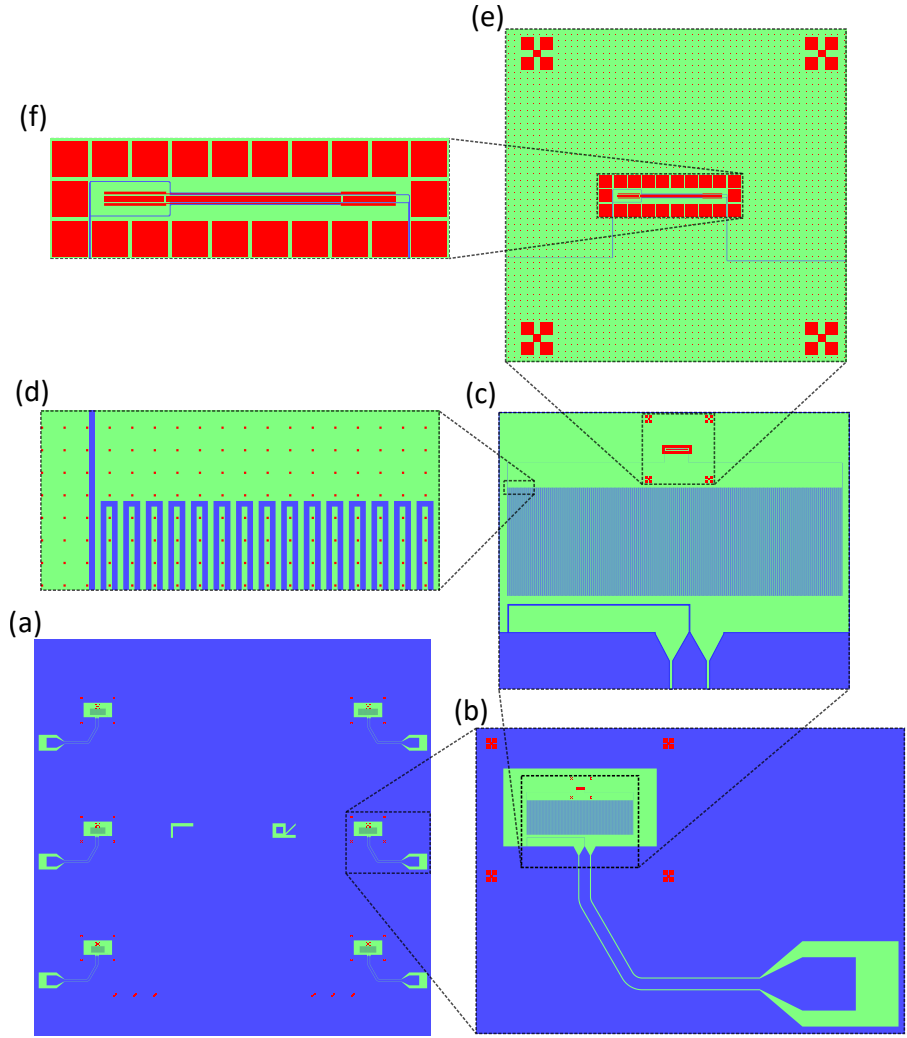}
    \caption[The three-layer design scheme]{The figure shows the three-layer design. Green, blue, and red colors indicate uncovered silicon layer, aluminum coated surfaces, and etched areas, respectively. (a) Each $10$ mm$\times10$ mm chip is designed to include six cavity electromechanical systems. (b) Each system is connected to a pad near the edge of the chip through a coplanar waveguide. (c) Each system contains a meander connected to an electromechanical device and capacitively coupled to the waveguide through a coupling wire. (d) The meander forms the microwave resonator due to its inductance and capacitance. (e) The electromechanical device is connected to the meander with wires. (f) Squares around the device release the thermal tension partially.}
    \label{fig:three-layer-desing}
\end{figure}

An alternative to a 3D microwave cavity is an integrated LC circuit on the chip (on-chip microwave cavity), close to the electromechanical device. Since all the electrical circuits have stray capacitance, to have an LC circuit, we only need to implement a self-inductor on the chip. Therefore, the LC circuit can be a meander \cite{cho2011novel, massel2011microwave}, made of aluminum, connected to the electromechanical device. Using on-chip cavities allows us to implement several standalone electromechanical devices on one single chip. Figure (\ref{fig:three-layer-desing}) demonstrates our design of on-chip systems.
There are six cavity electromechanical systems in the design, three on each side. Each system is connected to a pad designed to have $50$ $\Omega$ impedance, through a coplanar waveguide. The waveguide is coupled to the meander capacitively through the coupler wire, so we can control the extrinsic damping rate (coupling rate to the waveguide) by adjusting the distance of the coupler wire (the extension of the waveguide) and the meander. Similar to the two-layer design, we implement global and local markers, the periodic pattern of the holes that fill the area containing the meander and the electromechanical device, and the squares around the device for the same reasons as before. The meander is coupled to the device galvanically. In terms of fabrication, the main difference between the designs for a 3D cavity and an on-chip cavity is in the ground plane. A ground plane decreases ground noise and interference of various parts of the circuit \cite{wilson2017circuit}.\\
The ground plane occupies a relatively large area on the surface. Using an advanced EBL tool with a high voltage limit, writing the pattern of the ground plane can be feasible. However, the EBL tool (RAITH150 Two) that we have access to has a $30$ kV limit, and this limit makes writing large patterns like the ground plane extremely time-consuming. On the other hand, there are optical lithography tools that write large patterns extremely faster than EBL tools, with the cost of lower resolution. The optical lithography tool that we have access to (Heidelberg MLA150) can write features as small as $1$ $\mu$m \cite{HeidelbergMLA150}, and since the features of the ground plane and the waveguides are all larger than or equal to the size limit, MLA150 can be used to write the ground plane. Note that since the electromechanical device and the meanders have features in the nanometer scale, we still have to write them in another layer, using EBL.

The first layer is basically the same as the first layer of the two-layer design. We spin-coat the chip with CSAR60 and write the markers, electromechanical devices, squares, and holes. Exposure definitely takes a longer time than the two-layer design, and the area that should be exposed is larger, but it still should be done in a reasonable amount of time. Then we develop the pattern using AR 600-549 developer, etch the pattern with the ICP-RIE tool, remove the resist with remover PG, and do the undercut.

The second layer is dedicated to writing the ground plane, coplanar waveguides, and coupler wires. We start with spin-coating the chip with AZ 1529 photoresist \cite{Datasheets2023AZ1529, tilli2020handbook}. Then, using MLA150 we write the pattern on the resist.\\
MLA150 targets the surface of the chip with a focused beam of laser and hovers the beam spot on the surface to expose our desired pattern. Similar to E-beam, the laser changes the molecular properties of the photoresist and makes it solvable in the proper developer.\\
The global markers will be used at this step to align the ground plane and waveguides with the etched patterns. After the exposure, we develop the chip using AZ 400K developer \cite{AZelectronicMaterialsAZ400K}, coat the surface with aluminum by PVD method, and lift-off the resist with heated Remover PG.

The third layer is quite similar to the last layer of the two-layer design, except we have to write the meanders along with the electrodes of the capacitor and the wires, and there are six systems instead of only one. Aside from the features that should be written, the process is completely the same.

\section{Release}

After the last layer of the design (either two-layer or three-layer) is done, the etched patterns and the aluminum pieces are implemented in their places. However, since the entire silicon layer is stuck to the silicon dioxide underneath, no pieces (including the nano-strings) are free to oscillate. Furthermore, the quality factor of the microwave cavity can be affected negatively by the material loss of silicon dioxide.

\begin{figure}[!h]
    \centering
    \includegraphics[width=0.95\textwidth]{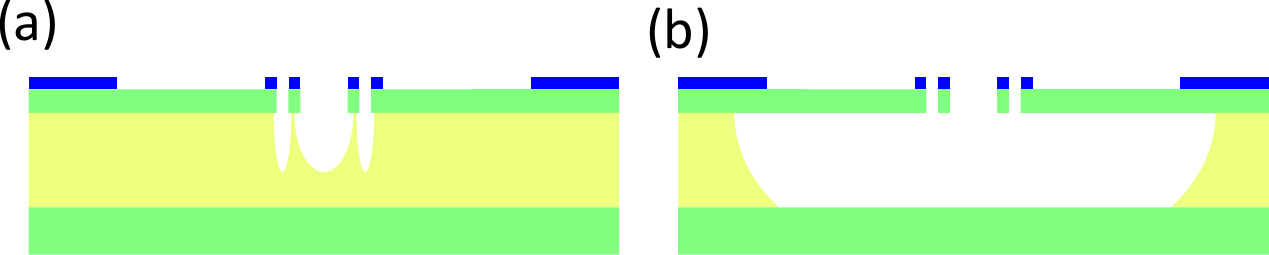}
    \caption[Releasing the nano-strings]{The scheme shows a part of the cross-section of the chip. Green and yellow colors indicate silicon and silicon dioxide, and blue is aluminum. For the last step, release, we should have (a) the chip with the etched pattern, undercut, and aluminum wires (and the ground plane for the three-layer design). Then, by exposing the chip to vapor HF, (b) the silicon dioxide will be etched away isotropically, and the nano-strings will be released.}
    \label{fig:release-step}
\end{figure}
After the last layer, we release the Silicon slab from the silicon dioxide by exposing the chip to anhydrous vapor HF \cite{tilli2020handbook}. We place the chip in the chamber and apply precisely adjusted flows of vapour HF and the catalyst (water or alcohol) under carefully maintained pressure and at a fixed temperature. The HF gas penetrates through the etched patterns, into the silicon dioxide layer. HF reacts with silicon dioxide and etches the oxide away. Therefore, there will be no silicon dioxide under the silicon layer, close to the etched patterns, and the nano-strings will be released.\\
One important point that should be mentioned is that anhydrous vapor HF can cause corrosion to aluminum as well \cite{hanestad2001stiction}. To prevent aluminum corrosion, the pressure and temperature of the chamber need to be adjusted. Otherwise, the metallic parts deposited on the chip will be etched as well.
  \chapter{Characterization}
As has been mentioned in chapter \ref{chp:theory}, we can identify microwave, mechanical, and electromechanical characteristics of a device from its microwave reflection or transmission. Furthermore, electromechanical cooling or heating can be investigated by measuring the output power spectrum of the device. For measuring the reflection or transmission of a microwave cavity, we use a Vector Network Analyzer (VNA), and for the power spectrum, we use a Spectrum Analyzer (SA).\\
In microwave quantum measurements, we deal with extremely low-energy quanta that can easily be dominated by the environment's noise at room temperature. To overcome the noise of the environment, we cool down the devices to an extremely low temperature, in a Dilution-Refrigerator (DR). To measure the response of the devices at room temperature, we amplify the signals inside the DR to amplitudes that are greater than the thermal noise at room temperature and guide the amplified signals to the measurement equipment out of the DR. Finally, by measuring the response of the devices and fitting them with the equations from the theory, we can characterize the devices.

In this chapter, we go through some details of our measurement equipment and the methods that we employ to characterize our microwave and electromechanical systems. Note that, the complete measurement of the electromechanical sample is an ongoing process and will not be reported here.

\section{Measurement setup}
Our measurement setup consists of the electronic equipment (e.g. VNA, SA, RF sources), and the DR for preparing a proper environment with a low noise level. We start with explaining the equipment and then explain about the fridge. In the end, we will see how we prepared the entire setup by bringing the electronic equipment and the DR together.

\subsection{Electronic equipment}
Since the reflection or transmission of a microwave device reveals many of its properties, the first device that we use in microwave measurements is VNA to measure reflection/transmission.
A typical VNA has three or four ports each of which can both generate a microwave coherent signal and measure the receiving microwave signal.
VNA performs a heterodyne measurement \cite{R&SZNB} on the received signal and extracts the amplitude and the phase of the signal at a specific frequency.
By sweeping the frequency of its local oscillator, VNA measures the amplitude and the phase of the received signal over a range of frequencies.
A four-port VNA measures the s-parameters which are complex elements of the scattering matrix, $\mathbf{S}$,
\begin{equation}
    \mathbf{S} = \begin{pmatrix}
        S_{11}(\omega) & S_{12}(\omega) & S_{13}(\omega) & S_{14}(\omega) \\
        S_{21}(\omega) & S_{22}(\omega) & S_{23}(\omega) & S_{24}(\omega) \\
        S_{31}(\omega) & S_{32}(\omega) & S_{33}(\omega) & S_{34}(\omega) \\
        S_{41}(\omega) & S_{42}(\omega) & S_{43}(\omega) & S_{44}(\omega)
    \end{pmatrix},
\end{equation}
that relates the received signals ($E_i^\text{out}$) to the generated signals ($E_j^\text{in}$).
\begin{equation}
    \begin{pmatrix}
        E_1^\text{out}(\omega) \\
        E_2^\text{out}(\omega) \\
        E_3^\text{out}(\omega) \\
        E_4^\text{out}(\omega)
    \end{pmatrix} = 
    \begin{pmatrix}
        S_{11}(\omega) & S_{12}(\omega) & S_{13}(\omega) & S_{14}(\omega) \\
        S_{21}(\omega) & S_{22}(\omega) & S_{23}(\omega) & S_{24}(\omega) \\
        S_{31}(\omega) & S_{32}(\omega) & S_{33}(\omega) & S_{34}(\omega) \\
        S_{41}(\omega) & S_{42}(\omega) & S_{43}(\omega) & S_{44}(\omega)
    \end{pmatrix}
    \cdot \begin{pmatrix}
        E_1^\text{in}(\omega) \\
        E_2^\text{in}(\omega) \\
        E_3^\text{in}(\omega) \\
        E_4^\text{in}(\omega)
    \end{pmatrix}.
\end{equation}
As an example, if one connects port-1 and port-2 of the VNA to a two-sided microwave cavity, $S_{11}(\omega)$ and $S_{22}(\omega)$ will be the reflection of the cavity at frequency $\omega$ from both sides and $S_{12}(\omega)$ and $S_{21}(\omega)$ will be the transmission of the cavity through both possible directions.

Another microwave equipment that we will need is SA to measure the output power spectrum of a microwave device. SA, compared to VNA, is a slightly simpler device, as it does not have any internal signal generator, and only measures the amplitude of its input signal, at a specific frequency, and by sweeping the frequency of measurement, it obtains the power spectrum of the signal.

The other microwave equipment that we frequently use in the measurements is Radio Frequency (RF) sources. RF sources generate coherent signals at any desired frequency within their designed range. To synchronize the RF sources and the VNA, we connect them to an atomic clock in the lab, as a reference.

\subsection{Low-temperature microwave measurement setup}

\begin{figure}[h!]
    \centering
    \begin{subfigure}{\textwidth}
            \refstepcounter{subfigure}\label{fig:low-temp-measurement-setup-a}
            \refstepcounter{subfigure}\label{fig:low-temp-measurement-setup-b}
        \end{subfigure}
    \includegraphics[height=3.5in]{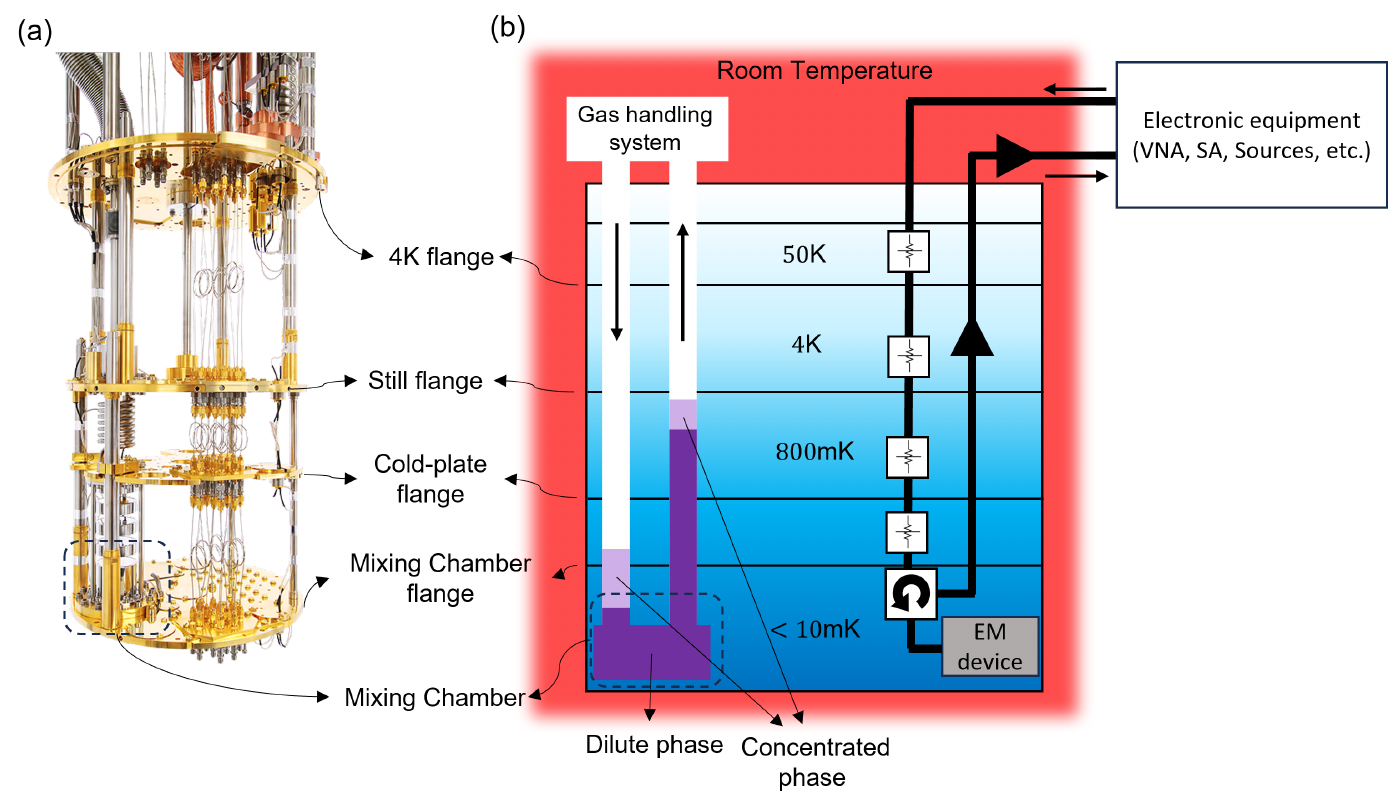}
    \caption[Dilution refrigerator for microwave measurements]{(a) The inner structure of a dilution refrigerator. DRs consist of several flanges, each maintained at a specific temperature while the fridge is operating. The pipes implemented for circulating Helium are visible on the left side of the image, while the RF coaxial cables are on the right side. Reprinted from Bluefors dilution refrigerator systems for quantum technology, retrieved from https://bluefors.com (b) A schematic of the fridge, showing the temperature of each section, a simplified demonstration of the cooling loop, and the input-output system.}
    \label{fig:low-temp-measurement-setup}
\end{figure}

As explained in Chapter \ref{chp:Introduction}, electromechanical systems should be cooled down to behave quantum mechanically. Dilution refrigerators (Figure \ref{fig:low-temp-measurement-setup-a}) are widely used to cool microwave devices down to their ground states, as they can reach temperatures below $10$ mK \cite{radebaugh1971dilution, uhlig20023he, zu2022development}.
Roughly speaking, we can split the operation of the dilution refrigerators into three stages. In the first stage, liquid Helium (a mixture of a rare Helium isotope, ${}^3\text{He}$, and the most common Helium isotope, ${}^4\text{He}$, with a carefully adjusted ratio) circulates in a loop that passes through the fridge and reduces the temperature to $\sim4.2$ K. Subsequently, in the second stage, the pressure on the liquid Helium will be increased to push it into the superfluid phase and turn it into a Bose-Einstein condensate. At this stage, evaporative cooling of Helium superfluid causes the temperature drop to $\sim800$ mK.\\
Helium condensation results in the accumulation of the heavier isotope, ${}^4\text{He}$, at the bottommost part of the loop in the fridge which is called the Mixing Chamber (MC) (See Figure \ref{fig:low-temp-measurement-setup}).
At temperatures below $800$ mK, the superfluid phase of ${}^4\text{He}$ tends to keep a specific percentage of mixed ${}^3\text{He}$ with itself. The rich ${}^4\text{He}$ (diluted with ${}^3\text{He}$) phase is called the dilute phase. The rest of ${}^3\text{He}$ form another phase known as the concentrated phase. At the last stage of cooling, a pump pulls the mixture from the outgoing channel in the fridge (shown in Figure \ref{fig:low-temp-measurement-setup-b}) and since the vapor pressure of ${}^3\text{He}$ is much larger than ${}^4\text{He}$, pumping causes ${}^3\text{He}$ demix from the dilute phase. As the dilute phase tends to keep the ${}^3\text{He}$ concentration fixed, ${}^3\text{He}$ will mix with the dilute phase in the incoming channel, to compensate for the demixing. Mixing ${}^3\text{He}$ with the dilute phase is an endothermic process, therefore, it cools the MC down. On the other hand, demixing, which is an exothermic process, happens at a higher level in the fridge. Therefore, the entire process takes heat from the MC to higher-level flanges, and then it will be extracted from the higher-level flanges of the fridge by evaporation and compression processes. The mixing-demixing cycles can bring the temperature of the fridge below $10$ mK.

The fridge provides an excellent environment for microwave quantum systems, but the electronic pieces of equipment required for the measurements operate at room temperature. To connect the cooled microwave devices to the electronics at room temperature, we use coaxial RF cables suitable for cryogenic wiring \cite{krinner2019engineering}. Each communication line made out of coaxial cables consists of one input and one output line. To prevent thermal noise from penetrating into the fridge through the input line, we implemented attenuators along the line. The attenuations are chosen $20$ dB on the $4$ K flange, $20$ dB on the still flange, $10$ dB on the Cold-plate, and $10$ dB on the Mixing Chamber flange, to suppress Johnson-Nyquist noise \cite{johnson1928thermal, nyquist1928thermal} corresponding to the temperature of the flanges \cite{krinner2019engineering}. At the end of the input line, a microwave circulator leads the input signals to the cavity, and the output of the cavity to the output line. The output line is equipped with a High Electron-Mobility Transistor (HEMT) \cite{mimura2002early} amplifier at the $4$ K level with a nominal $38$ dB gain at $10$ GHz, and a room temperature amplifier installed on top of the fridge with a nominal $40$ dB gain. Including the loss of the coaxial cables, the total loss of the input line is $\sim73$ dB and the net gain of the output line is $\sim75$ dB.

\section{Measurement steps}
To characterize the devices, we first need to prepare them so we can secure them in the DR and connect the RF cables to them. Then, we can characterize the devices by measuring their responses. In this section, we will see the details of the preparation of the devices and how we can characterize an electromechanical system.

\subsection{Device preparation}
The microwave cavities should be connected to the input-output lines of the fridge through standard connections, so we can characterize their electromechanical properties. There is a difference between preparing 3D and integrated resonators to connect them to the microwave lines of the fridge. We go through the details of both of them.

\subsubsection{3D cavities}
The electromechanical devices we fabricate for 3D cavities are capacitors on SOI chips that should be coupled to a microwave cavity. 3D microwave cavities are hollow spaces that are completely surrounded by conductor or superconductor materials. The surrounding material restricts and traps electromagnetic fields and forms stationary modes. The modes' frequencies of a 3D cavity depend entirely on the shape and the dimensions of the hollow space.
\begin{figure}
    \centering
    \includegraphics[width=0.8\textwidth]{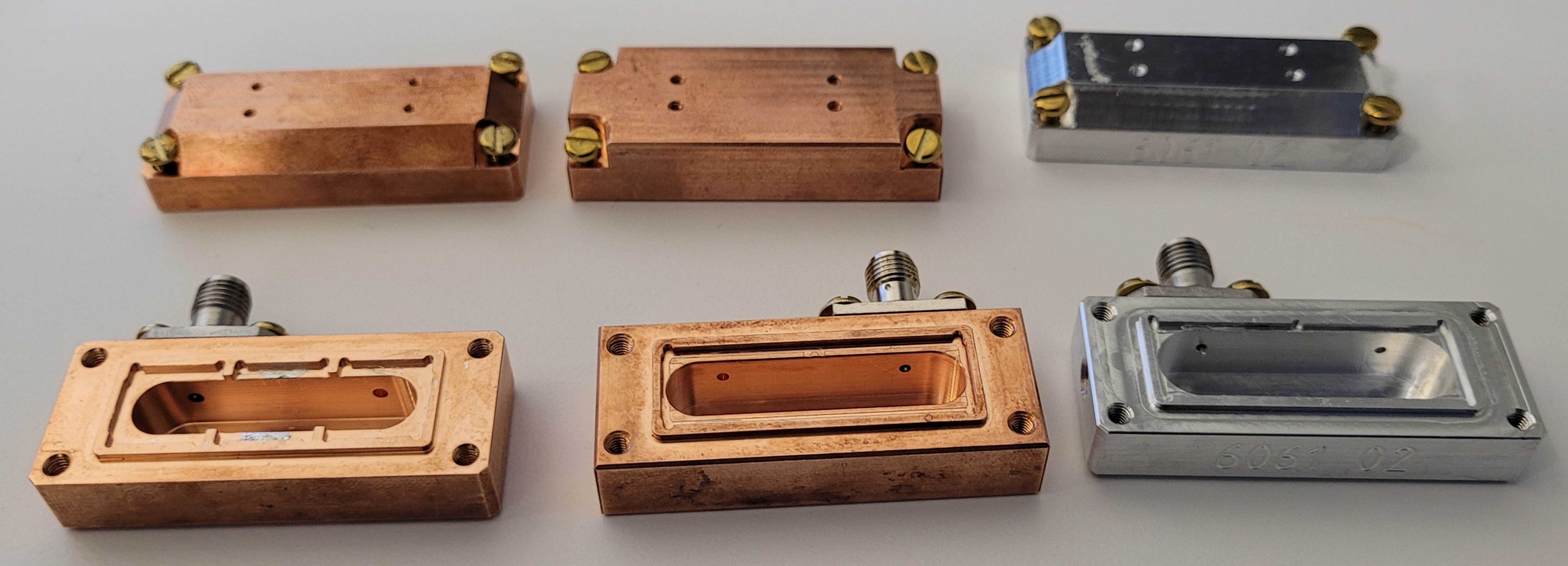}
    \caption[Semi-cuboid 3D cavities]{The figure shows two copper and one aluminum cavities. The cavities are designed to be compatible with SMA straight panel mount jacks.}
    \label{fig:3D-cavities}
\end{figure}
Our semi-cuboid cavities (Figure \ref{fig:3D-cavities}) are designed to have their first mode close to $10$ GHz. The intrinsic damping rate of these cavities is primarily influenced by the purity of the materials utilized. For our cavities, we employed oxygen-free copper and high-purity aluminum. Aluminum offers the advantage of superconductivity, resulting in a lower intrinsic damping rate and higher quality factor, particularly at cryogenic temperatures. On the other hand, aluminum is not ideal for experiments involving magnetic fields at cryogenic temperatures, as superconductors are magnetic shields. For experiments involving magnetic fields, copper cavities are preferred.\\
Each of the cavities can accommodate up to two SMA straight panel mount jacks, allowing for the flexible configuration of one-sided (coupled to the environment and one communication line) or two-sided (coupled to the environment and two distinct communication lines) cavities. The extrinsic damping rate of a 3D cavity primarily relies on the extent of overlap between the inner pin of the SMA jack, which enters the cavity, and the electric field of the mode. By increasing the length of the pin, we can enhance the extrinsic damping rate. Additionally, one of the halves of the cavity features designed steps along its edges to securely hold the chips in place.

\begin{figure}
    \centering
    \includegraphics[width=0.45\textwidth]{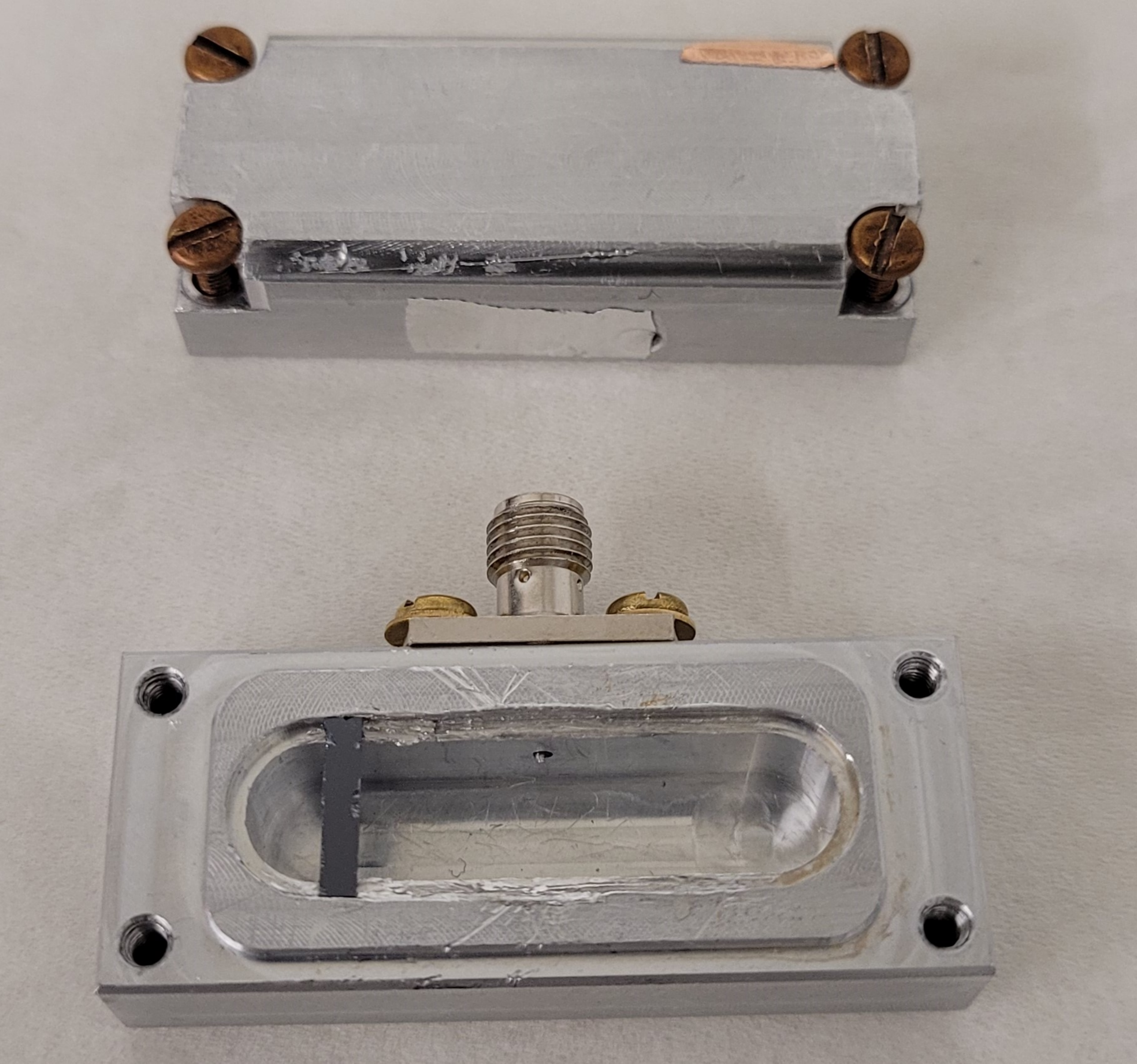}
    \caption[3D cavity and the electromechanical device]{The device is coupled to the cavity galvanically, using the flip-chip technique. The chip is placed upside down on the half of the cavity, so the pads implemented on the sides of the chip touch the cavity directly, and make a galvanic connection. Indium spread in the contact regions for better galvanic connection is also noticeable on the edges of the cavity, near the chip.}
    \label{fig:preparation-3D-cavity}
\end{figure}

There are two main ways to couple an electromechanical device fabricated on an SOI chip to a 3D cavity. We can either couple them capacitively or galvanically. We couple the devices to cavities galvanically, due to the higher efficiency. To galvanically couple, we need to connect the pads implemented on the chip directly to the sides of the cavity. This can be done by the flip-chip technique (see Figure \ref{fig:preparation-3D-cavity}): We put the chip on the steps around the cavity upside-down, so the pads touch the surface of the steps. To enhance the connection, we spread indium on the surface of the steps before placing the chip. The indium between the pads and the surface of the steps guarantees good contact. The indium also makes the chip stick to the cavity stronger, so it will not move or slip while measurements are running.

To facilitate the measurement of the cavity's reflection, we connect only one SMA jack to it, while the other hole initially intended for a second jack is covered with aluminum tape. The 3D cavity can be securely mounted on the cold finger within the DR. Subsequently, we establish a connection between the SMA jack and the microwave input-output lines using a coaxial RF cable.

\subsubsection{Integrated resonator}
\begin{figure}
    \centering
    \includegraphics[width=0.50\textwidth]{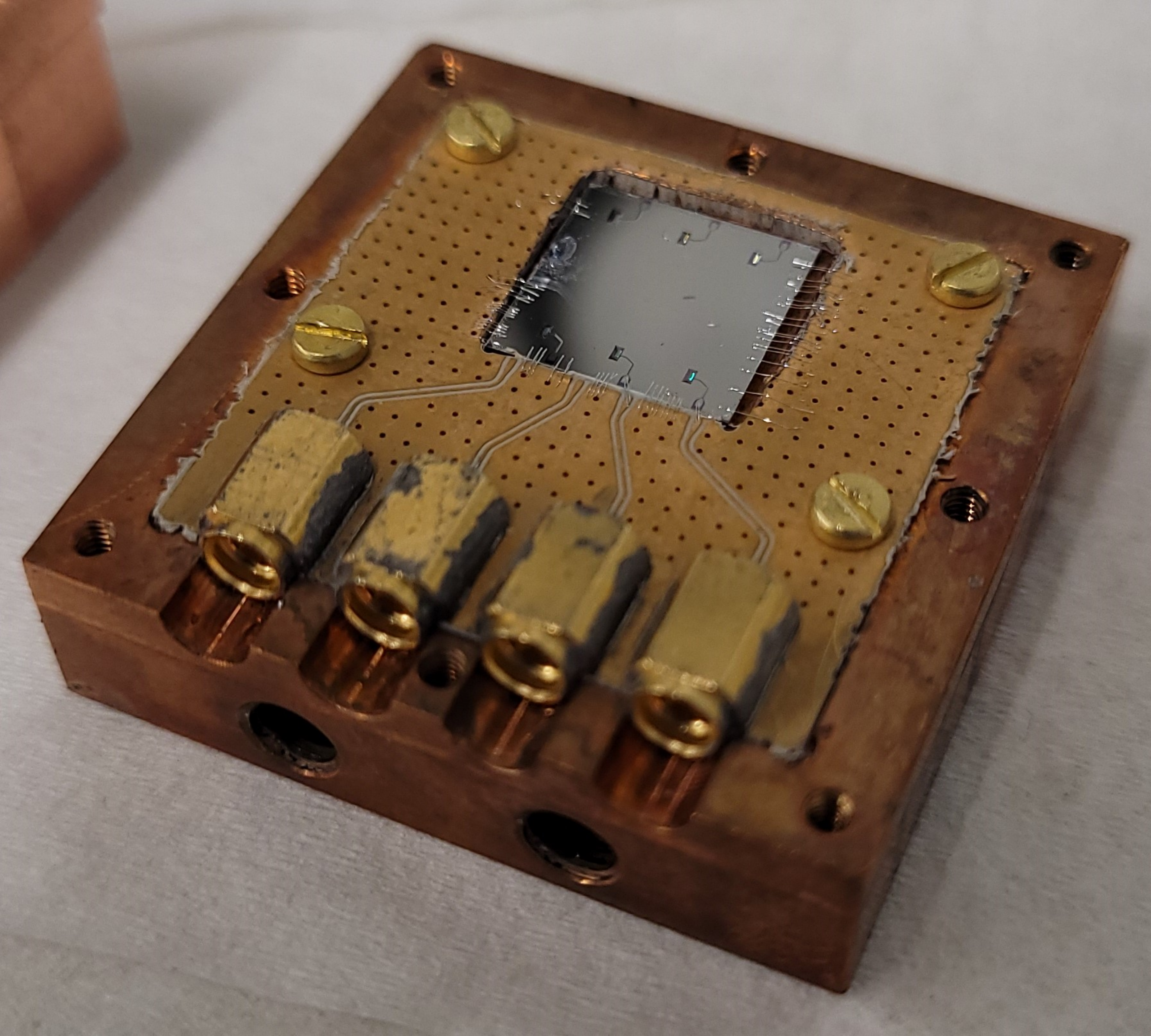}
    \caption[Integrated cavity design in the holder box]{The figure shows a copper PCB holder with a PCB mounted on it. The PCB is designed and made to keep a $10$ mm$\times10$ mm chip in itself. The pads on the chip are connected to the MMPX jack on the PCB through aluminum wire bonds and the PDB's coplanar waveguides. The ground plane of the chip is also connected to the ground plane of the PCB through wire bonds.}
    \label{fig:preparation-integrated-cavity}
\end{figure}

Unlike the 3D cavity design, the integrated resonator design contains the cavities in the form of an LC circuit on the chip. Therefore, to prepare the chip for measurement we need to connect the pads to the input-output lines. We have PCBs and copper PCB holders designed for $10$ mm$\times10$ mm SOI chips (Figure \ref{fig:preparation-integrated-cavity}). The pads on the chip
 will be connected to the pads on the PCB by aluminum wire bonds, and MMPX jacks will be mounted on the ports of the PCB. Eventually, using a MMPX-to-SMA cable, we can connect the bonded devices to the input-output lines.

\subsection{Measurement and post-processing}
\label{chp:post-processing}
After measuring the reflection of a cavity, we fit the data points to the equation we have from the cavity electromechanics theory. The first step is typically characterizing the microwave properties of the cavity which are the resonant frequency and the quality factor, by running a frequency sweep over a frequency span around the nominal frequency of the cavity and then fitting the data with the Equation (\ref{eq:reflection_one_sided_cavity}). It can also be helpful to have microwave properties as functions of the probe's power. To do that, we run frequency and power sweeps. An example of such a measurement is presented in the next section.\\

Once the microwave properties of the system are established, we introduce an RF source into the setup. By tuning the source within the red side-band region and sweeping its frequency around the value of $f_p = \omega_c/2\pi - \Omega/2\pi$, where $\omega_c$ represents the cavity's frequency and $\Omega$ is the nominal mechanical frequency of the system, our objective is to observe the Opto-Mechanically Induced Transparency (OMIT) effect in the reflection. The presence of OMIT confirms the existence of electromechanical interaction in the system. As soon as the OMIT effect becomes apparent, we can determine the actual mechanical frequency by calculating the difference between the frequency of the source and the peak that emerges in the cavity profile.

Subsequently, with the frequency of the source fixed, we perform a power sweep. The OMIT effect evolves as detailed in Chapter (\ref{chp:theory-OMIT}), and by fitting the obtained data points to Equation (\ref{eq:OMIT}), we can extract the electromechanical properties of the system.

\subsection{A result sample}
\label{chp:result-sample}
\begin{figure}
    \centering
    \begin{subfigure}[b]{0.48\textwidth}
        \includegraphics[width=\textwidth]{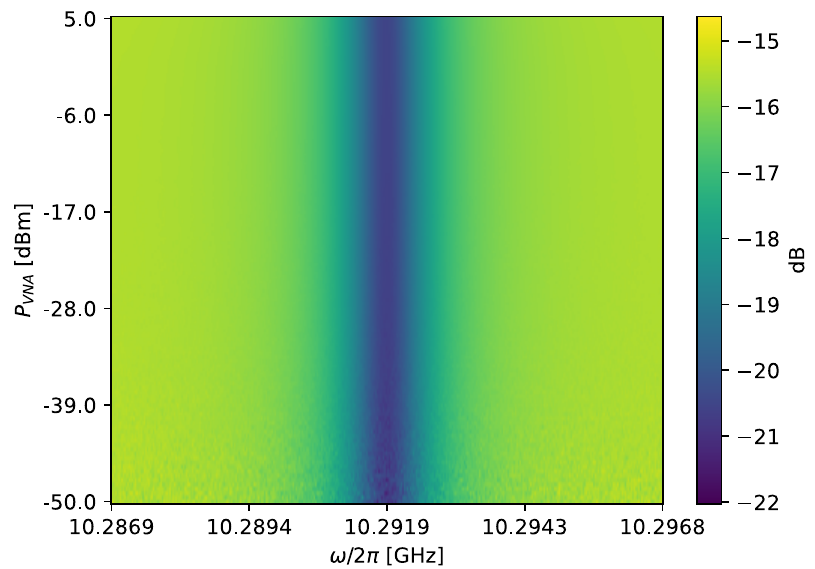}
        \caption{}
        \label{fig:cavity_reflection_measurement_power_sweep}
    \end{subfigure}
    \begin{subfigure}[b]{0.48\textwidth}
        \includegraphics[width=\textwidth]{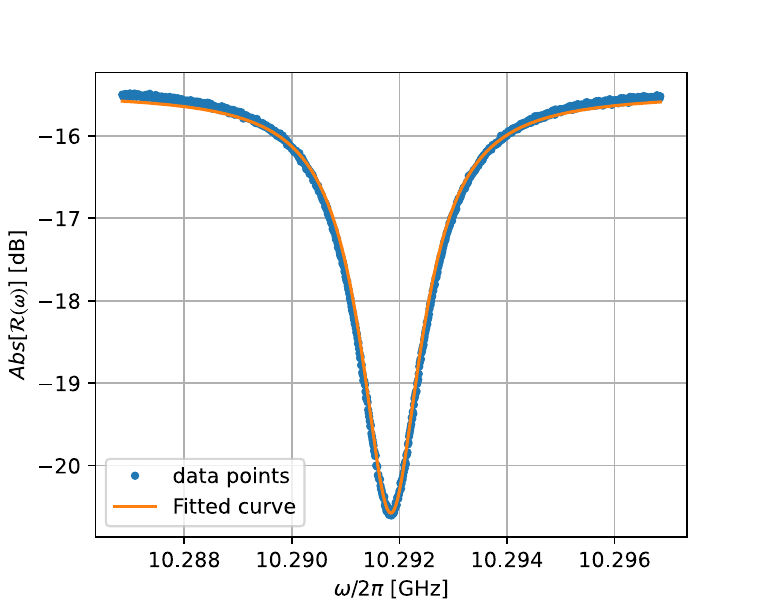}
        \caption{}
        \label{fig:cavity_reflection_measurement_one_power}
    \end{subfigure}
    \caption[Microwave resonator characterization]
    {The figure shows the reflection of a cavity while sweeping the frequency and power of the VNA output signal. (a) The reflection profile of the resonator vs. frequency and the power of the VNA. Note that the vertical axis shows the output power of the VNA. The total loss of the input line of the fridge should be subtracted from the output power of the VNA to find the power of the probe signal in the resonator. (b) The reflection of the cavity for $-20$ dBm VNA output power. The mode frequency and the intrinsic and extrinsic damping rates can be obtained by fitting the reflection equation to the data points.}
    \label{fig:cavity_reflection_measurement}
\end{figure}

Here we present the result of the characterization of a simple microwave resonator, as a basic example. We measured the reflection of a copper 3D microwave cavity at low temperature ($\sim7$ mK), over a frequency span and for different values of the VNA power. The result of the measurement is demonstrated in Figure (\ref{fig:cavity_reflection_measurement_power_sweep}).
To extract the characteristics of the microwave cavity, we fit the equation of reflection of a one-sided cavity to the data points. Equation (\ref{eq:reflection_one_sided_cavity}) describes the reflection of a one-sided cavity measured just next to the cavity. However, if there is a considerable distance between the cavity and the measurement device, the phase shift of the signal should be included in the equation. Furthermore, the baseline of the reflection trace might be slightly tilted around the frequency of the cavity, due to any reason. To overcome the non-horizontal baseline, we manually add a parameter in the reflection equation to impose the tilt in the baseline. Finally, the equation we use to fit the reflection of a cavity will be obtained as,
\begin{equation}
    \mathcal{R}(\omega) = A e^{-i\left(\omega\tau+\phi\right)}\left(-\frac{-i\left(\omega-\omega_c\right) + \frac{\kappa_\text{in}-\kappa_\text{ex}}{2} + i\delta}{-i\left(\omega-\omega_c\right) + \frac{\kappa_\text{in}+\kappa_\text{ex}}{2}}\right),
\end{equation}
where $A$ includes the total loss and gain through the communication lines, $\phi$ is a constant phase shift, $\tau$ indicates the transmission time of the probing signal, and $\delta$ is the parameter that makes the baseline tilted. The rest of the parameters are defined as before.
It is recommended to fit the function to the real and imaginary parts of the measured data points, instead fitting to the absolute value and the phase of them.\\
Figure (\ref{fig:cavity_reflection_measurement_power_sweep}) shows the data points of the reflection measurement for VNA power $-20$ dBm, along with the fitted curve. The results of fitting are summarized in Table (\ref{tab:fitting_results}).
\begin{table}[]
    \centering
    \begin{tabular}{|c|c|c|c|c|c|c|}
         \hline
         $\omega_c/2\pi$ & $\kappa_\text{in}/2\pi$ & $\kappa_\text{ex}/2\pi$ & $\tau$ & $\phi$ & $A$ & $\delta$ \\\hline
         $10.29184$ GHz & $0.41$ MHz & $1.45$ MHz & $63.51$ ns & $1.20$ rad & $0.168$ & $-20.36$ mHz\\
         \hline
    \end{tabular}
    \caption[Fitting results for a microwave resonator measurement]{\textbf{Fitting results for a microwave resonator measurement} - The parameters are extracted from fitting the reflection equation to the data points. The parameters $\omega_c$, $\kappa_\text{in}$, and $\kappa_\text{ex}$ are the characteristics of the cavity.}
    \label{tab:fitting_results}
\end{table}

\section{Discussion}
As we saw in this chapter, a cryogenic microwave measurement setup can be implemented to characterize an electromechanical device. Up to the date of writing this thesis, I have had characterization attempts on a few electromechanical systems we fabricated. The characterizations included source frequency and power sweeps while measuring the reflection of the microwave cavity. The purpose of the sweeps has been to find the frequency-power ranges that enable the OMIT effect. However, I have observed no OMIT behavior from the devices we fabricated. Therefore, I am optimizing the nanofabrication recipe to resolve possible issues caused by fabrication inaccuracies and make a working device.
  \chapter{Electromagnonic entanglement}

This chapter presents a theoretical model I developed for investigating tripartite electro-magno-mechanical systems. As will be mentioned, the nature of pure electromagnonic interaction does not allow any photon-magnon entanglement generation, however, the model shows that the existence of an electromechanical interaction in the system can lead to a tripartite entanglement, consequently, a photon-magnon entanglement.

\section{Electromagnonics}
Magnetic materials contain particles with a non-zero net spin, resulting in interactions with neighboring spins. This configuration gives rise to spin waves within the material, and the quantization of these waves leads to the concept of magnons, defined as quanta of spin waves \cite{kittel2018introduction}. Magnonic systems exhibit a remarkable ability to couple with various types of subsystems \cite{lachance2019hybrid}, including microwave \cite{bienfait2016reaching, bienfait2017magnetic, haigh2015dispersive, grezes2016towards, o2014interfacing, hisatomi2016bidirectional, li2018magnon}, optical \cite{graf2021design, sharma2017light, satoh2012directional, graf2018cavity, o2014interfacing}, and mechanical \cite{li2018magnon, zhang2016cavity, amazioug2023enhancement, li2019squeezed} systems. These systems offer high tunability and typically possess long coherence times \cite{li2020hybrid}. Due to these exceptional properties, magnonic systems have garnered significant interest in the field of quantum science and technology in recent years.

It can be shown that a spatial overlap of a magnetic material (under a magnetic bias) and the magnetic field of a cavity mode can couple the magnonic system to the microwave mode \cite{lachance2019hybrid}. The coupling Hamiltonian will be,
\begin{equation}
    \hat{\mathcal{H}}_\text{EM} = \hbar g_\text{EM} \left(\hat{a}^\dagger\hat{c}+\hat{a}\hat{c}^\dagger\right),
\end{equation}
where $\hat{a}$ and $\hat{c}$ are the annihilation operators of microwave and magnonic modes, respectively, and $g_\text{EM}$ is the electromagnetic coupling rate. The generic form of the electromagnetic coupling rate can be obtained as,
\begin{equation}
    g_\text{EM} = \frac{1}{2\hbar} \sqrt{2s} g^*\mu_B\int \delta\vec{B}(\vec{r})\cdot\vec{s}(\vec{r})d^3\vec{r},
\end{equation}
in which $s$ and $g^*$ are the spin and the g-factor of the magnetic particles, $\mu_B$ is Bohr magneton, $\delta\vec{B}(\vec{r})$ is the spatial distribution of the magnetic field of the cavity mode where $\vert\delta\vec{B}(\vec{r})\vert$ is the amplitude of the vacuum fluctuations of the mode i.e. $\vec{\hat{B}}(\vec{r}) = \delta\vec{B}(\vec{r})\left(\hat{a} + \hat{a}^\dagger\right)$, and $\vec{s}(\vec{r})$ is the spatial distribution of spins in the magnetic material.\\

\section{Developing the electro-magno-mechanical model}
\begin{figure}[!h]
    \centering
    \begin{subfigure}[b]{0.65\textwidth}
    \includegraphics[width=\textwidth]{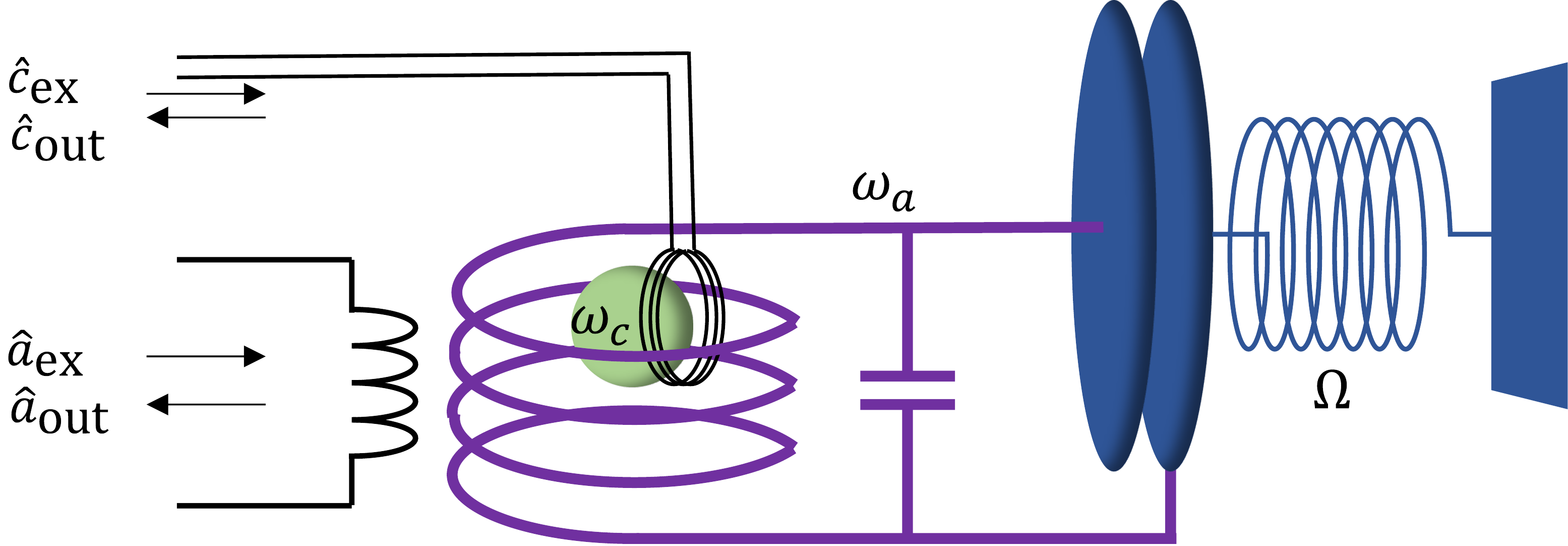}
        \caption{}
        \label{fig:tripartite-scheme-a}
    \end{subfigure}
    \begin{subfigure}[b]{0.8\textwidth}
        \includegraphics[width=\textwidth]{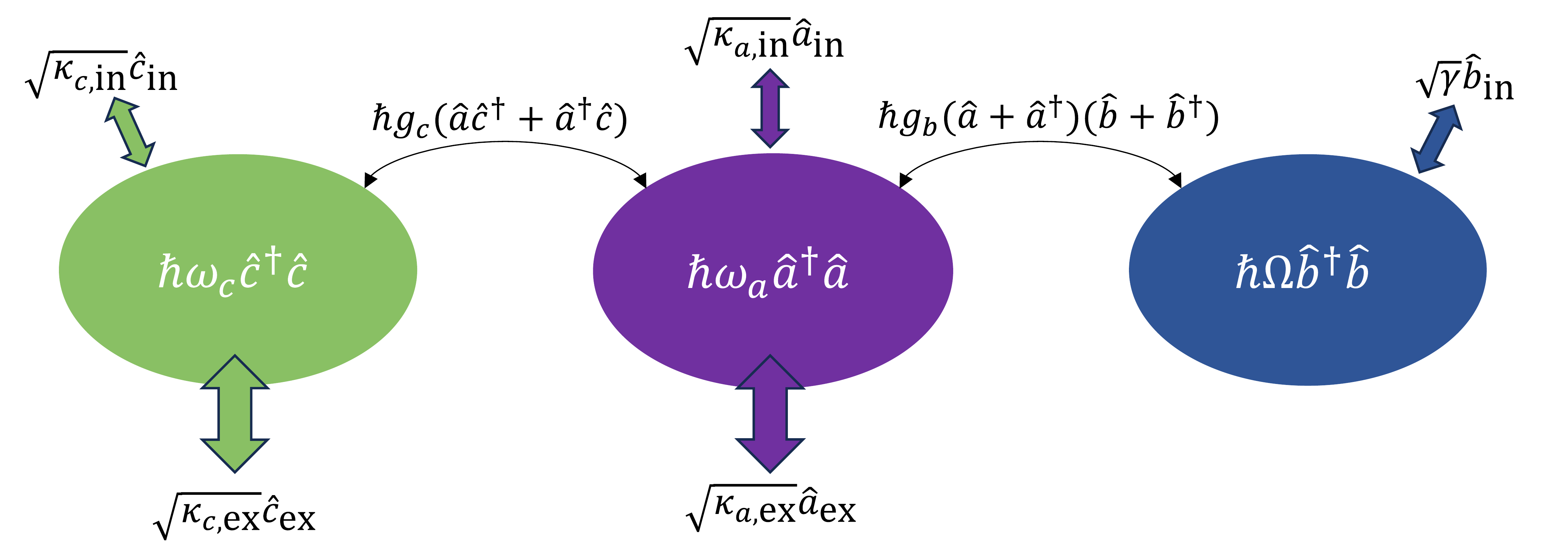}
        \caption{}
        \label{fig:tripartite-scheme-b}
    \end{subfigure}
    \caption[Eelectro-magno-mechanical system schematic]{The figure presents a schematic for the tripartite electro-magno-mechanical system. The magnonic system is coupled to the microwave cavity through the magnetic field of the cavity's mode, and the cavity is coupled to a mechanical oscillator electromechanically. (a) The concept of an implementation of the tripartite system. (b) The schematic showing the interactions of the subsystems.}
    \label{fig:tripartite-scheme}
\end{figure}
We start making the model by considering a microwave cavity coupled to a mechanical oscillator. There is also a magnetic material (e.g. YIG) in the system such that its magnonic field is coupled to the magnetic field of the microwave cavity. The system is schematically demonstrated in Figure (\ref{fig:tripartite-scheme-a}). The Hamiltonian of such a system in the frame rotating at the frequencies of the external pumps, $\omega_{p, a}$ (applied on microwave cavity) and $\omega_{p, c}$ (applied on the magnonic system), and after linearization (see section \ref{chp:Hamiltonian-linearization}) reads,
\begin{equation}
    \hat{\mathcal{H}} = \hbar\Delta_a \hat{a}^\dagger\hat{a} + \hbar\Omega\hat{b}^\dagger\hat{b} + \hbar\Delta_c\hat{c}^\dagger\hat{c} + \hbar g_b \left(\hat{a}^\dagger + \hat{a}\right)\left(\hat{b}^\dagger + \hat{b}\right) + \hbar g_c \left(\hat{a}^\dagger\hat{c} + \hat{a}\hat{c}^\dagger\right),
    \label{HMM}
\end{equation}
where $\hbar$ is the reduced Planck's constant, $\hat{a}$, $\hat{b}$ and $\hat{c}$ are the annihilation operators of the microwave, mechanical and magnonic subsystems respectively, $\Omega$ is the mechanical frequency, $\Delta_{a(c)} = \omega_{a(c)}-\omega_{p, a(c)}$ is the detuning of the microwave cavity (magnonic system) with frequency $\omega_{a(c)}$ with respect to the frequency of the external pump field while $g_b$ and $g_c$ are the electromechanical and electromagnonic coupling rates, respectively. As illustrated in Figure (\ref{fig:tripartite-scheme-b}), the Fourth term in the Hamiltonian (\ref{HMM}) describes the interaction between the microwave cavity and the mechanical oscillator. As we saw in Chapter \ref{chp:theory}, we can divide it into two groups with different physical behaviours by expanding it. Let us have a quick review of it:
\begin{equation}
    \hbar g_b\left(\hat{a}^\dagger + \hat{a}\right)\left(\hat{b}^\dagger + \hat{b}\right) = \hbar g_b\left(\hat{a}^\dagger\hat{b} + \hat{a}\hat{b}^\dagger\right) + \hbar g_b\left(\hat{a}^\dagger\hat{b}^\dagger + \hat{a}\hat{b}\right). 
\end{equation}
The first expression above describes the beam-splitter interaction, enabling photon-phonon transduction. However, the second expression represents the parametric interaction, which leads to amplification or the generation of continuous variable photon-phonon entanglement. Additionally, the last term in the Hamiltonian accounts for the electromagnonic interaction. As this interaction follows the beam-splitter form, it only facilitates photon-magnon energy exchange and does not generate entanglement. However, we will observe that the presence of the electromechanical interaction allows for the generation of tripartite entanglement, involving photon-magnon interactions.

To proceed, we will derive the Langevin equations of motion for the tripartite system 
\begin{align}
    \dot{\hat{a}} =& \frac{1}{i\hbar} \left[\hat{a}, \hat{\mathcal{H}}\right] - \frac{\kappa_a}{2}\hat{a} + \sqrt{\kappa_{a, \text{ex}}} \hat{a}_\text{ex} + \sqrt{\kappa_{a, \text{in}}} \hat{a}_\text{in}\nonumber\\
    =& \left(-i\Delta_a -\frac{\kappa_a}{2}\right)\hat{a} -i g_b \left(\hat{b}^\dagger + \hat{b}\right) - i g_c \hat{c} + \sqrt{\kappa_{a, \text{ex}}} \hat{a}_\text{ex} + \sqrt{\kappa_{a, \text{in}}} \hat{a}_\text{in},\\
    \dot{\hat{b}} =& \frac{1}{i\hbar}\left[\hat{b}, \hat{\mathcal{H}}\right] - \frac{\gamma}{2} \hat{b} + \sqrt{\gamma} \hat{b}_\text{in} \nonumber\\
    =& \left(-i\Omega - \frac{\gamma}{2}\right) \hat{b} - i g_b \left(\hat{a}^\dagger + \hat{a}\right) + \sqrt{\gamma} \hat{b}_\text{in},\\
    \dot{\hat{c}} =& \frac{1}{i\hbar}\left[\hat{c}, \hat{\mathcal{H}}\right] - \frac{\kappa_c}{2} \hat{c} + \sqrt{\kappa_{c, \text{ex}}} \hat{c}_\text{ex} + \sqrt{\kappa_{c, \text{in}}} \hat{c}_\text{in}\nonumber\\
    =& \left(-i\Delta_c -\frac{\kappa_c}{2}\right) \hat{c} - i g_c \hat{a} + \sqrt{\kappa_{c, \text{ex}}} \hat{c}_\text{ex} + \sqrt{\kappa_{c, \text{in}}} \hat{c}_\text{in}.
\end{align}
where $\kappa_{a(c), \text{ex}}$, $\kappa_{a(c), \text{in}}$, and $\kappa_{a(c)} = \kappa_{a(c), \text{ex}}+\kappa_{a(c), \text{in}}$ are the extrinsic, intrinsic, and the total damping rates of the microwave (magnonic) subsystem respectively, $\hat{a}(\hat{c})_\text{ex}$ and $\hat{a}(\hat{c})_\text{in}$ are the externally applied field and the environmental noise field operators on microwave (magnonic) subsystems, and $\gamma$ and $\hat{b}_\text{in}$ are the mechanical damping rate and the environmental noise field operator on the mechanical oscillator. Note that there is no externally applied field on the mechanical oscillator, as it only interacts with the environment (heat bath) and the microwave cavity. 

All the equations of motion can be rewritten in the following matrix form,
\begin{equation}
    \dot{\mathbf{\eta}} = \mathbf{A}\mathbf{\eta} + \mathbf{B}\mathbf{\eta}_\text{in},\label{eq:time-domain-equation-of-motion}
\end{equation}
where $\mathbf{\eta}=\left[\hat{a}, \hat{a}^\dagger, \hat{b}, \hat{b}^\dagger, \hat{c}, \hat{c}^\dagger\right]^T$ and $\mathbf{\eta}_\text{in}=\left[\hat{a}_\text{in}, \hat{a}_\text{in}^\dagger, \hat{a}_\text{ex}, \hat{a}_\text{ex}^\dagger, \hat{b}_\text{in}, \hat{b}_\text{in}^\dagger, \hat{c}_\text{in}, \hat{c}_\text{in}^\dagger, \hat{c}_\text{ex}, \hat{c}_\text{ex}^\dagger\right]^T$ with the superscript $T$ being the transposition operator, and the drift matrix, $\mathbf{A}$, is defined as,
\begin{equation}
    \mathbf{A} = \begin{pmatrix}
        -i\Delta_a - \kappa_a/2 & 0 & -ig_b & -ig_b & -ig_c & 0  \\
        0 & i\Delta_a - \kappa_a/2 & ig_b & ig_b & 0 & ig_c \\
        -ig_b & -ig_b & -i\Omega - \gamma/2 & 0 & 0 & 0 \\
        ig_b & ig_b & 0 & i\Omega - \gamma/2 & 0 & 0\\
        ig_c & 0 & 0 & 0 & -i\Delta_c - \kappa_c/2 & 0\\
        0 & ig_c & 0 & 0 & 0 & i\Delta_c - \kappa_c/2
    \end{pmatrix},
\end{equation}
and the matrix $\mathbf{B}$ is,
\begin{equation}
    \mathbf{B} = \begin{pmatrix}
        \sqrt{\kappa_{a, \text{in}}} & 0 & \sqrt{\kappa_{a, \text{ex}}} & 0 & 0 & 0 & 0 & 0 & 0 & 0 \\
        0 & \sqrt{\kappa_{a, \text{in}}} & 0 & \sqrt{\kappa_{a, \text{ex}}} & 0 & 0 & 0 & 0 & 0 & 0 \\
        0 & 0 & 0 & 0 & \sqrt{\gamma} & 0 & 0 & 0 & 0 & 0 \\
        0 & 0 & 0 & 0 & 0 & \sqrt{\gamma} & 0 & 0 & 0 & 0 \\
        0 & 0 & 0 & 0 & 0 & 0 & \sqrt{\kappa_{c, \text{in}}} & 0 & \sqrt{\kappa_{c, \text{ex}}} & 0 \\
        0 & 0 & 0 & 0 & 0 & 0 & 0 & \sqrt{\kappa_{c, \text{in}}} & 0 & \sqrt{\kappa_{c, \text{ex}}}
    \end{pmatrix}.
\end{equation}
Aside from the equations of motion, another utilization of the drift matrix, $\mathbf{A}$, is in stability determination. For the system to be stable, the real parts of all the eigenvalues of $\mathbf{A}$ should be negative. Therefore, we will use the drift matrix in the numerical analysis to find whether a set of parameters lead to a stable state of the system or not.

To analyze the system in the time domain, one can use Equation (\ref{eq:time-domain-equation-of-motion}) and proceed, but we are interested in the frequency domain behaviour of the system. Therefore, by applying a Fourier transform, we convert the equations of motion to the frequency domain.
\begin{equation}
    -i\omega\mathbf{\eta} = \mathbf{A}\mathbf{\eta} + \mathbf{B}\mathbf{\eta}_\text{in} \Rightarrow \mathbf{\eta} = \left[\left(-i\omega\mathbf{I}-\mathbf{A}\right)^{-1}\mathbf{B}\right]\mathbf{\eta}_\text{in}, \label{eq:frequency-domain-equation-of-motion}
\end{equation}
where $\mathbf{I}$ is an identity matrix. The above equation describes the intracavity field operators as functions of input field operators. However, we cannot measure the intracavity fields directly, in practice. Using the input-output theorem, we can also include the output field operators in the analysis and make predictions that are directly measurable.
\begin{equation}
    \begin{cases}
        \hat{a}_\text{out} = \sqrt{\kappa_{a, \text{ex}}} \hat{a} - \hat{a}_\text{ex}\\
        \hat{c}_\text{out} = \sqrt{\kappa_{c, \text{ex}}} \hat{c} - \hat{c}_\text{ex}
    \end{cases} \Rightarrow \mathbf{\eta}_\text{out} = \mathbf{C}\mathbf{\eta} - \mathbf{D}\mathbf{\eta}_\text{in}, \label{eq:input-output-theorem-second-eqaution}
\end{equation}
where $\mathbf{\eta}_\text{out} = \left[\hat{a}_\text{out}, \hat{a}_\text{out}^\dagger, \hat{c}_\text{out}, \hat{c}_\text{out}^\dagger\right]^T$ and the matrices $\mathbf{C}$ and $\mathbf{D}$ are defined as,
\begin{equation}
    \mathbf{C} = \begin{pmatrix}
        \sqrt{\kappa_{a, \text{ex}}} & 0 & 0 & 0 & 0 & 0\\
        0 & \sqrt{\kappa_{a, \text{ex}}} & 0 & 0 & 0 & 0\\
        0 & 0 & 0 & 0 & \sqrt{\kappa_{c, \text{ex}}} & 0\\
        0 & 0 & 0 & 0 & 0 & \sqrt{\kappa_{c, \text{ex}}}, \end{pmatrix}
\end{equation}
\begin{equation}
    \mathbf{D} = \begin{pmatrix}
        0 & 0 & 1 & 0 & 0 & 0 & 0 & 0 & 0 & 0 \\
        0 & 0 & 0 & 1 & 0 & 0 & 0 & 0 & 0 & 0 \\
        0 & 0 & 0 & 0 & 0 & 0 & 0 & 0 & 1 & 0 \\
        0 & 0 & 0 & 0 & 0 & 0 & 0 & 0 & 0 & 1
    \end{pmatrix}.
\end{equation}
Now, after plugging Equation (\ref{eq:input-output-theorem-second-eqaution}) into Equation (\ref{eq:frequency-domain-equation-of-motion}), we have,
\begin{equation}
    \mathbf{\eta}_\text{out} = \left[\mathbf{C}\left(-i\omega\mathbf{I}-\mathbf{A}\right)^{-1}\mathbf{B}-\mathbf{D}\right]\mathbf{\eta}_\text{in}.
\end{equation}
The matrix,
\begin{equation}
    \mathbf{S}_\eta = \mathbf{C}\left(-i\omega\mathbf{I}-\mathbf{A}\right)^{-1}\mathbf{B}-\mathbf{D},
\end{equation}
is the scattering matrix that relates the output operators to the input operators.\\
Our examination of entanglement involves studying the correlation between the quadrature operators. However, our current equations are expressed in terms of annihilation/creation operators. To obtain the equations of motion for quadrature operators, we can introduce specific rotation matrices and apply them to convert the previously derived equations. For the single-mode case, we arrive at the following expressions
\begin{equation}
    \begin{cases}
        \hat{X}_a = \frac{1}{\sqrt{2}}\left(\hat{a}+\hat{a}^\dagger\right)\\
        \hat{Y}_a = \frac{1}{i\sqrt{2}}\left(\hat{a}-\hat{a}^\dagger\right)
    \end{cases} \Rightarrow
    \begin{pmatrix}
        \hat{X}_a\\
        \hat{Y}_a
    \end{pmatrix} = \frac{1}{\sqrt{2}}
    \begin{pmatrix}
        1 & 1\\
        -i & i
    \end{pmatrix}
    \begin{pmatrix}
        \hat{a}\\
        \hat{a}^\dagger
    \end{pmatrix},
\end{equation}
where $\hat{X}_a$ and $\hat{Y}_a$ are the dimensionless quadrature operators of the field $\hat{a}$. Similarly, to translate $\mathbf{\eta}_\text{in}$ and $\mathbf{\eta}_\text{out}$ vectors to $\mathbf{q}_\text{in}$ and $\mathbf{q}_\text{out}$ vectors that we define as,
\begin{align*}
  \mathbf{q}_\text{in} =& \left[\hat{X}_{a,\text{in}}, \hat{Y}_{a,\text{in}}, \hat{X}_{a,\text{ex}}, \hat{Y}_{a,\text{ex}}, \hat{X}_{b,\text{in}}, \hat{Y}_{b,\text{in}}, \hat{X}_{c,\text{in}}, \hat{Y}_{c,\text{in}}, \hat{X}_{c,\text{ex}}, \hat{Y}_{c, \text{ex}} \right]^T,\\
  \mathbf{q}_\text{out} =& \left[\hat{X}_{a, \text{out}}, \hat{Y}_{a, \text{out}}, \hat{X}_{c, \text{out}}, \hat{Y}_{c, \text{out}}\right]^T,
\end{align*}
one can use rotations matrices $\mathbf{R}_5$ and $\mathbf{R}_2$,
\begin{align}
    \mathbf{R}_5 \equiv \mathbf{I}_5 \otimes \frac{1}{\sqrt{2}}\begin{pmatrix}
        1 & 1\\
        -i & i
    \end{pmatrix} &\Rightarrow \mathbf{q}_\text{in} = \mathbf{R}_5\mathbf{\eta}_\text{in},\\
    \mathbf{R}_2 \equiv \mathbf{I}_2 \otimes \frac{1}{\sqrt{2}}\begin{pmatrix}
        1 & 1\\
        -i & i
    \end{pmatrix} &\Rightarrow \mathbf{q}_\text{out} = \mathbf{R}_2\mathbf{\eta}_\text{out},
\end{align}
where $\mathbf{I}_n$ is an $n\times n$ identity matrix, and $\otimes$ denotes Kronecker product. Obviously, there should be a scattering matrix, $\mathbf{S}_q$, such that $\mathbf{q}_\text{out} = \mathbf{S}_q\mathbf{q}_\text{in}$.
\begin{equation*}
     \mathbf{\eta}_\text{out} = \mathbf{S}_\eta\mathbf{\eta}_\text{in} \Rightarrow \mathbf{R}_2^{-1}\mathbf{q}_\text{out} = \mathbf{S}_\eta\mathbf{R}_5^{-1}\mathbf{q}_\text{in} \Rightarrow \mathbf{q}_\text{out} = \mathbf{R}_2\mathbf{S}_\eta\mathbf{R}_5^{-1}\mathbf{q}_\text{in},
\end{equation*}
therefore,
\begin{equation}
    \mathbf{S}_q = \mathbf{R}_2\mathbf{S}_\eta\mathbf{R}_5^{-1}.
\end{equation}
Now, having the equations of motion of the quadrature operators, we can construct the covariance matrix of the output quadrature operators,
\begin{equation}
    \mathbf{V} = \frac{1}{2}\begin{pmatrix}
        2\langle \hat{X}_a^2\rangle & \langle \hat{X}_a\hat{Y}_a+\hat{Y}_a\hat{X}_a\rangle & \langle \hat{X}_a\hat{X}_c+\hat{X}_c\hat{X}_a\rangle & \langle \hat{X}_a\hat{Y}_c+\hat{Y}_c\hat{X}_a\rangle \\
        \langle \hat{Y}_a\hat{X}_a+\hat{X}_a\hat{Y}_a\rangle & 2\langle \hat{Y}_a^2\rangle & \langle \hat{Y}_a\hat{X}_c+\hat{X}_c\hat{Y}_a\rangle & \langle \hat{Y}_a\hat{Y}_c+\hat{Y}_c\hat{Y}_a\rangle\\
        \langle \hat{X}_c\hat{X}_a+\hat{X}_a\hat{X}_c\rangle & \langle \hat{X}_c\hat{Y}_a+\hat{Y}_a\hat{X}_c\rangle & 2\langle \hat{X}_c^2\rangle & \langle \hat{X}_c\hat{Y}_c+\hat{Y}_c\hat{X}_c\rangle\\
        \langle \hat{Y}_c\hat{X}_a+\hat{X}_a\hat{Y}_c\rangle & \langle \hat{Y}_c\hat{Y}_a+\hat{Y}_a\hat{Y}_c\rangle & \langle \hat{Y}_c\hat{X}_c+\hat{X}_c\hat{Y}_c\rangle & 2\langle \hat{Y}_c^2\rangle.
    \end{pmatrix}.
\end{equation}

\begin{align*}
    \mathbf{V} &= \frac{1}{2}\langle\mathbf{q}_\text{out}\mathbf{q}_\text{out}^T\rangle \\
    &= \frac{1}{2}\langle\left(\mathbf{S}_q\mathbf{q}_\text{in}\right) \left(\mathbf{S}_q\mathbf{q}_\text{in}\right)^T\rangle\\
    &= \frac{1}{2}\langle\mathbf{S}_q\mathbf{q}_\text{in}\mathbf{q}_\text{in}^T\mathbf{S}_q^T\rangle\\
    &= \frac{1}{2}\mathbf{S}_q\langle\mathbf{q}_\text{in}\mathbf{q}_\text{in}^T\rangle\mathbf{S}_q^T,
\end{align*}
therefore,
\begin{equation}
    \mathbf{V} = \left(\mathbf{R}_2\mathbf{S}_\eta\mathbf{R}_5^{-1}\right)\mathbf{N}\left(\mathbf{R}_2\mathbf{S}_\eta\mathbf{R}_5^{-1}\right)^T,
\end{equation}
where $\mathbf{N}$ is the input noise matrix which reads,
\begin{align}
    \mathbf{N} &= \frac{1}{2}\langle\mathbf{q}_\text{in}\mathbf{q}_\text{in}^T\rangle\nonumber\\
    &= \left(\operatorname{diag}\left[\bar{n}_{a,\text{in}}, \bar{n}_{a,\text{ex}}, \bar{n}_{b,\text{in}},
    \bar{n}_{c,\text{in}},
    \bar{n}_{c,\text{ex}}\right]+\frac{1}{2}\mathbf{I}_5\right)\otimes\mathbf{I}_2.
\end{align}
Note that to find the elements of the $\mathbf{N}$, we assumed all the inputs are thermal noises, thus,
\begin{align}
    2\langle\hat{X}^2_{a(c), \text{in}}\rangle =&  2\langle\hat{Y}^2_{a(c), \text{in}}\rangle =
    2 \bar{n}_{a(c), \text{in}} + 1 = 2 \frac{1}{e^{\hbar\omega_{a(c)}/k_BT} - 1} + 1,\\
    2\langle\hat{X}^2_{a(c), \text{ex}}\rangle =&  2\langle\hat{Y}^2_{a(c), \text{ex}}\rangle =
    2 \bar{n}_{a(c), \text{ex}} + 1 = 2 \frac{1}{e^{\hbar\omega_{a(c)}/k_BT} - 1} + 1,\\
    2\langle\hat{X}^2_{b, \text{in}}\rangle =&  2\langle\hat{Y}^2_{b, \text{in}}\rangle =
    2 \bar{n}_{b, \text{in}} + 1 = 2 \frac{1}{e^{\hbar\Omega/k_BT} - 1} + 1
\end{align}
where $T$ is the temperature, $k_B$ is the Boltzmann constant, and we used the Bose-Einstein distribution \cite{greiner2012thermodynamics} to find the last terms in each equation. All the other elements of the matrix vanish, as there is no correlation between different quadrature operators of the same input fields, and between different input fields, for the case of thermal noise.\\
The covariance matrix $\mathbf{V}$ appears as a block matrix,
\begin{equation}
    \mathbf{V} = \begin{pmatrix}
        \mathbf{V}_a & \mathbf{V}_{ac} \\
        \mathbf{V}_{ac}^T & \mathbf{V}_c
    \end{pmatrix},
\end{equation}
in which $\mathbf{V}_a$, $\mathbf{V}_c$, and $\mathbf{V}_{ac}$ are $2\times 2$ matrices. The smallest symplectic eigenvalue of the covariance matrix can be obtained as \cite{vidal2002computable, adesso2004extremal, barzanjeh2011entangling},
\begin{equation}
    \zeta^- = \left(\frac{\Sigma-\sqrt{\Sigma^2-4\left|\mathbf{V}\right|}}{2}\right)^{1/2}
\end{equation}
in which $\Sigma = \left|\mathbf{V}_a\right| + \left|\mathbf{V}_c\right| - 2\left|\mathbf{V}_{ac}\right|$.
The lower bound imposed by classical physics for $\zeta^-$ is $1/2$, hence, any value below $1/2$ indicates a non-classical correlation, meaning quantum entanglement. The logarithmic negativity which quantifies the entanglement is defined as,
\begin{equation}
    E_N = \max\left[0, -\log 2\zeta^-\right].\label{eq:Logarithmic_Negativity}
\end{equation}

\section{Numerical results}\label{sec3}

Using Equation (\ref{eq:Logarithmic_Negativity}), one can perform numerical analysis to determine the presence of entanglement for various sets of parameters. We conducted an extensive numerical investigation of the model and observed that measurable entanglement can be achieved in the output fields for typical physical parameters that are attainable in laboratory conditions. In the subsequent sections, we will present some of the numerical results to illustrate this phenomenon.\\
We have considered a microwave cavity, with the frequency $\omega_a/2\pi 
= 10$ GHz, coupled to a mechanical oscillator and a magnonic system with frequencies $\Omega/2\pi = 4$ MHz and $\omega_c/2\pi \sim 10$ GHz. Note that we can tune the magnonic system's frequency over a wide range, by changing the applied constant magnetic field. The intrinsic damping rates of the microwave, magnonic, and mechanical systems are $\kappa_{a, \text{in}}/2\pi = 0.8$ MHz, $\kappa_{c, \text{in}}/2\pi = 0.8$ MHz, and $\gamma/2\pi = 100$ Hz. The extrinsic damping rates of the microwave and magnonic subsystems are $\kappa_{a, \text{ex}}/2\pi = \kappa_{c, \text{ex}}/2\pi = 1.2$ MHz.

We investigate the change in the logarithmic negativity in response to the shifts in the electromechanical coupling, $g_b$, the electromagnonic coupling, $g_c$, and the microwave, and the magnonic frequency detunings, $\Delta_a$ and $\Delta_c$. Practically, the electromechanical coupling rate can be controlled by changing the power of the input field. The electromagnonic coupling rate depends on the overlap of the microwave magnetic field and the magnetic material, therefore, the electromagnonic coupling can be controlled by an accurate adjustment of the position of the magnetic material with respect to the microwave resonator. The microwave detuning can simply be adjusted by tuning the input field frequency, and finally, the magnonic detuning can be tuned by changing the magnonic frequency which depends on the amplitude of the bias magnetic field.

\begin{figure}[!h]
    \centering
    \begin{subfigure}[b]{0.45\textwidth}
        \includegraphics[width=\textwidth]{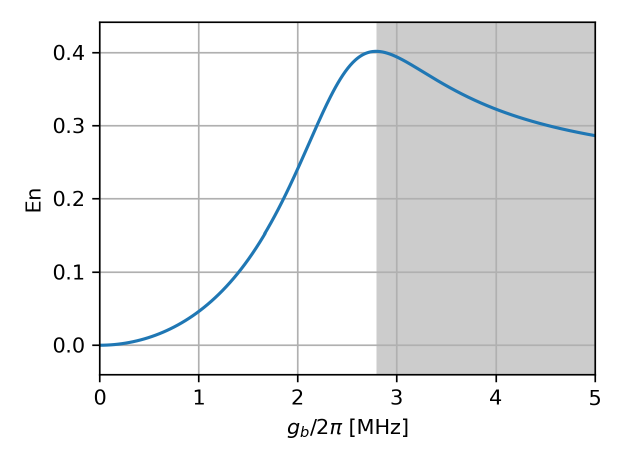}
        \caption{}
    \end{subfigure}
    \hfill
    \begin{subfigure}[b]{0.45\textwidth}
        \includegraphics[width=\textwidth]{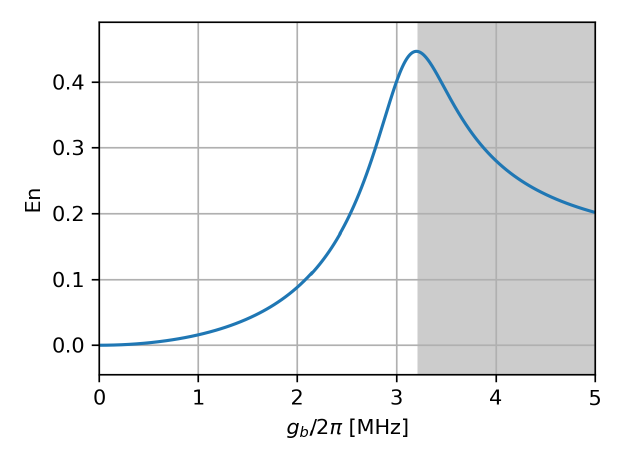}
        \caption{}
    \end{subfigure}
    \caption[Logarithmic negativity vs. electromagnonic coupling for $\Delta_c = -\Omega$]{The figures show logarithmic negativity versus electromechanical coupling for $\Delta_c = -\Omega$. The shaded regions show the unstable region. As can be seen, with no electromechanical coupling, no entanglement exists in the output fields. As the electromechanical coupling increases, the electromechanical interaction leads to electromagnetic entanglement. (a) Fixed parameters: $g_c/2\pi = 6.43$ MHz, $\Delta_a = -\Omega$. (b) Fixed parameters: $g_c/2\pi = 4.89$ MHz, $\Delta_a = \Omega$.}
\end{figure}

As we have seen previously, the electromagnonic coupling is in the beam-splitting form which cannot generate entanglement. However, as we include the electromechanical coupling, phonon-photon entangled pairs will be generated, and due to the photon-magnon excitation exchange in the system, the entanglement partially redistributes to the magnonic system as well. Ultimately, the tripartite intracavity entanglement leads to entangled output fields.\\
As mentioned previously, instability occurs when the real part of at least one eigenvalue of the drift matrix, $\mathbf{A}$, becomes positive. Further investigation of the effect of photonic and magnonic subsystems on the mechanical damping rate can shed light on the conditions of instability.

\begin{figure}[!h]
    \centering
    \begin{subfigure}[b]{0.45\textwidth}
        \includegraphics[width=\textwidth]{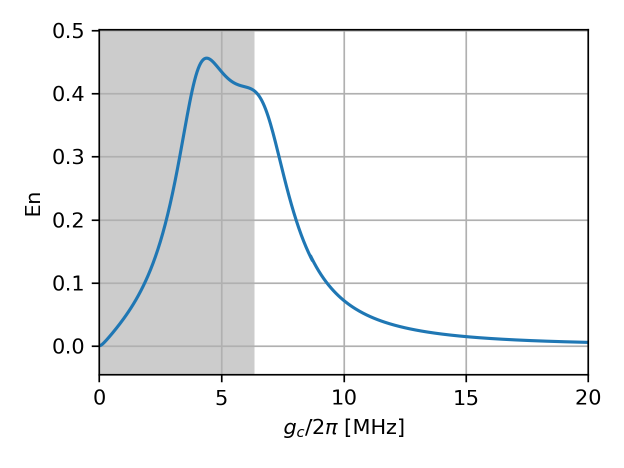}
        \caption{}
        \label{fig:gc_sweep_Delta_a=-Omega}
    \end{subfigure}
    \hfill
    \begin{subfigure}[b]{0.45\textwidth}
        \includegraphics[width=\textwidth]{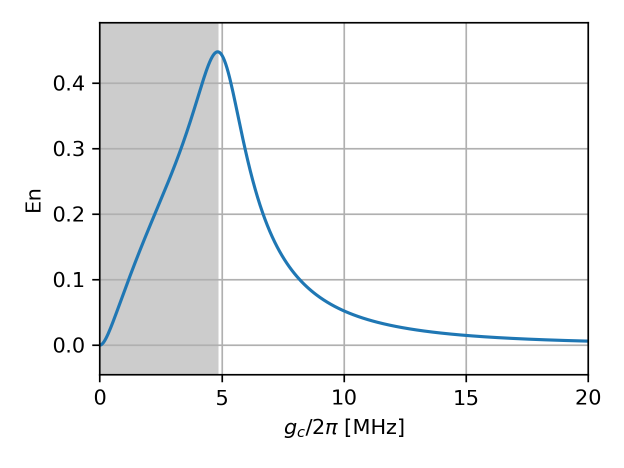}
        \caption{}
        \label{fig:gc_sweep_Delta_a=Omega}
    \end{subfigure}
    \caption[Logarithmic negativity vs. electromagnonic coupling for $\Delta_c = -\Omega$]{The figures show logarithmic negativity versus electromagnonic coupling for $\Delta_c = -\Omega$. The shaded regions show the unstable region. (a) Fixed parameters: $g_b/2\pi = 2.78$ MHz, $\Delta_a = -\Omega$ (b) Fixed parameters: $g_b/2\pi = 3.20$ MHz, $\Delta_a = \Omega$}
\end{figure}
\begin{figure}[!h]
    \centering
    \includegraphics[width=0.7\textwidth]{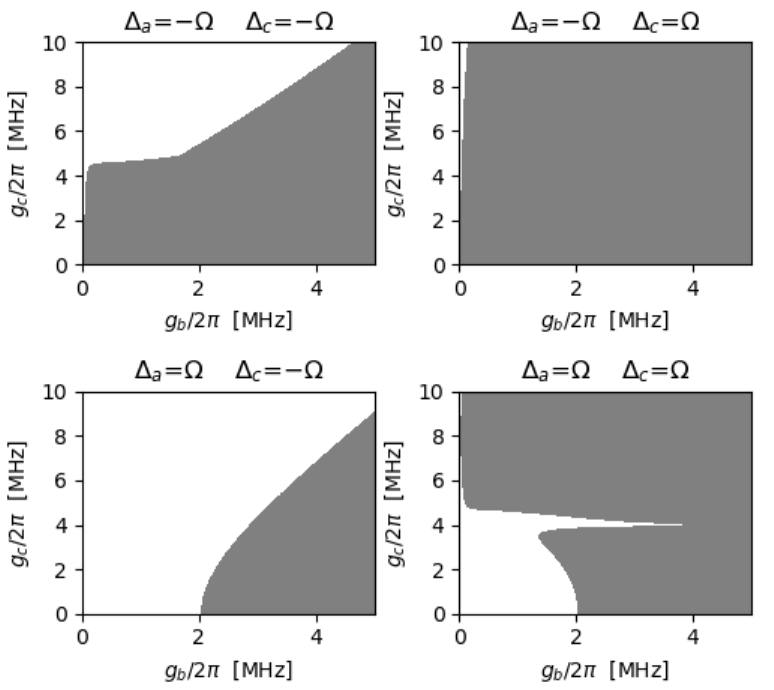}
    \caption[Stability vs. coupling rates]{The figures show the stability regions for different ranges of electromechanical and electromagnonic coupling rates. The shaded areas indicate unstable regions.}
    \label{fig:instability2}
\end{figure}

The diagrams presented in figures (\ref{fig:gc_sweep_Delta_a=-Omega}) and (\ref{fig:gc_sweep_Delta_a=Omega}) show that the system is unstable in the absence of the electromagnonic interaction, with the indicated fixed detunings and electromechanical coupling rate. As the electromagnonic coupling increases, the magnonic system enhances mechanical cooling by adding a second energy-damping source and makes the system stable. Figure (\ref{fig:instability2}) provides more insight into the stability behaviour of the system. As can be seen, generally, with the microwave blue-detuned ($\Delta_a = -\Omega$) condition the critical electromechanical coupling rate (over which the system enters the unstable regime) is smaller compared to the microwave red-detuned ($\Delta_a = \Omega$) condition, which means the stability of the system is more robust to electromechanical coupling increment under the red-detuned condition. It is worth mentioning that a simple numerical analysis showed that the critical electromechanical coupling increases with the intrinsic mechanical damping rate, $\gamma$.

  \chapter{Conclusion and future scope}

To summarize the thesis, we reviewed the basic concepts and definitions of opto/electromechanics in Chapter 1 and went through the theoretical model describing linear opto/electromechanical interaction in Chapter 2.\\
Then, in Chapter 3, we introduced the nano-string concept, explained our electromechanical nano-string device, and showed how we simulated such a device. The final results of the simulations were also presented for silicon and diamond substrates.\\
In Chapter 4, the fabrication method of the electromechanical device was demonstrated and explained. Specifically, two types of designs for fabrication were introduced, one for 3D microwave cavities which contained one device on a single chip, and the other with on-chip integrated resonators which contained more than one device on a single chip and needed a ground plane.\\
Chapter 5 was dedicated to the measurement setup and method. As we have already mentioned before, no electromechanical device was characterized, however, the measurement setup is implemented and the procedure is clearly stated.\\
Finally, in Chapter 6, a theoretical model based on electromechanical interaction was introduced. I developed the theoretical model to show how an electromechanical interaction can enrich electromagnetic interaction, and lead to photon-magnon entanglement generation.

The future steps following this thesis would be optimizing and improving the fabrication recipe, and fabricating a well-functioning electromechanical device. Then, after confirming and characterizing the electromechanical interaction, more subsystems can be implemented beside the device, to make more complicated systems. An example would be implementing a photonic crystal coupled to the nano-string, to make a microwave-optical transducer. Another avenue would be the coupling of a magnetic material such as a YIG sphere with the microwave resonator, either the integrated LC circuit or 3D cavity. This allows us to test and confirm the theoretical model developed and presented in the last chapter of the thesis. 


  \cleardoublepage\phantomsection
  \addcontentsline{toc}{chapter}{Bibliography}

\bibliography{MSc-thesis-Armin-Tabesh}
\bibliographystyle{unsrt}


  \appendix             

  \chapter{Mathematical tools}

\section{Operators in frequency domain}
\label{chp:frequency_domain_analysis}
In quantum mechanics, we generally define the operators in the time domain. However, sometimes we are interested in analyzing the physical systems in the frequency domain; we want to understand their behaviors in different frequencies. Therefore, we transfer the equations and relations from time domain to frequency domain by applying Fourier transform.\\
We define the Fourier transform and its inverse on an arbitrary function $x(t)$ as follows.
\begin{align}
    x(\omega) =& \mathcal{F}\left\{x(t)\right\} = \frac{1}{\sqrt{2\pi}}\int_{-\infty}^\infty x(t) e^{i\omega t}dt,\\
    x(t) =& \mathcal{F}^{-1}\left\{x(\omega)\right\} = \frac{1}{\sqrt{2\pi}}\int_{-\infty}^\infty x(\omega) e^{-i\omega t}d\omega.
\end{align}
The Fourier transform of an operator is exactly the same as the scalar function x(t).
\begin{align}
    \hat{a}(\omega) =& \mathcal{F}\left\{\hat{a}(t)\right\} = \frac{1}{\sqrt{2\pi}} \int_{-\infty}^\infty \hat{a}(t) e^{i\omega t} dt,\label{eq:Fourier_transform_a}\\
    \hat{a}(t) =& \mathcal{F}^{-1}\left\{\hat{a}(\omega)\right\} = \frac{1}{\sqrt{2\pi}}\int_{-\infty}^\infty \hat{a}(\omega) e^{-i\omega t}d\omega.
\end{align}
Using the integration by parts method, and the assumption that $\hat{a}(t)$ vanishes at $t\rightarrow\pm\infty$, it is easy to show that $\mathcal{F}\left\{\dot{\hat{a}}(t)\right\}=-i\omega\hat{a}(\omega)$.
\begin{align}
    \mathcal{F}\left\{\dot{\hat{a}}(t)\right\} =& \frac{1}{\sqrt{2\pi}}\int_{-\infty}^\infty \frac{d\hat{a}(t)}{dt}e^{i\omega t} dt =\nonumber\\
    &\frac{1}{\sqrt{2\pi}}\left(\hat{a}(t)e^{i\omega t}\biggr|_{-\infty}^\infty - \int_{-\infty}^\infty \hat{a}(t)\frac{de^{i\omega t}}{dt} dt\right) = \nonumber\\
    & \frac{-i\omega}{\sqrt{2\pi}}\int_{-\infty}^\infty\hat{a}(t)e^{i\omega t} dt = -i\omega\mathcal{F}\left\{\dot{\hat{a}}(t)\right\} = -i\omega\hat{a}(\omega).
\end{align}
Similar equations can be defined and obtained for $\hat{a}^\dagger$;
\begin{gather}
    \hat{a}^\dagger(\omega) = \mathcal{F}\left\{\hat{a}^\dagger(t)\right\} = \frac{1}{\sqrt{2\pi}} \int_{-\infty}^\infty \hat{a}^\dagger(t) e^{i\omega t} dt,\\
    \hat{a}^\dagger(t) = \mathcal{F}^{-1}\left\{\hat{a}^\dagger(\omega)\right\} = \frac{1}{\sqrt{2\pi}}\int_{-\infty}^\infty \hat{a}^\dagger(\omega) e^{-i\omega t}d\omega,\\
    \mathcal{F}\left\{\dot{\hat{a}}^\dagger(t)\right\} = -i\omega\hat{a}^\dagger(\omega).
\end{gather}
We know $\left(\hat{a}(t)\right)^\dagger = \hat{a}^\dagger(t)$ by definition, but to find a similar relations for $\hat{a}(\omega)$ and $\hat{a}^\dagger(\omega)$, we attempt to find the Hermitian conjugate of the Equation (\ref{eq:Fourier_transform_a}). 
\begin{align}
    \left(\hat{a}(\omega)\right)^\dagger =& \left(\frac{1}{\sqrt{2\pi}}\int_{-\infty}^\infty \hat{a}(t) e^{i\omega t}dt\right)^\dagger = \frac{1}{\sqrt{2\pi}}\int_{-\infty}^\infty \hat{a}^\dagger(t) e^{-i\omega t}dt = \nonumber\\
    &\frac{1}{\sqrt{2\pi}}\int_{-\infty}^\infty \hat{a}^\dagger(t) e^{i(-\omega)t}dt = \hat{a}^\dagger(-\omega) \Rightarrow\nonumber\\
    \left(\hat{a}(\omega)\right)^\dagger =& \hat{a}^\dagger(-\omega).
\end{align}
The last relation to find is the commutation relation in the frequency domain, $\left[\hat{a}(\omega), \hat{a}^\dagger(\omega^\prime)\right]$.
\begin{align}
    \left[\hat{a}(\omega), \hat{a}^\dagger(\omega^\prime)\right] =& \left[\frac{1}{\sqrt{2\pi}}\int_{-\infty}^\infty \hat{a}(t) e^{i\omega t}dt, \frac{1}{\sqrt{2\pi}}\int_{-\infty}^\infty \hat{a}^\dagger(t^\prime) e^{i\omega^\prime t^\prime}dt^\prime\right] =\nonumber\\
    &\frac{1}{2\pi}\int_{-\infty}^\infty dt\int_{-\infty}^\infty dt^\prime \left[\hat{a}(t), \hat{a}^\dagger(t^\prime)\right]e^{i(\omega t + \omega^\prime t^\prime)} = \nonumber\\
    &\frac{1}{2\pi}\int_{-\infty}^\infty dt\int_{-\infty}^\infty dt^\prime\delta(t-t^\prime)e^{i(\omega t + \omega^\prime t^\prime)} = \nonumber\\
    &\frac{1}{2\pi}\int_{-\infty}^\infty dt e^{i(\omega + \omega^\prime)t} = \delta(\omega+\omega^\prime) \Rightarrow \nonumber\\
    \left[\hat{a}(\omega), \hat{a}^\dagger(\omega^\prime)\right] =& \delta(\omega+\omega^\prime).
\end{align}
In the recent calculations, the definition of the delta function,
\begin{equation}
    \delta(\omega) = \frac{1}{2\pi}\int_{-\infty}^\infty e^{i\omega t} dt,
\end{equation}
and the fundamental commutation relation,
\begin{equation}
    \left[\hat{a}(t), \hat{a}^\dagger(t^\prime)\right] = \delta(t-t^\prime),
\end{equation}
are used.

\section{Input-output theorem}
\label{appendix:input-output-theorem}
The input-output theorem is a mathematical model that describes the intracavity and the output fields of a cavity when it is subjected to an externally applied field. To build the model, we start by deriving the Langevin equation, and then we find the equations of interest.

\subsection{Quantum Langevin equations}
We assume a single-mode system in contact with a heat bath with infinitely many modes. The Hamiltonian reads,
\begin{equation}
    \hat{\mathcal{H}} = \hat{\mathcal{H}}_\text{sys} + \hat{\mathcal{H}}_\text{bath} + \hat{\mathcal{H}}_\text{int},
\end{equation}
where $\hat{\mathcal{H}}_\text{sys}$ is the generic Hamiltonian of the system,
\begin{equation}
    \hat{\mathcal{H}}_\text{bath} = \sum_n \hbar\omega_n\hat{b}^\dagger\hat{b},
\end{equation}
is the Hamiltonian of the heat bath, and
\begin{equation}
    \hat{\mathcal{H}}_\text{int} = i\hbar\sum_n g_n \left(\hat{a}\hat{b}_n^\dagger - \hat{a}^\dagger\hat{b}_n\right),
\end{equation}
describes the interaction between the system and the heat bath. $\hat{a}$ and $\hat{b}_n$ are the annihilation operators of the system and the $n$th mode of the heat bath, and $g_n$ indicates the coupling of the system with the $n$th mode of the heat bath. To find the equation of motion of the field of the system, we use the Heisenberg equation.
\begin{equation}
    \dot{\hat{a}} = \frac{1}{i\hbar}\left[\hat{a}, \hat{\mathcal{H}}\right] = \frac{1}{i\hbar}\left(\left[\hat{a}, \hat{\mathcal{H}}_\text{sys}\right] - i\hbar\sum_n g_n\hat{b}_n\right) = \frac{1}{i\hbar}\left[\hat{a}, \hat{\mathcal{H}}_\text{sys}\right] - \sum_n g_n \hat{b}_n.\label{eq:equation_of_motion_a}
\end{equation}
Now, we need to find an equation for $\hat{b}_n$.
\begin{align}
    \dot{\hat{b}}_n =& \frac{1}{i\hbar}\left[\hat{b}_n, \hat{\mathcal{H}}\right] = \frac{1}{i\hbar}\left(\left[\hat{b}_n, \hat{\mathcal{H}}_\text{bath}\right] + \left[\hat{b}_n, \hat{\mathcal{H}}_\text{int}\right]\right) = \nonumber\\
    &\frac{1}{i\hbar}\left(\sum_m\hbar\omega_m\left[\hat{b}_n, \hat{b}_m^\dagger\hat{b}_m\right]+i\hbar\sum_mg_m\left[\hat{b}_n, \hat{a}\hat{b}_m^\dagger\right]\right)\nonumber\\
    &\frac{1}{i\hbar}\left(\hbar\omega_n\hat{b}_n+i\hbar g_n\hat{a}\right) = -i\omega_n\hat{b}_n + g_n\hat{a}.
    \label{eq:operator_b_differential_equation}
\end{align}
Solving the differential equation, with the initial condition $\hat{b}_n(t_0)$ for $t_0<t$, we have,
\begin{equation}
    \hat{b}_n(t) = e^{-i\omega_n(t-t_0)}\hat{b}_n(t_0) + g_n \int_{t_0}^t dt^\prime e^{-i\omega_n(t-t^\prime)}\hat{a}(t^\prime),
\end{equation}
and plugging the result in Equation (\ref{eq:equation_of_motion_a}), we obtain,
\begin{equation}
    \dot{\hat{a}} = \frac{1}{i\hbar}\left[\hat{a}, \hat{\mathcal{H}}_\text{sys}\right] - \sum_n g_n e^{-i\omega_n(t-t_0)}\hat{b}_n(t_0) - \sum_n g_n^2\int_{t_0}^t dt^\prime e^{-i\omega_n(t-t^\prime)}\hat{a}(t^\prime).\label{eq:equation_of_motion_a2}
\end{equation}
At this point, we make an approximation to simplify the expression. We assume $g_n$ does not depend on $n$ and is a constant,
\begin{equation}
    g_n = \sqrt{\frac{\kappa}{2\pi}}.
\end{equation}
Therefore, the last term in Equation (\ref{eq:equation_of_motion_a2}) can be simplified as follows.
\begin{align}
    \sum_n g_n \int_{t_0}^t dt^\prime e^{-i\omega_n(t-t^\prime)}\hat{a}(t^\prime) \simeq& \frac{\kappa}{2\pi}\int_{t_0}^t dt^\prime\sum_ne^{-i\omega_n(t-t^\prime)}\hat{a}(t^\prime) = \nonumber\\
    &\frac{\kappa}{2\pi}\int_{t_0}^t dt^\prime 2\pi\delta\left(t-t^\prime\right)\hat{a}(t^\prime) = \frac{\kappa}{2}\hat{a}(t),
\end{align}
in which we use the identity $\sum_n \exp[-i\omega_nt] = 2\pi\delta(t)$, with $\delta(t)$ being the delta function. Furthermore, we define a new operator, $\hat{b}_\text{in}$, as
\begin{equation}
    \hat{b}_\text{in}(t) = -\frac{1}{\sqrt{2\pi}}\sum_n e^{-i\omega_n(t-t_0)}\hat{b}_n(t_0),
\end{equation}
which holds the commutation relation of the ladder operators, $\left[\hat{b}_\text{in}(t), \hat{b}_\text{in}^\dagger(t^\prime)\right] = \delta(t-t^\prime)$. Due to the initial condition we chose previously, $\hat{b}_\text{in}$, in fact, represents the collective external field applied to the system as an input field. Finally, with the approximation and the new definition, the simplified form of Equation (\ref{eq:equation_of_motion_a2}) reads,
\begin{equation}
    \dot{\hat{a}} = \frac{1}{i\hbar}\left[\hat{a}, \hat{\mathcal{H}}_\text{sys}\right] - \frac{\kappa}{2}\hat{a} + \sqrt{\kappa}\hat{b}_\text{in}. \label{eq:first_Langevin_equation}
\end{equation}
The recent equation is one of the two quantum Langevin equations and demonstrates fluctuation ($\sqrt{\kappa}\hat{b}_\text{in}$) and dissipation ($\frac{\kappa}{2}\hat{a}$) in the system. To find the second equation, we have to consider the other possibility for the initial condition in solving differential Equation (\ref{eq:operator_b_differential_equation}), i.e. $\hat{b}_n(t_1)$ for $t_1>t$.
\begin{equation}
    \hat{b}_n(t) = e^{-i\omega_n(t-t_1)}\hat{b}_n(t_1) - g_n \int_{t_1}^t dt^\prime e^{-i\omega_n(t-t^\prime)}\hat{a}(t^\prime).
\end{equation}
Now, with defining $\hat{b}_\text{out}(t)$, as,
\begin{equation}
    \hat{b}_\text{out}(t) = -\frac{1}{\sqrt{2\pi}}\sum_n e^{-i\omega_n(t-t_1)}\hat{b}_n(t_1),
\end{equation}
and similar analysis as for $\hat{b}_\text{in}$, we find the equation,
\begin{equation}
    \dot{\hat{a}} = \frac{1}{i\hbar}\left[\hat{a}, \hat{\mathcal{H}}_\text{sys}\right] + \frac{\kappa}{2}\hat{a} + \sqrt{\kappa}\hat{b}_\text{out}.
\end{equation}
Considering the initial condition of the differential equation, $\hat{b}_\text{out}$ is the collective field operator corresponding to the output field of the system. The recent equation and the first quantum Langevin equation (Eq. \ref{eq:first_Langevin_equation}), and after some algebra, one can simply find the second quantum Langevin equation,
\begin{equation}
    \hat{b}_\text{out} + \hat{b}_\text{in} = \sqrt{\kappa}\hat{a}.\label{eq:second_Langevin_equation}
\end{equation}
Having equations \ref{eq:first_Langevin_equation} and \ref{eq:second_Langevin_equation} and the explicit form of $\mathcal{H}_\text{sys}$, one can fully analyze the response of the system to the input fields. Using analogous analysis, it is easy to show that in the case that the system is coupled to $N$ reservoirs, the Langevin equations can be found as,
\begin{align}
    &\dot{\hat{a}} = \frac{1}{i\hbar}\left[\hat{a}, \hat{\mathcal{H}}_\text{sys}\right] + \sum_{n=1}^N -\frac{\kappa_n}{2}\hat{a} + \sqrt{\kappa_n}\hat{b}_\text{in}^{(n)},\\
    &\hat{b}_\text{out}^{(n)} + \hat{b}_\text{in}^{(n)} = \sqrt{\kappa_n}\hat{a},
\end{align}
where $\kappa_n$ is the coupling rate of the system with the $n$th reservoir, and $\hat{b}_\text{in(out)}^{(n)}$ is the annihilation operator of the input (output) field from (to) the $n$th reservoir.

\subsection{One-sided cavity}
A simple example of the utilization of input-output theorem is finding the reflection of a singe-mode ($\omega_c$) one-sided cavity. A one-sided cavity is a cavity that is coupled to one communication line extrinsically (with the extrinsic coupling rate $\kappa_\text{ex}$), and to a heat bath intrinsically (with the intrinsic coupling rate $\kappa_\text{in}$). Having the Hamiltonian of the cavity, $\mathcal{H}_\text{sys} = \hbar\omega_c\hat{a}^\dagger\hat{a}$, we can easily write the equation of motion of the annihilation operator of the cavity, $\hat{a}$.
\begin{align}
    \dot{\hat{a}} =& \frac{1}{i\hbar}\left[\hat{a}, \hat{\mathcal{H}}_\text{sys}\right] - \frac{\kappa_\text{in}+\kappa_\text{ex}}{2}\hat{a} + \sqrt{\kappa_\text{in}}\hat{a}_\text{in} + \sqrt{\kappa_\text{ex}}\hat{a}_\text{ex} =\nonumber\\
    & -i\omega_c\hat{a} - \frac{\kappa}{2}\hat{a} + \sqrt{\kappa_\text{in}}\hat{a}_\text{in} + \sqrt{\kappa_\text{ex}}\hat{a}_\text{ex},
\end{align}
where $\kappa = \kappa_\text{in}+\kappa_\text{ex}$ is the total damping rate of the cavity, and $\hat{a}$, $\text{in}$, and $\hat{a}_\text{ex}$ are the annihilation operators of the intracavity, heat bath, and the externally applied fields, respectively. It is not practically possible to receive and analyze the output field of the cavity to the heat bath, so we only care about the output field of the cavity to the communication line which we define by the annihilation operator $\hat{a}_\text{out}$. Therefore,
\begin{equation}
    \hat{a}_\text{out} + \hat{a}_\text{ex} = \sqrt{\kappa_\text{ex}} \hat{a}.
\end{equation}

To understand the response of the cavity in frequency domain, we apply Fourier transformation to the equation of motion.
\begin{align}
    -i\omega\hat{a} =& \left(-i\omega_c- \frac{\kappa}{2}\right)\hat{a} + \sqrt{\kappa_\text{in}}\hat{a}_\text{in} + \sqrt{\kappa_\text{ex}}\hat{a}_\text{ex}\Rightarrow \nonumber\\
    \left(-i(\omega-\omega_c) + \frac{\kappa}{2}\right)\hat{a} =& \sqrt{\kappa_\text{in}}\hat{a}_\text{in} + \sqrt{\kappa_\text{ex}}\hat{a}_\text{ex}.
\end{align}
At this point, one can find an equation for the number of photons in the cavity ($\bar{n}_c = \langle\hat{a}^\dagger\hat{a}\rangle$) when the temperature of the heat bath is around the cavity's ground state ($\langle\hat{a}_\text{in}^\dagger\hat{a}_\text{in}\rangle\simeq 0$).
\begin{align}
    \hat{a} =& \frac{\sqrt{\kappa_\text{in}}\hat{a}_\text{in} + \sqrt{\kappa_\text{ex}}\hat{a}_\text{ex}}{-i(\omega-\omega_c) + \frac{\kappa}{2}} \Rightarrow\nonumber\\
    \langle\hat{a}^\dagger\hat{a}\rangle =& \frac{\kappa_\text{ex}\langle\hat{a}_\text{ex}^\dagger\hat{a}_\text{ex}\rangle}{\left(\omega-\omega_c\right)^2 + \frac{\kappa^2}{4}} \Rightarrow \\
    \bar{n}_c =& \frac{\kappa_\text{ex}\frac{P_\text{ex}}{\hbar\omega_p}}{\left(\omega-\omega_c\right)^2 + \frac{\kappa^2}{4}},
\end{align}
where $P_\text{ex}$ and $\omega_p$ are the power and frequency of the externally applied field, and $\langle\hat{a}_\text{ex}^\dagger\hat{a}_\text{ex}\rangle = P_\text{ex}/\hbar\omega_p$ indicates the number of photons per unit time passing through the communication line.\\
Furthermore, we can find the reflection of the cavity, $\mathcal{R} = \langle\hat{a}_\text{out}\rangle/\langle\hat{a}_\text{ex}\rangle$, using Langevin equations in the frequency domain.
\begin{align}
    \mathcal{R} = \frac{\langle\hat{a}_\text{out}\rangle}{\langle\hat{a}_\text{ex}\rangle} = -\frac{-i(\omega-\omega_c) + \frac{\kappa_\text{in}-\kappa_\text{ex}}{2}}{-i(\omega-\omega_c) + \frac{\kappa_\text{in}+\kappa_\text{ex}}{2}}. \label{eq:reflection_one_sided_cavity}
\end{align}




  \chapter{Simulation instructions}

\section{The electromechanical model}
\label{appendix:simulation-electromechanical}

\subsection{Generating the geometry}
\label{chp:Simulation_Appendix-generating_the_geometry}
We start with a `Blank Model'. The first step of the simulation is to generate the geometry of the design. The most organized strategy is to define all the geometrical parameters such as the thickness of the slab and the Aluminum electrodes, the length of the nano-strings, the gap, etc. in the `Parameters' window. You can open this window by clicking on the `Parameters 1' node in the `Global definition' node in the `Model builder' window. Enter the values as appear in the following figure.

\begin{figure}[h]
    \centering
    \includegraphics[width=\textwidth]{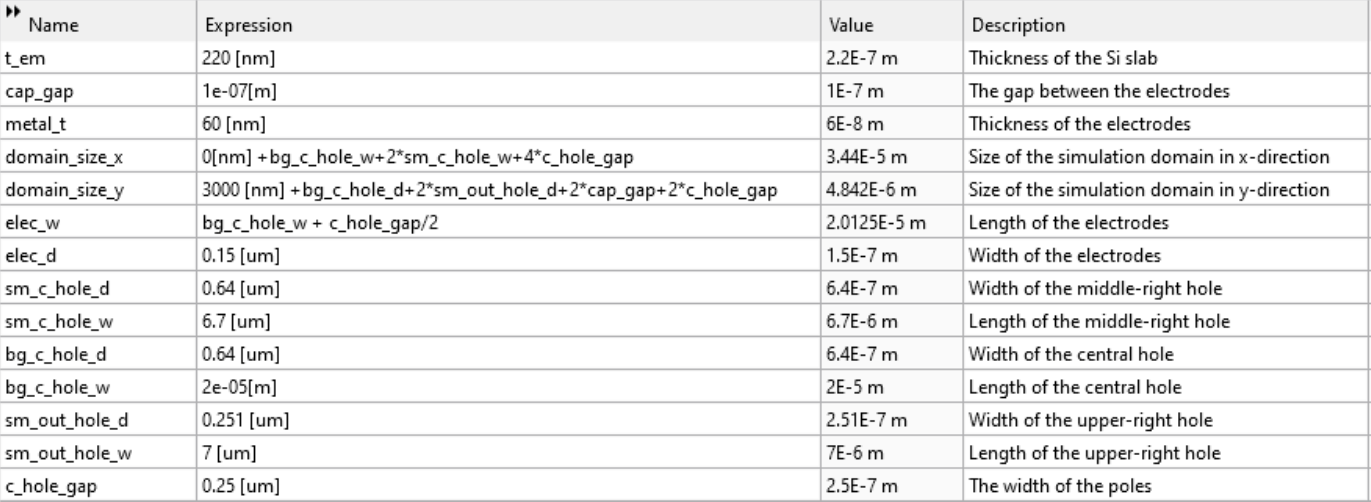}
    \caption[The parameter table in COMSOL]{The figure shows the table made in COMSOL to define parameters that will be used to generate the geometry.}
    \label{fig:COMSOL_paramter_table}
\end{figure}

\begin{figure}
    \centering
    \includegraphics[width=0.8\textwidth]{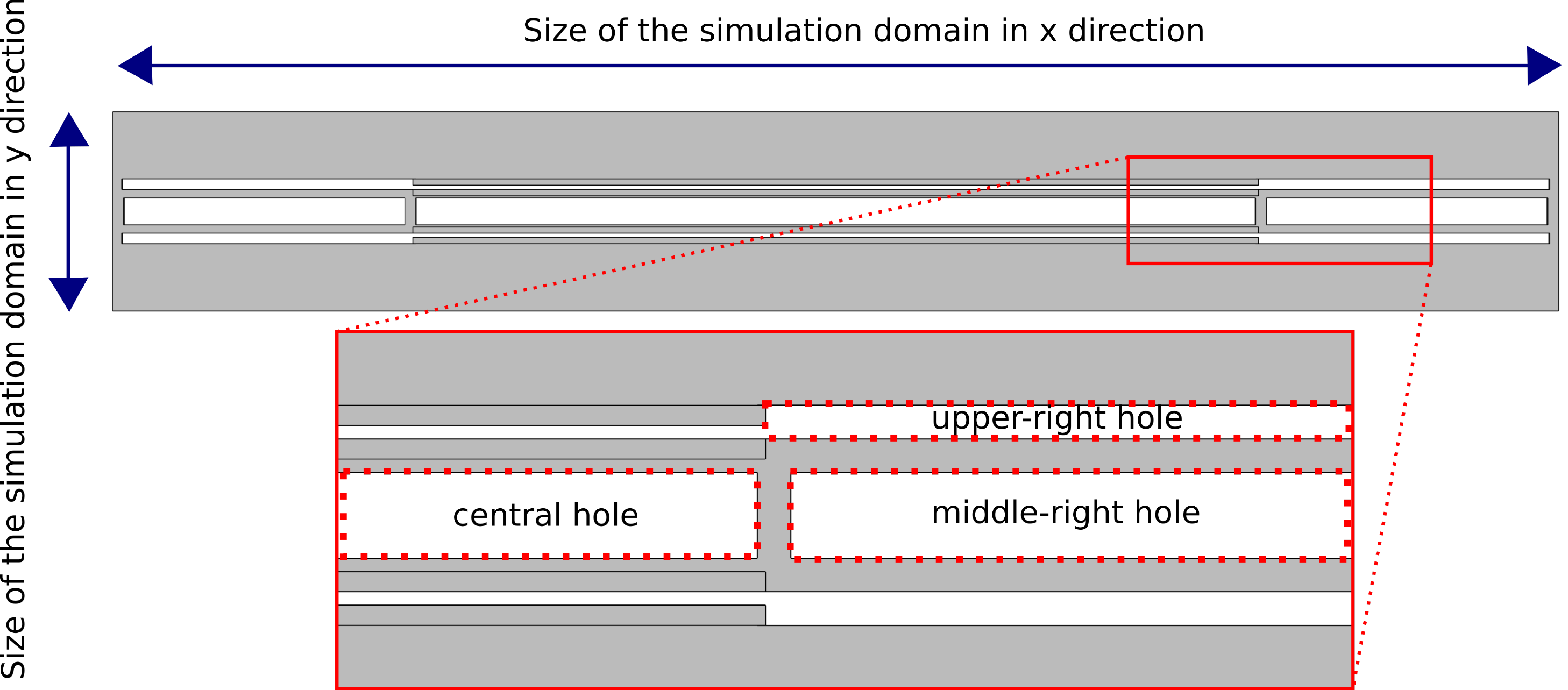}
    \caption[The cuts in the substrate]{To make the geometry, one can subtract nine rectangular shapes from the Silicon slab. The upper-right, upper-left, lower-right, and lower-left holes are identical in size. The central cut is the biggest hole. The middle-right and middle-left cuts are identical in size. The upper-middle and lower-middle cuts are identical and make the capacitors' gaps.}
    \label{fig:EM_topview9}
\end{figure}

After setting the parameters, we add a component by clicking on the `Add component' icon in the `Home' bar and choosing the `3D' option.
Then in the `Geometry 1' node, we make a block by right-clicking on the node. The width, depth and height of the block should be `domain\_size\_x', `domain\_size\_y', and `t\_em' respectively.
The block should be centred at the origin.
The next step is to remove some specific parts of the slab to form the structure.
To do so, we generate nine more blocks in the place of holes and then we subtract the blocks from the slab.
According to figure \ref{fig:EM_topview9} and using the defined parameters, one can generate the blocks correctly.
The four blocks on the four corners are identical in size and their dimensions are set as `sm\_out\_hole\_d' and `sm\_out\_hole\_w'.
The middle-right and middle-left blocks are identical in size and their dimensions are set as `sm\_c\_hole\_d' and `sm\_c\_hole\_w'.
The upper-middle and lower-middle blocks will make the gap, so their dimensions are `bg\_c\_hole\_w' and `cap\_gap', and finally, the central block should be made with the parameters `bg\_c\_hole\_w' and `bg\_c\_hole\_d'.
The position of each block can be determined using relevant parameters and after simple calculations.

After generating the nine blocks, we should add a `Difference' object, by right-clicking on the `Geometry 1' node, hovering `Boolean and Partitions' and then `Difference'. By selecting the `Objects to add' and `Objects to subtract' and 'Build All Objects' we will have all the holes carved in the slab.

The next step is to make the electrodes. Again, the electrodes are the same size, so we only use `elec\_w' and `elec\_d' and `metal\_t' parameters for their length, width, and thickness, and find and set their positions according to the design.
After making the electrodes, the geometry should look like figure \ref{fig:EM1}.

For simulating the electrical behaviour of the system, we need to define the space around the structure. Therefore, in addition to the slab and the electrodes, we define an airbox around the structure. The width and length of the airbox are obviously the same as the initial slab. Its thickness should be a few times larger than the slab thickness. In our simulation, we set it to `6*t\_em', but it can be larger. Figure \ref{fig:EM_with_airbox2} shows what the geometry looks like in transparent mode, at the end.

Up to this point, the structure has been made completely, but there is one more step to take in the geometry part. As we proceed to the next steps, we often need to select a group of domains or boundaries. To make a more organized simulation, we use the selection options in the geometry to make specific groups of domains or boundaries.
Selection objects can be made using `Explicit Selection', `Box Selection', `Adjacent Selection', `Difference Selection', `Intersection Selection', and `Union Selection' options. All the needed options can be found by hovering on `Selections' after right-clicking on the `Geometry 1' node.

The selections of domains that will be used and should be made are as follows.
\begin{itemize}
    \item The slab
    \item Four electrodes
    \item The airbox
    \item The inner electrodes
    \item The outer electrodes
    \item The slab and four electrodes altogether
\end{itemize}

The selection of boundaries that will be used and should be made are as follows.
\begin{itemize}
    \item The surfaces of the slab
    \item The surfaces of four electrodes
    \item The surfaces of the inner electrodes
    \item The surfaces of the outer electrodes
    \item The surfaces of the inner electrodes
    \item The airbox-slab intersection surfaces
    \item The airbox-electrodes intersection surfaces
    \item The slab-electrodes intersection surfaces
    \item The four side-surface of the slab
\end{itemize}

\subsection{Material, physics, mesh, and study setting}
The next steps after making the geometry start with setting the Materials. By choosing `Add Material from Library' after right-clicking on the `Materials' node, a new window opens where you can search for materials. Search for `Air', `Silicon', and `Aluminum [solid,bulk]', and add them to the model. The software automatically assigns the first added material to all the domains in the geometry. So we start with the second one, `Silicon', and in the `Selection' combo box we select the slab selection object that we have already made in the geometry step. A similar process should be done for assigning Aluminum to the electrodes selection object.

Now, we add the physics. In the `Physics' bar on top of the software's window click on the `Add Physics' icon. In the newly opened window, search for `Solid Mechanics' and `Electrostatics', and add them to the model. As soon as we add the physics, red x-marks appear on `Air' and `Aluminum' materials, because of missing physics-related parameters.
To solve the issues, we first exclude the airbox from the `Solid Mechanics' physics by clicking on the `Solid Mechanics' node and choosing the slab-and-electrodes selection object in the `Selection' combo box. We are interested in the mechanical modes of the slab and the electrodes only, and not the airbox around the structure.

The second material issue emerges because the relative permittivity of Aluminum is not defined. Although the electric field applied to the system, in reality, will be an AC field (microwave), to extract the electric and electromechanical characteristics of the system from the simulation, we apply a DC electric field in the simulation. Therefore, we should consider the electrostatic properties of the materials. We know that the relative permittivity of perfect conductors for static electric fields is infinity, so, we set the relative permittivity of Aluminum, which is a superconductor, to an extremely large number. To do so, we click on `Aluminum' and scroll to the `Material Contents' section. There is a row in the table for the `Relative permittivity' of the material, which appeared right after we added `Electrostatics' physics to the model. By entering a large number such as `1e12' in the `Value' column, the issue will be solved. 

To set the `Solid Mechanics' physics properly, the only change we need to make in its default settings is to fix the side surfaces of the slab, so only the nano-strings can oscillate. To fix the side surfaces, we add a `Fixed Constraint' by right-clicking on the `Solid Mechanics' node, and adding the constraint. Then click on `Fixed Constraint 1' and select the side-surfaces-of-the-slab boundary selection object. If you want to find only the in-plane mechanical modes, you can also set a `Symmetry' condition on the upper or lower surface of the slab. But we do not impose the in-plane constraint on the model.

Setting the `Electrostatic' physics is also easy. All the settings should be generated automatically when creating the physics, except the fixed voltages on the electrodes. In fabrication, we connect the inner electrodes to each other, and the outer ones to each other. Therefore, to find the capacitance of the capacitor and the electric field distribution in the case of a DC voltage, we apply $1$V to the outer electrodes and set the inner electrodes to the ground. To do so, we add a `Ground' boundary condition and select the boundaries of the inner electrodes, and add a `Terminal' boundary condition and select the boundaries of the outer electrodes. Note that when you right-click on the `Electrostatic' physics, you see two `Terminal' options. One is a domain condition and the other is a boundary condition. We should select the boundary condition option. Then, in the `Terminal' section of the `Terminal 1' window, change `Terminal Type' to `Voltage' and set `$V_0=1$V'.

\begin{figure}
    \centering
    \begin{subfigure}[b]{0.45\textwidth}
        \centering
        \includegraphics[width=\textwidth]{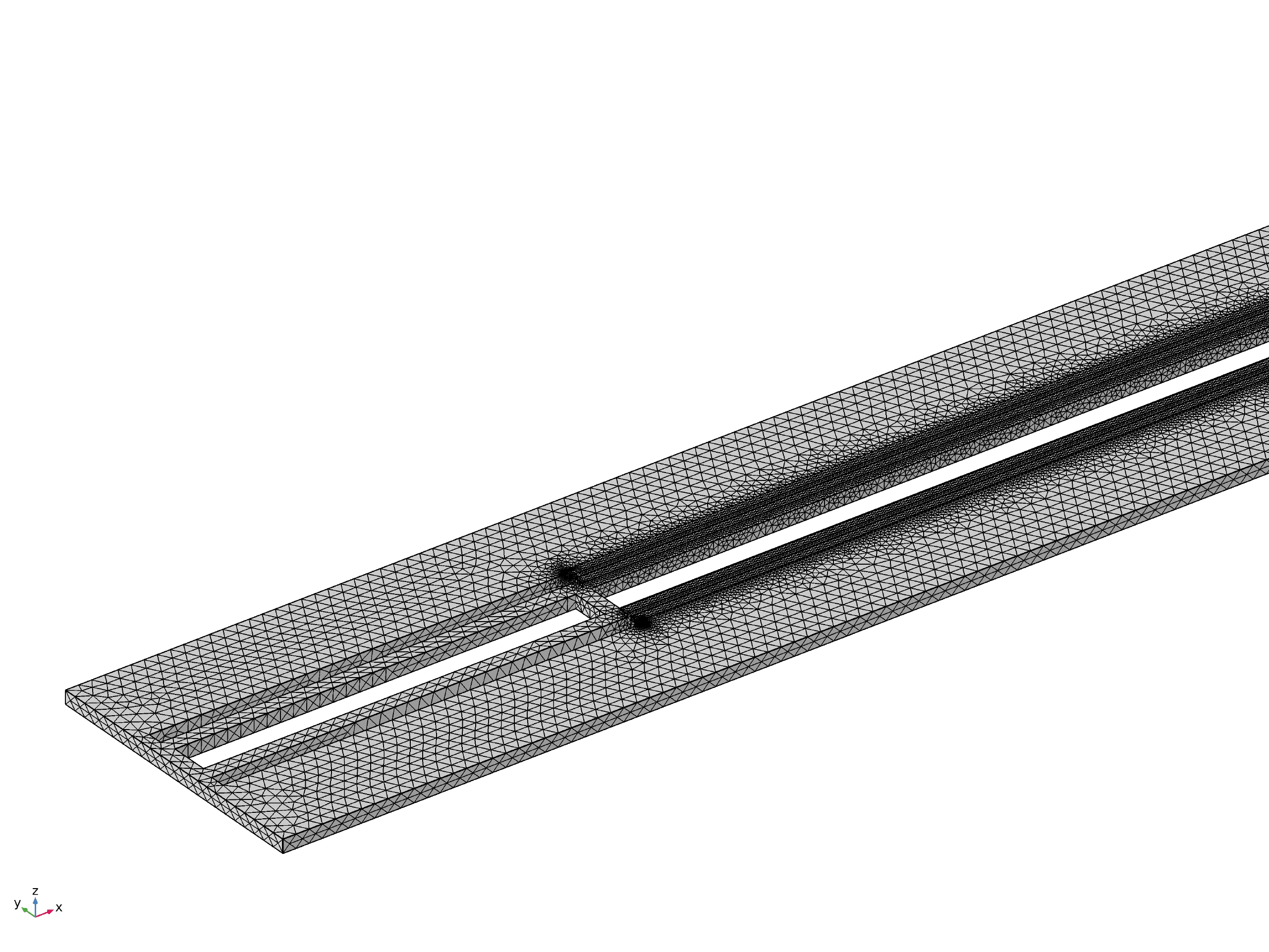}
    \end{subfigure}
    \begin{subfigure}[b]{0.45\textwidth}
        \centering
        \includegraphics[width=\textwidth]{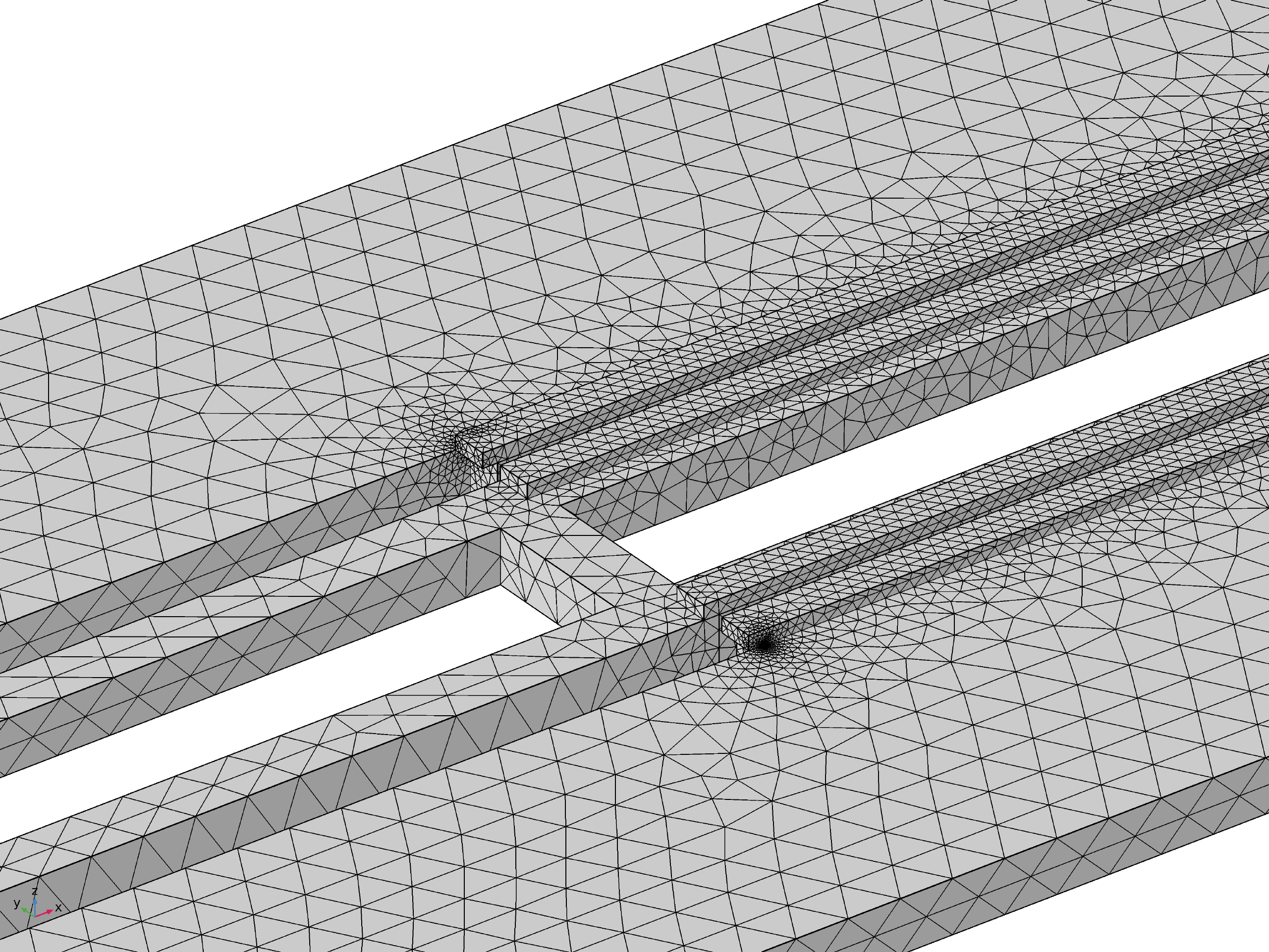}
    \end{subfigure}
    \caption[The final mesh of the geometry]{The images show the final mesh generated on the entire geometry.}
    \label{fig:EM_final_mesh}
\end{figure}

After setting the physics, we have to set the mesh. The purpose of setting the mesh is to make a fine enough mesh to have precise results while keeping it not too fine so the calculation takes place in a reasonable amount of time and with accessible RAM. To start making the mesh, in the `Mesh' bar on top of the screen, click on `Add Mesh'. Change the `Sequence Type' to the `User-controlled mesh' option. In the `Size' sub-node, click on the `Custom' radio button and set `Maximum element size', `Maximum growth rate', `Curvature factor', and `Resolution of narrow regions' to `2E-6 m', `1.3', `0.2', and `1', respectively. Then, set the `Minimum element size' to `min(5E-9, cap\_gap/20)'. The mentioned factors set general limits for all the mesh of the domain in the geometry.\\
The gap between the plates of the capacitors is much smaller than the other parts of the structure. To find the electric field distribution within the gaps with acceptable precision, we need to keep the mesh in the gap regions fine. However, we should note that we do not need a fine mesh for the rest of the air box domain. To impose a finer mesh to only the gap regions, we set a maximum limit to the mesh size of the electrodes, and it automatically imposes the mesh size in the regions around the electrodes, including the gaps. To do so, add one `Size' to the `Mesh 1' node, select the electrode domains, set the `Element size to custom', and set only `Maximum element size' manually to `cap\_gap/2'. This guarantees that the mesh size of the electrodes, and consequently, the gap regions will not be greater than half of the gap size, therefore, there will be at least two mesh elements across each gap. Of course, a finer gap yields more precise results, but at the same time increases the total number of mesh elements in the model, and at some point, the computational resources do not suffice. Our previous trials showed that having two mesh elements across the gap is enough to have reliable results.\\
The `Free Tetrahedral 1' that should be already generated by default will generate the mesh elements over all the domains of the geometry, based on the sizes we have set. As we care about the mesh in the vicinity of the electrodes more than any other regions, it is better to add one more `Free Tetrahedral', select the electrode domains, and make the mesh elements of the electrodes before the rest of the geometry. This last step is for making sure that no constraints will be imposed on the meshing of the electrodes from the other regions. To finalize the mesh, click on `Mesh1', and then `Build All'. Figure \ref{fig:EM_final_mesh} shows how the final mesh should appear.

After setting the mesh, the studies should be added. First, we add a `Stationary' study by clicking on `Add study' in the `Study' bar on top of the screen. After the study is added, `Solid Mechanics (solid)' should be unchecked from the `Physics and Variable Selection' section. We want to find the stationary solution for the Electrostatic physics only. Then, we click on `Compute' on top of the `Study' window, and COMSOL calculates the solution.\\
Then, we add an `Eigenfrequency' study in a similar way. In the `Study Setting' section of the recently added study, the search method should be set to `Manual'. The desired number of eigenfrequencies can be set arbitrarily. The field for `Search for eigenfrequencies around' indicates the initial guess for finding the solutions. The closer the initial guess is to the final solutions, the faster and better the simulation works. We know that the first in-plane mechanical frequency should be around $4$MHz, therefore, we enter $4$MHz in that field. `Eigenfrequency search method around shift' should be set to `Closest in absolute value'. In the next section, `Physics and variables Selection', `Electrostatic (es)' should be unchecked as we are looking for the mechanical modes. Again, by clicking on `Compute' on top of the study window, COMSOL finds the mechanical modes.

\subsection{Post-processing}
After running studies, the solutions will appear in the `Results' node. Up to this point, we have $\Omega$ for any desired mechanical mode without any post-processing. However, to find the rest of the parameters, post-processing is required.\\
The results of Stationary and Eigenfrequency simulations are saved in separate Datasets. As we will see shortly, we can use the separate Datasets directly to find $C_m$, $m_\text{eff}$, and $x_\text{ZPF}$, and $\int\varepsilon(\vec{r})\left\vert\vec{E}(\vec{r})\right\vert^2d^3\vec{r}$. But to calculate the integrals in Equation (\ref{eq:g_mb_electrical}) and find $g_0$, we need to make new Datasets that contain solutions from both Electrostatic and Mechanical simulations.\\
The integrals in the numerator of Equation (\ref{eq:g_mb_electrical}) are all surface integrals. To calculate them, we first need to extract the Electrostatic and Mechanical solutions only on the relevant boundaries (They originally present the results for the entire 3D geometry). To do so, we generate new Datasets, by right-clicking on the `Datasets' node and then choosing the option `Solution'. Then we click on the newly generated Solution, and in the Solution window, select the correct solution. For mechanical solutions, the Frame should be set on `Material (X, Y, Z)', and for the electrostatic solution Frame should be `Spatial (x, y, z)'. Then we right-click on the added solution sub-node, and click on `Selection'. In the newly appeared `Selection' sub-node, choose one of the three boundary selection objects made for intersections in Section (\ref{chp:Simulation_Appendix-generating_the_geometry}) (The airbox-slab intersection surfaces, The airbox-electrodes intersection surfaces, and The slab-electrodes intersection surfaces). So, six new `Solution's should be added in total. It is recommended to change the name of each `Selection' from the default name to a name that shows the corresponding Physics and Surface clearly.\\
After making the six needed solutions, we make three `Join' sub-nodes by right-clicking on the `Datasets' node. Then, in each `Join' sub-node, select Mechanical and Electrostatic solutions for the same surface selection, for `Data 1' and `Data 2', respectively. The method should be set to `Explicit'. After all three `Join's are made and set, we have the combined Datasets and we can proceed to set the post-processing.

We first add a `Volume Maximum' in the `Derived Values' node, and select the main mechanical Dataset. Change the label to `max\_disp', select `All domains', and enter `solid.disp' in the first row of the `Expressions' table. This finds the maximum displacement of the structure corresponding to mechanical modes and saves the number in an automatically generated table. We will need this value to normalize the mechanical deformations.\\
Then, we add a `Volume Integration', change the Label to `m\_eff', select `All domains', and enter `solid.rho*solid.disp\^{}2' in the `Expressions' table. According to Equation (\ref{eq:effective_mass}), this evaluation leads to the effective mass corresponding to the mechanical mode.\\
Now, we add another `Volume integral' for evaluating $\int\varepsilon(\vec{r})\left\vert\vec{E}(\vec{r})\right\vert^2d^3\vec{r}$. So, the Dataset should be set to the main Electrostatic solution. `All domains' should be selected, and in the `Expressions' table, either `2*es.W' or `es.normE*es.normD' can be entered. Both expressions give twice the electric energy stored in the structure, and are equal to the integral that we want to calculate. Note that this expression also equals the capacitance of the system, based on Equation (\ref{eq:electric_energy_of_a_capacitor}), and the fact that $V_C=1$.\\
In the next step, we add three `Surface Integration's. Each integration corresponds to one of the three joined Datasets. Therefore, we select the correct Dataset, and enter the following expression in the first row of the `Expressions' table.
\begin{lstlisting}[breaklines]
    data1(u*nx+v*ny+w*nz) * ( (mat1_repsilon-mat2_repsilon)*epsilon0_const * data2(abs(ny*es.Ez-nz*es.Ey)^2 + abs(nz*es.Ex-nx*es.Ez)^2 + abs(nx*es.Ey-ny*es.Ex)^2) - (1/mat1_repsilon-1/mat2_repsilon)/epsilon0_const * data2(abs(nx*es.Dx)^2 + abs(ny*es.Dy)^2 + abs(nz*es.Dz)^2) )
\end{lstlisting}
\lstinline{data1} and \lstinline{data2} indicate the first (Mechanical) and the second (Electrostatic) Datasets specified while joining. \lstinline{u}, \lstinline{v}, and \lstinline{w} represent spatial components of $\vec{Q}(\vec{r})$, \lstinline{nx}, \lstinline{ny}, and \lstinline{nz} are the components of the normal vector to the surface of integration. \lstinline{mat1_repsilon} and \lstinline{mat2_repsilon} are the relative electric permittivities of the materials involved in the integration. \lstinline{es.Ex}, \lstinline{es.Ey}, and \lstinline{es.Ez} are the components of the electric field, and \lstinline{es.Dx}, \lstinline{es.Dy}, and \lstinline{es.Dz} are the components of the electric displacement field. The recent expression represents the integrands of the numerator of Equation (\ref{eq:g_mb_electrical}).

After making all the required evaluations in the `Derived Values' node, we right-click on the node, and click on `Evaluate All'. The expressions will be calculated on the desired domain, and COMSOL generates tables and saves the evaluation results in them. One can use the results stored in the tables to calculate the physical parameters manually, however, it is also possible to make MATLAB scripts that extract the numbers and calculate the parameters automatically. We will see how we can use LiveLink with MATLAB to generate MATLAB scripts that automize the last step of post-processing.

\section{LiveLink for MATLAB}
\label{chp:LiveLink_instruction}
LiveLink for MATLAB \cite{COMSOLLiveLink} provides a link between COMSOL simulations and MATLAB scripts. In fact, we can make a MATLAB script that generates a COMSOL model and perform the simulation, without using the Graphical User Interface (GUI). Then, by calling the simulation script in other MATLAB scripts, we can extract the numerical results and complete the post-processing.\\
There is a detailed User's Guide \cite{LiveLinkInstruction} on how to use LiveLink for MATLAB to make COMSOL simulations, but making a COMSOL model from scratch using MATLAB is a tedious task. Instead, we can use COMSOL GUI to make a model, and then generate the MATLAB code, and then change the details as desired.\\
To generate a MATLAB script (.m file) from a COMSOL model, after making the model using the GUI and running the simulation, we can save the file as a `Model File for MATLAB (*.m)'. The generated .m file can be opened using MATLAB, but to run the simulation through MATLAB, one needs to run the `COMSOL Multiphysics with MATLAB' software first, and then open the .m file through the newly opened MATLAB window.\\
An important point to make is that every action we take in COMSOL GUI will be saved in the script, even if we overwrite the action and change the details later. So, for example, if one enters an expression with a syntax error in a text field, and then fixes it, what appears in the script is the definition of the incorrect expression first, and then it will be replaced with the correct expression. In the process of running the .m file script, MATLAB raises an error on the line with the syntax error and will not proceed. To overcome such issues, we need to clear the history of the model and save only the latest form of the model into a MATLAB script. To do so, before saving it as a .m file, we click on `Compact History' in the `File' tab.

The user's guide is a great resource for understanding the model script, but most of the syntax is understandable. With a bit of reviewing the script, one can understand its structure and then change the details so the script generates the exact COMSOL model of interest. The only alterations that scripts need are transforming the script into a MATLAB function by adding the function definition line in the beginning and returning the object \lstinline{model} at the end of the script. The following piece of code can be also added at the beginning of the script, optionally, so the progress bar will be shown while running the simulation.
\begin{lstlisting}
    ModelUtil.showProgress(true);
\end{lstlisting}
The following pseudo-code shows how a script should look in general.
\begin{center}
\noindent\fbox{
    \begin{minipage}{0.98\textwidth}
        \texttt{\textcolor{blue}{function} out = model\_builder($\cdots$ \textit{inputs} $\cdots$)}\\
        \null\qquad \texttt{import \textcolor{Rhodamine}{com.comsol.model.*}}\\
        \null\qquad \texttt{import \textcolor{Rhodamine}{com.comsol.umodel.util.*}}\\
        \null\qquad \texttt{model = ModelUtil.create(\textcolor{Rhodamine}{'Model'});}\\
        \null\qquad \texttt{ModelUtil.showProgress(true);}\\ \\
        \null\qquad \{A few blocks of code to make the parameters, tables, geometry, and\\
        \null\qquad\qquad selections. Any desired change based on the \textit{inputs} can be carefully made here.\}\\
        \null\qquad\texttt{model.component(component\_label).geom(geometry\_label).run;}\\ \\
        \null\qquad \{A few blocks of code to make the physics, mesh, evaluation expressions,\\
        \null\qquad\qquad materials, and studies.\}\\
        \null\qquad\texttt{model.sol(solution\_label).runAll;}\\ \\       
        \null\qquad \{A few blocks of code to make the Datasets and joined Datasets.\}\\
        \texttt{out = model;}
    \end{minipage}}
\end{center}

Having the \texttt{model\_builder} function defined in a .m file, we can call it in any other MATLAB script and receive the \texttt{model} object it generates to extract the parameters. Furthermore, using the command,
\texttt{mphsave(model, name\_of\_the\_file);}, we can save the COMSOL model returned by \texttt{model\_builder} function in mph format, so we can open it with the GUI later.\\
To access the numbers saved in tables, we first need to find the labels of the tables, and then we can extract the values using the command \texttt{mphtable(model, label\_of\_the\_table);}. The following pseudo-code is an example of how we can call the \texttt{model\_builder} function, extract the numbers, and perform post-processing.

\begin{center}
\noindent\fbox{
    \begin{minipage}{0.98\textwidth}
    \{Defining parameters that are going to be used as the \textit{inputs} of the \texttt{model\_builder} function.\}\\
    \texttt{model = model\_builder($\cdots$\textit{inputs}$\cdots$);}\\
    \texttt{mphsave(model, \textcolor{Rhodamine}{name\_of\_the\_file});}\\\\
    \texttt{table1 = mphtable(model, \textcolor{Rhodamine}{table1\_label});}\\
    \texttt{param1 = table1.data[$\cdots$\textit{indices}$\cdots$];}\\\\
    \texttt{table2 = mphtable(model, \textcolor{Rhodamine}{table2\_label});}\\
    \texttt{param2 = table2.data[$\cdots$\textit{indices}$\cdots$];}\\\\
    \texttt{table3 = mphtable(model, \textcolor{Rhodamine}{table3\_label});}\\
    \texttt{param3 = table3.data[$\cdots$\textit{indices}$\cdots$];}\\
    \null\qquad$\vdots$\\\\
    \{A block of code performing the post-processing mathematically using \texttt{param1}, \texttt{param2}, \texttt{param3}, etc. variables.\} \\
    \end{minipage}}
\end{center}

The presented pseudo-code is a simple example. More complicated tasks such as optimization can be also done as long as the optimization algorithm is written in MATLAB.
  \chapter{Fabrication recipe}

\section{E-beam lithography}
Cleaning the chips before spin-bake is quite important. If possible, Piranha cleaning is usually the best option. If the chips contain metal already and cannot be Piranha cleaned, Acetone and IPA are the next options. After cleaning and before starting the spin-bake process, it is good to make sure there is no contaminant on the chips, using an optical microscope.\\
The numbers that will be reported in the following can vary in different situations. For the best results, it is important to run a complete dose test to find the best set of parameters.\\
Before the `Resist descum' step, the plasma chamber should be cleaned for at least $30$ minutes and conditioned for at least $20$ minutes.
\begin{enumerate}
    \item Spin-baking the chip
    \begin{itemize}
        \item Prebake chips / Temperature $\ge 100^\circ$C / Duration $>5$ minutes.
        \item Cover chips with resist / CSAR60 / Ramp $3\text{K}\frac{\text{rpm}}{\text{s}}$ / Velocity $6$K rpm / Duration $60$ seconds
        \item Bake the chips / Temperature $160^\circ$C / Duration $5$ minutes
    \end{itemize}
    \item E-Beam exposure
    \begin{itemize}
        \item Voltage $15$kV / Aperture $10\mu$m / Base dose $50\frac{\mu\text{C}}{\text{cm}^{-2}}$
        \item Dose factor $0.7$ for smaller features (holes, nano-strings, electrodes, etc.)
        \item Dose factor $0.5$ for larger features (markers, square cuts, etc.)
    \end{itemize}
    \item Develop
    \begin{itemize}
        \item Submerge in AR600-549 / Room temperature / Duration $85$ seconds
        \item Submerge in IPA (stopper) / Room temperature / Duration $30$ seconds
        \item Rinse with DI water / Dry with Nitrogen
    \end{itemize}
    \item Resist descum
    \begin{itemize}
        \item O2 flow $98.0$sccm / Duration $10$ seconds 
    \end{itemize}
\end{enumerate}

\section{Optical lithography}
\begin{enumerate}
    \item Hexamethyldisilazane (HMDS) treatment ($20 $ minutes in total)
    \begin{itemize}
        \item Prebake chips / Temperature $\ge 100^\circ$C / Duration $>5$ minutes.
        \item Cover chips with resist / AZ 1529 / Velocity $3000$ rpm / Duration $40$ seconds
        \item Bake the chips / Temperature $100^\circ$C / Duration $1$ minute
    \end{itemize}
    \item Optical exposure
    \begin{itemize}
        \item Rectangular optical alignment
        \item Dose 330
        \item Defoc -2
    \end{itemize}
    \item Develop
    \begin{itemize}
        \item Submerge in AZ 400K / Room temperature / Duration $2$ minutes
        \item Submerge in DI water / Duration $15$ seconds
        \item Dry carefully with Nitrogen (The resist can be splashed if N2 is blown strongly)
    \end{itemize}
\end{enumerate}

\section{Silicon etching}
$1$ hour chamber cleaning and $30$ minutes chamber conditioning should be done before inserting the chips in the chamber and starting the etching process.\\
It is always good to use a microscope (SEM or Optical, depending on the features' size) to make sure the Silicon layer is etched completely through, before cleaning the resist. If the layer is not etched, a few more seconds of etching should be done.
\begin{enumerate}
    \item ICP-RIE
    \begin{itemize}
        \item C4F8 500 flow $14$sccm
        \item SF6 1000 flow $14$sccm
        \item Table temperature $15^\circ$C
        \item Duration $70$ seconds
    \end{itemize}
    \item Resist cleaning
    \begin{itemize}
        \item UV exposure for $6$ minutes
        \item Submerge in Remover PG for $2$ minutes
        \item Submerge in IPA for $1$ minute
        \item Rinse with DI water and Dry with Nitrogen
    \end{itemize}
\end{enumerate}

\section{Aluminum deposition}
Titanium getter can help to accelerate decreasing the pressure of the chamber.\\
The rate has been chosen for a $60nm$ thick layer of Aluminum ($5$ minutes). For a thicker layer, one might need a higher rate, as a considerably longer total deposition time might burn the resist and fail lift-off.
\begin{enumerate}
\item PVD (E-beam Aluminum evaporation)
\begin{itemize}
    \item Base pressure $\sim 10^{-7}$Torr (The lower pressure, the smoother metal surface)
    \item Deposition rate $2\text{\AA}/\text{s}$
\end{itemize}
\item Lift-off
\begin{itemize}
    \item Submerge chips in heated Remover PG ($\sim 90^\circ$C) / Duration $>1.5$ hour
    \item After being submerged for a long enough time, blow bubbles to the surface of the chip with a pipette.
\end{itemize}
\end{enumerate}

\section{Release}
\begin{enumerate}
    \item Etching step 1 / HF flow $40$sccm / Vaporiser N2 flow $20$sccm / H$_2$O $5$mgPm / Pressure $15$Torr / Temperature $25^\circ$C / Duration $4800$ seconds
    \item Etching step 2 / HF flow $40$sccm / Vaporiser N2 flow $20$sccm / H$_2$0 $5$mgPm / Pressure $17$Torr / Temperature $25^\circ$C / Duration $4800$ seconds
\end{enumerate}

\end{document}